\documentclass[floatfix,amsmath,amssymb,superscriptaddress,nofootinbib,longbibliography,twocolumn, prb]{revtex4-2}

\usepackage{lipsum}
\usepackage{verbatim}
\usepackage[colorlinks=true,linkcolor=blue,citecolor=blue]{hyperref}
\usepackage[utf8]{inputenc}
\usepackage[T1]{fontenc}

\usepackage{array}
\usepackage{mathtools}
\usepackage{bm}

\usepackage{physics}
\usepackage{algorithmic}
\usepackage{algorithm}
\usepackage{siunitx}

\usepackage{graphicx}
\usepackage{tikz}

\usepackage{tabularx}
\usepackage{booktabs}
\newcolumntype{L}[1]{>{\raggedright\let\newline\\\arraybackslash\hspace{0pt}}m{#1}}
\newcolumntype{C}[1]{>{\centering\let\newline\\\arraybackslash\hspace{0pt}}m{#1}}
\newcolumntype{R}[1]{>{\raggedleft\let\newline\\\arraybackslash\hspace{0pt}}m{#1}}
\newcolumntype{x}[1]{>{\centering\arraybackslash\hspace{0pt}}p{#1}}

\usepackage[caption=false]{subfig}
\newcommand{\phantomsubfloat}[1]{
    {%
        \captionsetup[subfigure]{labelformat=empty}
        \subfloat[][]{#1}
    }%
}

\usepackage{cleveref}
\crefname{equation}{Eq.}{Eqs.}
\crefname{section}{Sec.}{Secs.}
\crefname{subsection}{Sec.}{Secs.}
\crefname{appendix}{Appendix}{Appendices}
\crefname{figure}{Fig.}{Figs.}
\crefname{table}{Table}{Tables}
\creflabelformat{equation}{#2\textup{#1}#3}

\DeclareMathOperator{\diag}{diag}
\DeclareMathOperator{\sgn}{sgn}
\newcommand{\normord}[1]{:\mathrel{\mkern2mu #1 \mkern2mu}:}
\newcommand{\trans}[1]{#1^{\mathsf{T}}}

\usepackage{orcidlink}

\usepackage{pifont}
\newcommand{\cmark}{\ding{51}}
\newcommand{\xmark}{\ding{55}}

\newcommand\ucdot{\ensuremath{{}\cdot{}}}
\DeclareSIUnit\unitcellarea{A_{uc}}

\usepackage{xspace}
\newcommand{\hBN}{\textit{h}-BN\xspace}
\newcommand{\ie}{\textit{i}.\textit{e}.}
\newcommand{\eg}{\textit{e}.\textit{g}.}
\newcommand{\etc}{\textit{etc}.}

\newcommand{\icontext}[1]{\hspace{-3pt}\smash{\raisebox{-4.5pt}{\includegraphics[height = 13pt]{#1}}}\hspace{-1pt}}

\begin{document}

\title{Correlated Phases in Spin-Orbit-Coupled  Rhombohedral Trilayer Graphene}

\author{Jin Ming Koh\,\orcidlink{0000-0002-6130-5591}}
\affiliation{Department of Physics, California Institute of Technology, Pasadena, California 91125, USA}

\author{Jason Alicea\,\orcidlink{0000-0001-9979-3423}}
\affiliation{Department of Physics, California Institute of Technology, Pasadena, California 91125, USA}
\affiliation{Institute for Quantum Information and Matter, California Institute of Technology, Pasadena, California 91125, USA}

\author{\'Etienne Lantagne-Hurtubise\,\orcidlink{0000-0003-0417-64521}}
\affiliation{Department of Physics, California Institute of Technology, Pasadena, California 91125, USA}
\affiliation{Institute for Quantum Information and Matter, California Institute of Technology, Pasadena, California 91125, USA}

\begin{abstract}
Recent experiments indicate that crystalline graphene multilayers exhibit much of the richness of their twisted counterparts, including cascades of symmetry-broken states and unconventional superconductivity. Interfacing Bernal bilayer graphene with a WSe$_2$ monolayer was shown to dramatically enhance superconductivity—suggesting that proximity-induced spin-orbit coupling plays a key role in promoting Cooper pairing. Motivated by this observation, we study the phase diagram of spin-orbit-coupled rhombohedral trilayer graphene via self-consistent Hartree-Fock simulations, elucidating the interplay between displacement field effects, long-range Coulomb repulsion, short-range (Hund's) interactions, and substrate-induced Ising spin-orbit coupling. In addition to generalized Stoner ferromagnets, we find various flavors of intervalley coherent ground states distinguished by their transformation properties under electronic time reversal, $\text{C}_3$ rotations, and an effective anti-unitary symmetry. We pay particular attention to broken-symmetry phases that yield Fermi surfaces compatible with zero-momentum Cooper pairing, identifying promising candidate orders that may support spin-orbit-enhanced superconductivity.
\end{abstract}

\maketitle
\date{\today}

\section{Introduction}

\begin{figure} 
    \centering
    \includegraphics[width = 1\linewidth]{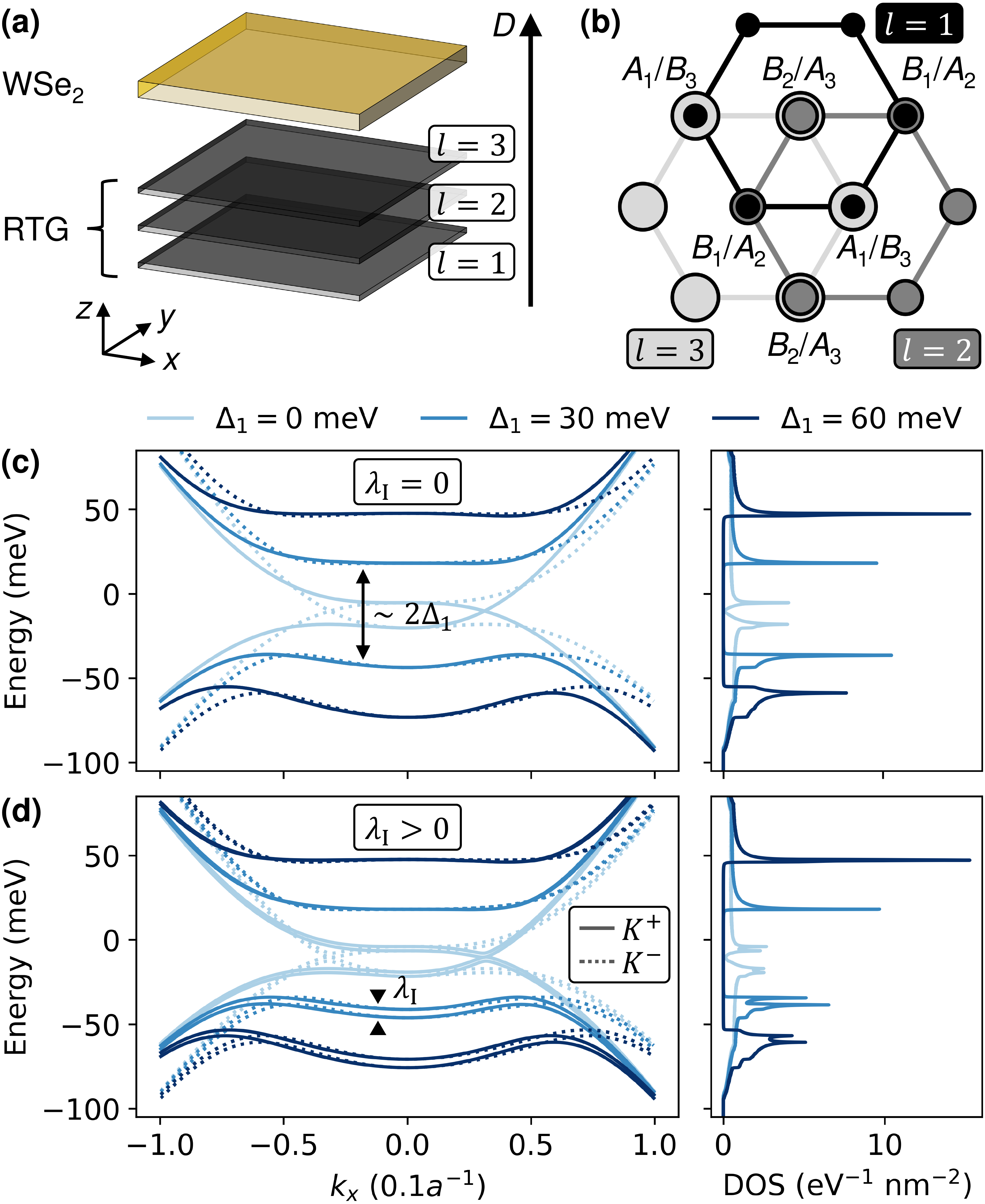}
    \phantomsubfloat{\label{fig:schematics-stack}}
    \phantomsubfloat{\label{fig:schematics-stacking}}
    \phantomsubfloat{\label{fig:schematics-band-structure-wo-soc}}
    \phantomsubfloat{\label{fig:schematics-band-structure-w-soc}}
    \vspace{-1.6\baselineskip}
    \caption{\textbf{Schematic of RTG/WSe$_2$ system.} \textbf{(a)} A WSe$_2$ monolayer placed in proximity to RTG induces spin-orbit coupling (SOC) on the meV scale. A perpendicular displacement field $D$ generates a potential difference $\Delta_1$ between adjacent layers through \cref{eq:Delta1-def}. \textbf{(b)} Top-down view of RTG showing its stacking configuration, with sublattices labeled $A_l$, $B_l$ for layer $l$. \textbf{(c--d)} Low-energy band structure for $k_y = 0$ (left panels) and corresponding DOS (right panels) for different $\Delta_1$, in the absence \textbf{(c)} and presence \textbf{(d)} of Ising SOC of strength $\lambda_{\text{I}}$. The $D$ field gaps out the valence/conduction band touchings and enhances divergences in the DOS (\ie~van Hove singularities) that facilitate strong interaction effects. We used an exaggerated $\lambda_{\text{I}} = \SI{10}{\milli\electronvolt}$ in \textbf{(d)} for visual clarity. The SOC-induced spin splitting appears mostly in the valence bands because the corresponding wavefunctions are pushed towards WSe$_2$ at $D > 0$; the spin splitting moves to the conduction bands for the opposite direction of the field, $D < 0$.}
    \label{fig:schematic}
\end{figure}

Rhombohedral graphene multilayers---for which graphene sheets are stacked in an `ABC' pattern---provide an attractive playground to study electronic correlations in ultraclean crystalline environments largely free of inhomogeneities present in bulk materials and twisted superlattices.  The low-energy physics of rhombohedral graphene multilayers can be tuned by applying a perpendicular displacement field $D$, which opens a spectral gap at charge neutrality and locally flattens the bands near the Brillouin zone corners (see \cref{fig:schematic}). The correspondingly enhanced density of states near the conduction and valence band edges suggests a nontrivial interplay between band structure and interaction effects for lightly doped systems.  Indeed, experiments on rhombohedral graphene multilayers have uncovered rich phase diagrams, comprising a wealth of symmetry-broken correlated insulating and metallic phases as well as unconventional superconductivity~\cite{Weitz2010, Shi2020, zhou2021half, zhou2021superconductivity, Kerelsky2021, zhou2022isospin, Seiler2022, delaBarrera2022, Zhang2023, Holleis2023, Han2023, liu2023interactiondriven, han2023orbital, lu2023fractional}.

Striking behavior arises already in AB-stacked Bernal bilayer graphene (BLG): weak in-plane magnetic fields stabilize superconductivity (albeit with a low critical temperature $T_{\text{c}} \sim \SI{30}{\milli\kelvin}$) near the phase boundary to a symmetry-broken metal~\cite{zhou2022isospin}. Remarkably, the observed superconducting state is likely spin-triplet in character and resides deep in the clean limit, with mean-free paths far exceeding the superconducting coherence length---a clear testament to the exceptional sample purity.  Moreover, pairing is dramatically enhanced~\cite{Zhang2023, Holleis2023} when BLG sits proximate to monolayer tungsten diselenide (WSe$_2$), which imparts $\si{\milli\electronvolt}$-scale spin-orbit coupling (SOC) into the graphene sheets. Specifically, superconductivity in BLG/WSe$_2$ sets in even at zero magnetic field, exhibits an order-of-magnitude larger $T_{\text{c}}$, and descends from a parent symmetry-broken normal state over a broad density range (as opposed to being confined to the vicinity of a phase transition).  These discoveries have spurred intense theoretical efforts aimed at understanding the origin of unconventional superconductivity in `pure' BLG~\cite{Chou2022, Dong2023, dong2023signatures, Shavit2023, Wagner2023, Jimeno-Pozo2022, li2023charge, Dong2023superconductivity, szabo2022competing}  as well as the influence of WSe$_2$ on its phase diagram, \eg~due to induced SOC~\cite{Zhang2023, Curtis2023, Wagner2023, Jimeno-Pozo2022, Ming2023,Shavit2023,li2023charge} or virtual tunneling events~\cite{Yangzhi2022}. 

Moving up one layer, rhombohedral trilayer graphene (RTG) also hosts a family of symmetry-broken correlated metallic phases as well as superconductivity~\cite{zhou2021half,zhou2021superconductivity}---though the latter requires neither magnetic fields nor SOC, in contrast to BLG. Two distinct superconducting regions are observed: the first (SC1) has $T_{\text{c}} \sim \SI{105}{\milli\kelvin}$ and is consistent with spin-singlet pairing, while the second (SC2) has weaker $T_{\text{c}} \sim \SI{30}{\milli\kelvin}$ and is likely of spin-triplet character. A flurry of theoretical activity has proposed pairing mechanisms for RTG~\cite{chou2021acoustic, Cea2022, You2022, Jimeno-Pozo2022,  Berg2021, dong2021superconductivity,  Chatterjee2022, Szab2022, Qin2023, Lu2022, li2023charge, Dai2023,  Dong2023superconductivity} including acoustic phonons~\cite{chou2021acoustic}, over-screened Coulomb interactions (\ie~Kohn-Luttinger physics)~\cite{Berg2021, Cea2022, You2022, Jimeno-Pozo2022, li2023charge}, and order parameter fluctuations~\cite{dong2021superconductivity, Chatterjee2022, Dong2023superconductivity}. Experiments on RTG/WSe$_2$ have not yet been reported but are extremely interesting to consider in light of the dramatic influence of WSe$_2$ on BLG phenomenology. For instance, can WSe$_2$ qualitatively alter the symmetry-broken metallic phases observed in RTG? And can WSe$_2$ similarly enhance RTG superconductivity? These questions are intimately related, since the band structure and symmetries of correlated normal states influence not only the nesting condition for forming Cooper pairs, but also their resilience against order parameter fluctuations and disorder.  

Motivated by the preceding questions, we investigate the phase diagram of RTG both with and without an adjacent WSe$_2$ layer using self-consistent Hartree-Fock techniques.  Our calculations incorporate realistic RTG band structure, a displacement field, screened long-range Coulomb interactions, short-range `Hund's coupling' (named in analogy to Hund's rules in atomic physics 
due to its tendency to align spins in the two valleys), and Ising-type SOC induced by WSe$_2$ (or some other transition metal dichalcogenide).  We consider a large family of candidate symmetry-broken orders and pay special attention to the role played by SOC in stabilizing correlated states conducive to Cooper pairing. Aside from generalized Stoner ferromagnets, wherein a subset of the four spin and valley flavors are spontaneously polarized, our analysis also captures `intervalley coherent' (IVC) metallic states that spontaneously hybridize the two valleys of graphene---thus breaking translation symmetry on the atomic scale. Indeed, recent STM experiments have directly imaged the atomic-scale reconstruction characteristic of IVC states in monolayer graphene in the quantum Hall regime~\cite{Liu2022, Coissard2022} and in twisted graphene superlattices~\cite{Nuckolls2023, Kim2023}. From the viewpoint of superconducting instabilities, IVC states are interesting because they can be compatible with zero-momentum Cooper pairing depending on the symmetries they preserve---in contrast to, \eg~valley-polarized states. We find several IVC states distinguished by their spin structure as well as symmetry properties.  In particular, experimentally relevant ferromagnetic Hund's coupling favors spin-polarized and spin-triplet IVC states, whereas Ising SOC tilts the balance towards IVC orders that preserve an anti-unitary operation $\mathcal{T}'$ corresponding to electronic time-reversal composed with a valley rotation. See \cref{tab:symmetry-orders-legends} for details and symmetry properties of the ground states captured by our treatment.

We further investigate the tendency of the various (Stoner-like and IVC) symmetry-breaking phases toward secondary nematic instabilities~\cite{Jung2015, Dong2021, Huang2023} whereby small Fermi pockets, either centered around $\text{C}_3$-related locations in the Brillouin zone or along a thin annulus, spontaneously reorganize in a rotation symmetry-breaking manner. This phenomenon is also referred to as `momentum flocking' or `momentum polarization'. Our analysis here is motivated by quantum oscillations~\cite{Zhang2023, Holleis2023} and transport measurements~\cite{Lin2023} reporting that the number of Fermi pockets in certain polarized phases (including the parent state of superconductivity in BLG/WSe$_2$~\cite{Zhang2023, Holleis2023}) is not consistent with preserved $\text{C}_3$ symmetry. Interestingly, we find that induced Ising SOC enhances tendencies toward nematic ordering in RTG.  

Collectively, our results uncover a rich competition between interactions and induced SOC and provide guiding principles for future experiments combining RTG and transition metal dichalcogenides. The richness and tunability of the phase diagram of RTG/WSe$_2$ could potentially be leveraged to create devices with novel properties, such as purely electrical control of orbital and spin magnetism as proposed in a recent related Hartree-Fock study~\cite{Wang2023}, or gate-defined Josephson junctions that host topological superconductivity~\cite{Xie2023}.  More broadly, we expect that our systematic study of trilayers, in conjunction with earlier work on bilayers, will help shed light on correlated phenomena in the wider family of crystalline graphene multilayers.  

The rest of this paper is organized as follows. In \cref{sec:model-methods} we introduce the non-interacting model describing RTG in the presence of induced SOC, discuss screened Coulomb interactions, and describe our self-consistent Hartree-Fock procedure. In \cref{sec:phase-diagram-wo-soc} we consider RTG without SOC and investigate the competition between long-range Coulomb interactions, which preserve an enhanced ${\text{SU}(4)}$ symmetry group, and an intervalley interaction term (or Hund's coupling) that partially breaks the resulting degeneracy. In \cref{sec:phase-diagram-w-soc} we explore the effects of induced Ising SOC on the phase diagram of RTG, and its subtle interplay with both Hund's coupling and nematic ordering tendencies. In \cref{sec:benchmarking} we benchmark our phase diagrams against experimental results, allowing an estimation of the strength of the two types of interactions considered. Finally, in \cref{sec:outlook} we summarize our results and provide insights for future experiments.

\section{Model and methods}
\label{sec:model-methods}

Rhombohedral trilayer graphene consists of three graphene layers stacked in the ABC configuration shown in \cref{fig:schematics-stacking}. Beginning with pure RTG (without an adjacent WSe$_2$ layer), the symmetry group in the presence of a displacement field $D$ contains three-fold rotations $\text{C}_{3}$, mirror symmetries, translations, time reversal $\mathcal{T}$, as well as---to an excellent approximation---${\text{SU}}(2)_{\text{s}}$ spin rotations. Additionally, in the low-energy limit the system exhibits approximate $\text{U}(1)_{\text{v}}$ valley conservation.  

The tight-binding Hamiltonian of pure RTG can be expanded near the two valleys $\tau \in \{\pm 1\}$ of graphene as
\begin{equation}\begin{split}
    \hat{H}_{\text{B}} = \sum_{\vb{k}} \sum_{\tau s \sigma \sigma'} 
        h(\tau \vb{K} + \vb{k})_{\sigma \sigma'} 
        c_{\tau s \sigma \vb{k}}^\dag 
        c_{\tau s \sigma' \vb{k}},
\end{split}\end{equation}
where the fermion operator $c_{\tau s \sigma \vb{k}}$ annihilates an electron at momentum $\vb{k}$ for valley index $\tau$, spin index $s \in \smash{\{\uparrow, \downarrow\}}$ and sublattice index $\sigma \in \{A_1, B_3, B_1, A_2, B_2, A_3\}$. Henceforth, we also use a combined flavor index $\alpha = (\tau, s, \sigma)$ for notational simplicity. The matrix $h$ is detailed in \cref{app-sec:hamiltonian/tight-binding-soc}, and retains the three leading-order tunneling matrix elements between adjacent layers~\cite{zhang2010band, jung2013gapped}.

Near charge neutrality, the low-energy conduction and valence bands in each valley touch at three Dirac points positioned at $\text{C}_3$-symmetric locations around the Brillouin zone corners. Under an applied perpendicular displacement field $D$, these Dirac points are gapped out and acquire non-trivial Berry curvature distributions~\cite{Koshino2009} that integrate to Berry phases of $3 \pi \tau \sgn{(D)}$. The resulting low-energy bands become locally flat (see \cref{fig:schematics-band-structure-wo-soc}), leading to build-ups in the density of states (DOS) near van Hove singularities that dramatically enhance interaction effects. We convert the displacement field $D$ to an interlayer potential difference $\Delta_1$ entering the non-interacting Hamiltonian $\hat{H}_{\text{B}}$ through 
\begin{equation}
     \Delta_1 = q_e d^\perp D / \epsilon_{\text{r}}^\perp,
     \label{eq:Delta1-def}
\end{equation} 
with $q_{\text{e}}$ the electron charge, $\epsilon_{\text{r}}^\perp = 4.4$ the dielectric constant of \hBN (the usual dielectric spacer layer between gates) and $d^\perp \approx \SI{3.3}{\angstrom}$ the interlayer distance in RTG. 

Coulomb interactions between electrons are included using a decomposition into long- and short-range components. The long-range component
\begin{equation}\begin{split}
    \hat{H}_{\text{C}} &= \frac{1}{2A} \sum_{\vb{q}} V_{\text{C}}(\vb{q}) 
        \normord{\rho(\vb{q}) \rho(-\vb{q})}
\end{split}\end{equation}
couples to the slowly varying part of the electronic density, $\smash{\rho(\vb{q})} = \smash{\sum_{\vb{k}, \alpha} c_{\alpha \vb{k}}^\dag c_{\alpha (\vb{k} + \vb{q})}}$. We use the dual-gated screened Coulomb potential
\begin{equation}
    V_{\text{C}}(\vb{q}) = \frac{q_{\text{e}}^2}{2 \epsilon_{\text{r}} \epsilon_0 q} \tanh{(q d)},
\end{equation}
with the screening length $d$ taken as the distance from RTG to the gates, $\epsilon_{\text{r}}$ the relative permittivity, and $\epsilon_0$ the permittivity of free space. In typical \textit{h}-BN-encapsulated devices, the dielectric environment contributes $\epsilon_{\text{r}} \approx 4.4$. To also account for screening originating from electrons in the graphene sheets, we treat $\epsilon_{\text{r}} > 4.4$ as a phenomenological parameter that controls the strength of the gated Coulomb potential $V_{\text{C}}(\vb{q})$.  Such a density-density interaction is invariant under an $\text{SU}(4)$ symmetry acting in spin-valley space.  The kinetic energy, however, partially breaks this $\text{SU}(4)$ symmetry (due to the $\tau$ dependence in the $h$ matrix from $\smash{\hat{H}_{\text{B}}}$). Consequently, the interacting model $\smash{\hat{H}_{\text{B}}} + \smash{\hat{H}_{\text{C}}}$ preserves a non-generic $\text{SU}(2) \times \text{SU}(2)$ symmetry corresponding to a pair of spin rotations that can be enacted separately in each valley.  

\begin{figure}
	\centering
	\includegraphics[width = 1\linewidth]{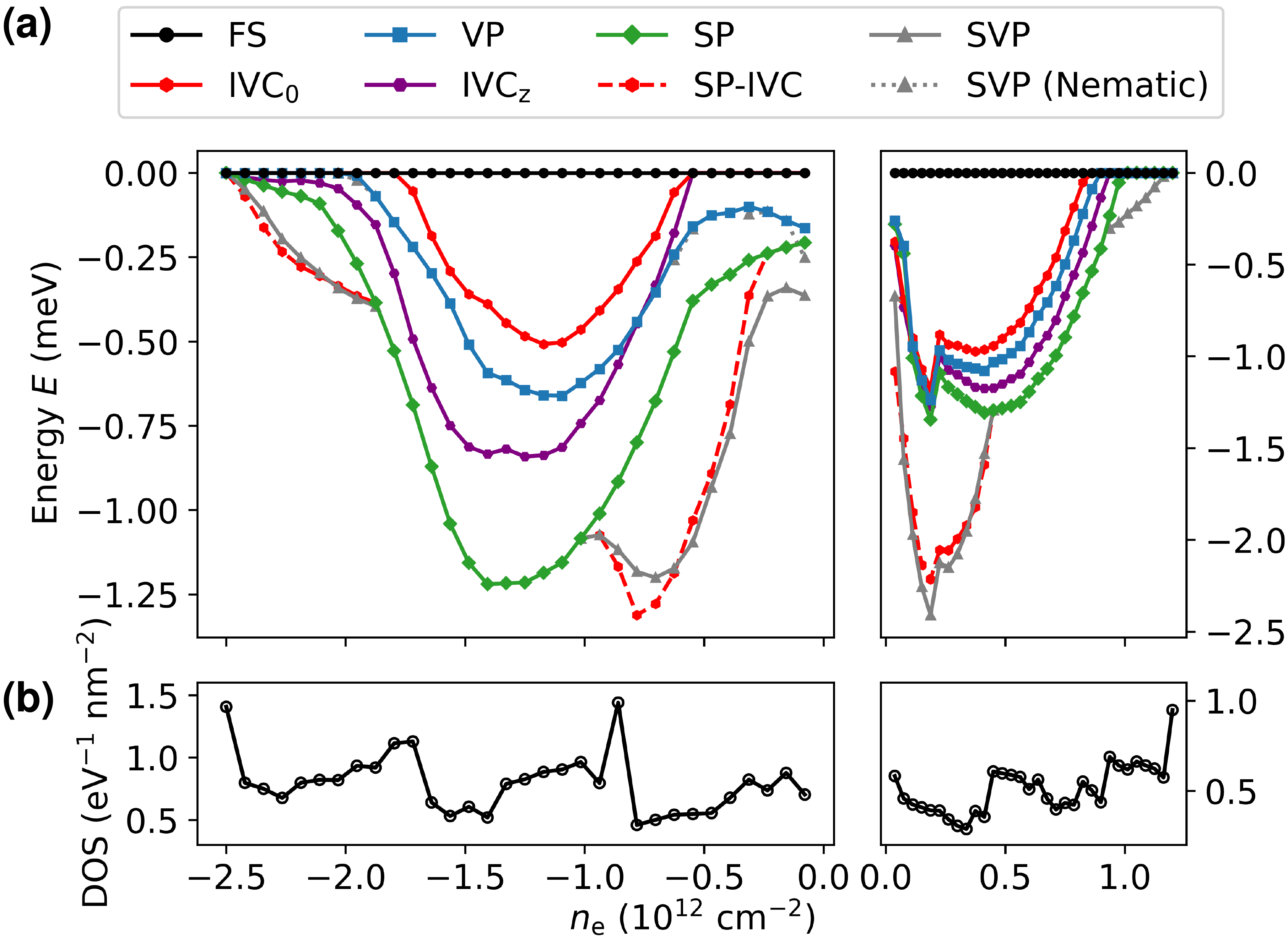}
	\caption{\textbf{Typical phase diagram slice from self-consistent Hartree-Fock.} \textbf{(a)} Hartree-Fock energy $E$ per carrier (see \cref{app-eq:hartree-fock-energy}) across electron density $n_{\text{e}}$. Line colors and styles denote various symmetry-restricted solutions listed in \cref{tab:symmetry-orders-legends}; energies are measured relative to the fully symmetric metal. The lowest-energy solution represents the best mean-field ground state. \textbf{(b)} Corresponding Fermi-level DOS. Sharp variations in DOS are observed at transitions between different ground-state orders, and when the Fermi surface topology changes within the same phase (\ie~through Lifshitz transitions). Here $\epsilon_{\text{r}} = 20$, $J_{\text{H}} = \SI{4}{\electronvolt\ucdot\unitcellarea}$, $\Delta_1 = \SI{40}{\milli\electronvolt}$ and $\lambda_{\text{I}} = 0$. Only a subset of relevant symmetry-restricted solutions is shown for visual clarity.}
	\label{fig:energy-slice}
\end{figure}

\begin{table*}
    \centering
    \newcommand{\legendicon}[1]{\smash{\raisebox{-5pt}{\includegraphics[height = 15pt]{#1}}}}
    \begin{tabular}{p{7.6cm} p{1.7cm} p{2.4cm} C{0.6cm} C{0.7cm} C{0.6cm} | C{1cm} C{0.8cm} | C{0.5cm} p{0.6cm}}
        \toprule 
        Order Description & Symbol & Order Operators & $\mathcal{T}$ & \hspace{-0.2cm} $\text{U}(1)_{\text{v}}$ & $\mathcal{T}'$ & $\text{SU}(2)_{\text{s}}$ & $\text{U}(1)_{\text{s}}$ & $g$ & Leg. \\
        \midrule 
        Fully symmetric & FS
            & - 
            & \cmark & \cmark & \cmark & \cmark & \cmark 
            & $4$
            & \legendicon{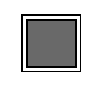} 
            \\
        Valley-polarized & VP 
            & $\tau^z s^0$ 
            & \xmark & \cmark & \xmark & \cmark & \cmark 
            & $2$ 
            & \legendicon{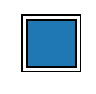} 
            \\
        Spin-polarized & $\text{SP}$ 
            & $\tau^0 s^z$ 
            & \xmark & \cmark & \xmark & \xmark & \cmark 
            & $2$
            & \legendicon{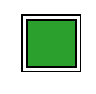} 
            \\
        Spin-valley-locked & $\text{SVL}$ 
            & $\tau^z s^z$ 
            & \cmark & \cmark & \cmark & \xmark & \cmark 
            & $2$
            & \legendicon{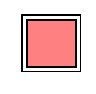} 
            \\
        Spin-valley locked + in-plane spin-polarized & $\text{SVL}$+$\text{SP}_{\pi}$
            & $\tau^z s^z, \tau^0 s^{x/y}$ 
            & \xmark & \cmark  & \xmark & \xmark & \xmark 
            & $2$
            & \legendicon{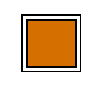}
            \\
        Intervalley-coherent spin-singlet & $\text{IVC}_{\text{0}}$ 
            & $\tau^x s^0$
            & \cmark & \xmark & \xmark & \cmark & \cmark 
            & $2$
            & \legendicon{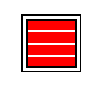} 
            \\
        Intervalley-coherent spin-triplet & $\text{IVC}_{\text{z}}$ 
            & $\tau^x s^z$ 
            & \xmark & \xmark & \cmark & \xmark & \cmark 
            & $2$
            & \legendicon{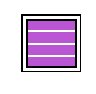} 
            \\
        Intervalley-coherent spin-triplet + spin-valley-locked & SVL-$\text{IVC}_{\text{z}}$ 
            & $\tau^x s^z, \tau^z s^z$ 
            & \xmark & \xmark & \cmark & \xmark & \cmark 
            & $2$
            & \legendicon{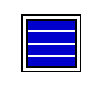} 
            \\
           Spin-valley-polarized  & SVP
            & $\tau^z s^0, \tau^0 s^z$ 
            & \xmark & \cmark & \xmark & \xmark &  \cmark 
            & $1$
            & \legendicon{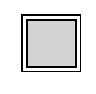} 
            \\
        Spin-polarized intervalley-coherent & $\text{SP}$-$\text{IVC}$
            & $\tau^x s^0, \tau^x s^z, \tau^0 s^z$ 
            & \xmark & \xmark & \xmark & \xmark & \cmark 
            & $1$
            & \legendicon{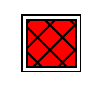} 
            \\
        Spin-valley-locked intervalley-coherent & $\text{SVL}$-$\text{IVC}$
            & $\tau^x s^x, \tau^z s^z$ 
            & \xmark & \xmark & \cmark & \xmark & \xmark 
            & $1$
            & \legendicon{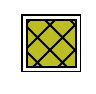}
            \\
        \bottomrule
    \end{tabular}
    \caption{\textbf{Symmetry classification of ground states found in self-consistent Hartree-Fock.} For each ground state, a minimal set of symmetry-breaking order operators in the spin-valley subspace is listed along with their transformation properties under electronic time-reversal $\mathcal{T} = \tau^x s^y \mathcal{K}$, valley charge conservation $\text{U}(1)_{\text{v}}$, and the effective anti-unitary symmetry $\mathcal{T}' = \tau^y s^y \mathcal{K}$. The transformation properties under full $\text{SU}(2)_{\text{s}}$ spin rotations, relevant in the SOC-free problem, and under $\text{U}(1)_{\text{s}}$ spin rotation around the $z$ axis, relevant in the case with Ising SOC, are also shown. The corresponding spin-valley degeneracy of the Fermi surfaces is denoted by the integer $g$, with $g = 4$ describing a fully symmetric metal. Last column presents the color and hatching scheme used in phase diagrams throughout this work. Nematicity (\ie~momentum polarization) is denoted with an overlaid circle hatching (\icontext{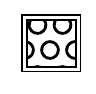}) when present in each phase.}
    \label{tab:symmetry-orders-legends}
\end{table*}

The short-range component $\smash{\hat{H}_{\text{V}}}$, with coupling strength $J_{\text{H}}$, captures scattering of electrons between different valleys and effectively encodes a Hund's coupling interaction (see \cref{app-sec:hamiltonian/coulomb-hunds} for details). Such a term breaks down the enlarged $\text{SU}(2) \times \text{SU}(2)$ symmetry group to physical global spin rotations $\text{SU}(2)_{\text{s}}$ by providing an energetic preference for aligning/anti-aligning the electron spins in the two valleys (for ferromagnetic/anti-ferromagnetic Hund's coupling respectively). We estimate the relevant regimes for the interaction strength parameters $\epsilon_{\text{r}}$ and $J_{\text{H}}$ from benchmarking to experimental results~\cite{zhou2021half, zhou2021superconductivity} (see \cref{sec:benchmarking,app-sec:results/benchmarking}).

The addition of an adjacent WSe$_2$ monolayer, in the configuration shown in \cref{fig:schematics-stack}, breaks $\text{SU}(2)_{\text{s}}$ spin rotation symmetry by inducing Ising- and Rashba-type SOC in the top layer of RTG~\cite{Gmitra2016, Yang2017, Zihlmann2018, Wang2019, Island2019, Wang2020, Amann2022, Sun2022determining}:
\begin{equation}\begin{split}
    \hat{H}_{\text{I}} &= \frac{\lambda_{\text{I}}}{2} \sum_{\vb{k}} 
        \vb{c}^\dag_{\vb{k}} \left( \tau^z s^z \mathbb{P}_3 \right) \vb{c}_{\vb{k}} , \\
    \hat{H}_{\text{R}} &= \frac{\lambda_{\text{R}}}{2} \sum_{\vb{k}} 
        \vb{c}_{\vb{k}}^\dag \left( \tau^z s^y \sigma^x - s^x \sigma^y \right) \mathbb{P}_3 \vb{c}_{\vb{k}}.
\end{split}\end{equation}
Here $\trans{\vb{c}}_{\vb{k}} = \mqty[ c_{+ \uparrow A_1 \vb{k}} & \ldots & c_{- \downarrow A_3 \vb{k}} ]$ is a vector of fermion operators enumerated over valley, spin, and sublattice indices. The Ising and Rashba SOC energy scales are respectively denoted $\lambda_{\text{I}}$ and $\lambda_{\text{R}}$, while $\mathbb{P}_3$ projects onto the top RTG layer. Throughout we use $\tau^\mu$, $s^\mu$, and $\sigma^\mu$ to label Pauli matrices acting on the valley, spin and sublattice degrees of freedom, respectively. Due to the layer polarization of the low-energy bands of RTG under an applied displacement field $D$, Ising SOC primarily leads to a band splitting in the valence (conduction) band~\cite{Gmitra2017, Khoo2017} for $D>0$ ($D<0$), as shown in \cref{fig:schematics-band-structure-w-soc}. 

The relative twist angle of WSe$_2$ and RTG provides a knob to tune the ratio of Ising and Rashba SOC~\cite{Li2019, David2019, Naimer2021, Chou2022}. However, sublattice polarization of the low-energy wavefunctions of RTG at large $D$ fields~\cite{Zaletel2019} effectively suppresses Rashba SOC; hence we focus on Ising SOC and set $\lambda_{\text{R}} = 0$ throughout for simplicity.
In this limit the interacting Hamiltonian preserves global U$(1)_{\text{s}}$ spin rotations along the Ising ($z$) axis. We briefly discuss effects of re-introducing Rashba SOC---thereby breaking the U$(1)_{\text{s}}$ symmetry---in the Outlook (\cref{sec:outlook}). 

We implement a self-consistent Hartree-Fock procedure, whereby a trial Slater-determinant ansatz $\ket{\Phi_{\text{HF}}}$ for the many-electron ground state is first chosen, usually respecting a certain set of symmetries. This trial ground state is characterized by the covariance matrix 
\begin{equation}
    \Delta(\vb{k})_{\alpha \alpha'}
    = \mel{\Phi_{\text{HF}}}{c^\dag_{\alpha \vb{k}} c_{\alpha' \vb{k}}}{\Phi_{\text{HF}}},
\end{equation}
which is then input into the mean-field decomposition of the Hamiltonian $\smash{\hat{H}_{\text{HF}}[\Delta]}$---see \cref{app-sec:hartree-fock} for details. A new ground state is then obtained by diagonalizing $\smash{\hat{H}_{\text{HF}}[\Delta]}$ until convergence is attained. In practice, many iterations of this procedure are performed for ansatzes exhibiting different sets of broken symmetries, and the best ground state is identified as the trial state with the lowest energy; see \cref{fig:energy-slice} for an example of the comparison between various trial states. As the tight-binding Hamiltonian $\smash{\hat{H}_{\text{B}}}$ is fitted to ab-initio (DFT) data, it already includes to an extent interaction effects at charge neutrality. Thus, to avoid double-counting interactions, we subtract the contribution from reference mean-field $\smash{\hat{H}_{\text{C}}}$ and $\smash{\hat{H}_{\text{V}}}$ constructed with the fully symmetric $\ket{\Phi_{\text{HF}}}$ at charge neutrality (see \cref{app-sec:hartree-fock/setup}).

\begin{figure*}
    \centering
    \includegraphics[width = \linewidth]{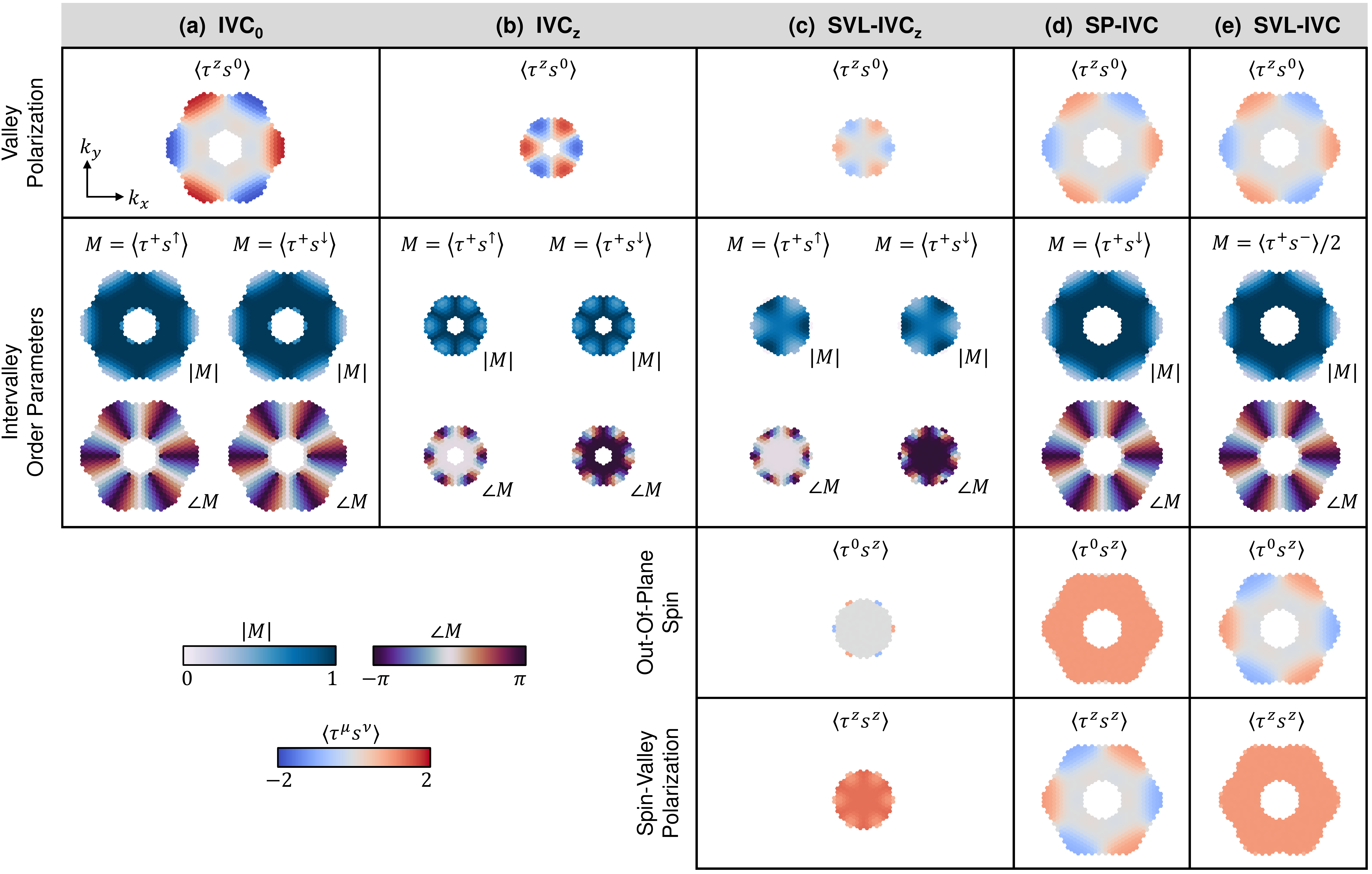}
    \phantomsubfloat{\label{fig:results-ivc-textures-[IVC-0]}}
    \phantomsubfloat{\label{fig:results-ivc-textures-[IVC-Z]}}
    \phantomsubfloat{\label{fig:results-ivc-textures-[SVL-IVC-Z]}}
    \phantomsubfloat{\label{fig:results-ivc-textures-[SP-IVC]}}
    \phantomsubfloat{\label{fig:results-ivc-textures-[SVL-IVC]}}
    \vspace{-1.8\baselineskip}
    \caption{\textbf{Momentum-resolved structure of IVC states.} Columns \textbf{(a)} through \textbf{(e)} illustrate the order parameters characterizing various IVC ground states obtained in this study, projected to the occupied (hole) bands. The first row shows the valley polarization, which averages to zero over the Brillouin zone but takes advantage of the trigonal-warping-induced energy difference between the two valleys. The second row depicts the in-plane components of the valley pseudospin, parametrized by $\tau^+ = \tau^x + i \tau^y$, accompanied by the relevant spin operators that differentiate each state. The third and fourth rows show the spin and spin-valley-locked polarizations, which respectively benefit from Hund's coupling and Ising SOC. The $\text{IVC}_0$ and $\text{IVC}_{\text{z}}$ states develop intervalley coherence for both spin projections, albeit with a relative $\pi$ phase shift for $\text{IVC}_{\text{z}}$. The closely related SVL-$\text{IVC}_{\text{z}}$ state additionally develops a large spin-valley-locked polarization ($\tau^z s^z$). The SP-IVC and SVL-IVC states exhibit a single Fermi surface corresponding to a definite spin projection (SP-IVC) or spin-valley locking (SVL-IVC).}
    \label{fig:results-ivc-textures}
\end{figure*}

The canonical approach in determining $\Delta(\vb{k})$ from $\smash{\hat{H}_{\text{HF}}[\Delta]}$ involves the filling of electronic states up to a given electron density; but the naïve way of doing so, which disregards degeneracies at the Fermi level, can lead to anomalous symmetry-breaking artifacts. We address this issue through a fractional filling scheme, which considers an ensemble-averaged $\Delta(\vb{k})$ free from such symmetry-breaking anomalies (see \cref{app-sec:hartree-fock/ground-state}). To reduce computational costs, we employ a semi-adaptive momentum grid with resolution and momentum cutoff chosen based on the non-interacting Fermi surfaces (see \cref{app-sec:hartree-fock/momentum-grid}). The phase diagrams presented in this work were computed on momentum grids comprising ${\sim} 1800$ points. We verify the convergence of our Hartree-Fock solutions by comparison against results at larger momentum grid resolution and cutoffs; moreover, for each ground state identified, we repeatedly impose random symmetry-breaking perturbations and run until convergence, to check that no lower-energy solutions exist (see \cref{app-sec:hartree-fock/convergence-stability-checks}).

\cref{tab:symmetry-orders-legends} lists all the symmetry-broken ground states obtained in this work, along with abbreviations and color schemes used to label them in the text and in phase diagrams. The transformation properties of the ground states under various symmetries, as well as their Fermi surface degeneracy, is also tabulated for future reference. \cref{tab:symmetry-orders-legends} notably includes five families of IVC orders; \cref{fig:results-ivc-textures} contrasts these IVC states by plotting their valley and spin textures projected to the active band of interest.

\section{Phase diagram of RTG without spin-orbit coupling}
\label{sec:phase-diagram-wo-soc}

We first explore the correlated physics of RTG in the absence of induced Ising SOC, $\lambda_{\text{I}} = 0$. We fix the Coulomb interaction strength by taking $\epsilon_{\text{r}} = 20$, and consider the cases with $J_{\text{H}} = 0$ and $J_{\text{H}} \neq 0$ in turn.

\subsection{Zero Hund's coupling}
\label{sec:phase-diagram-wo-soc/wo-hunds}

\begin{figure*}
	\centering
	\includegraphics[width = 1\linewidth]{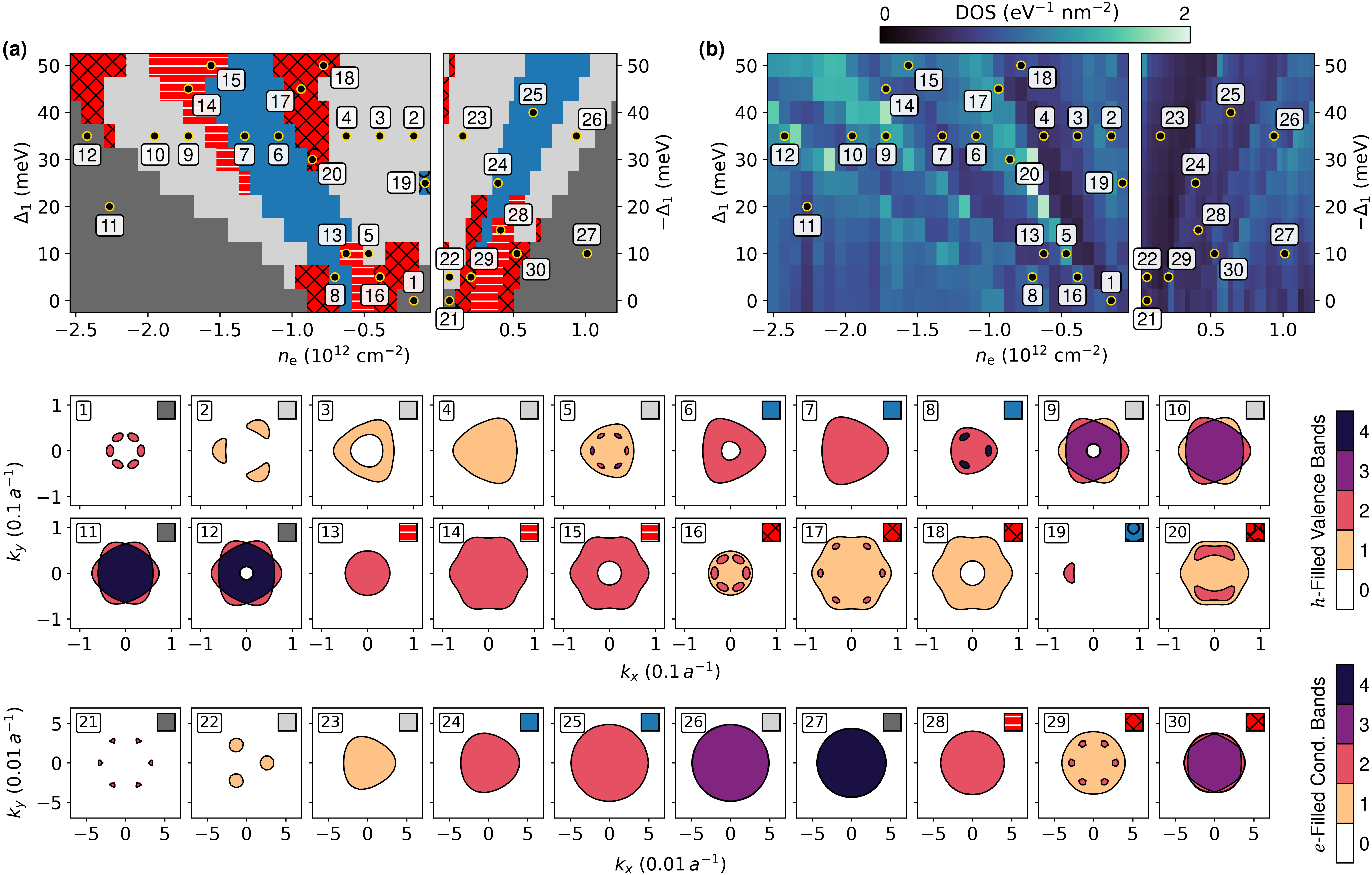}
    \phantomsubfloat{\label{fig:results-wo-hunds-wo-soc-phasediag}}
    \phantomsubfloat{\label{fig:results-wo-hunds-wo-soc-dos}}
    \vspace{-1.8\baselineskip}
	\caption{\textbf{RTG phases without SOC and Hund's coupling.} \textbf{(a)} Phase diagram of RTG in the electron density--interlayer potential ($n_{\text{e}}$--$\Delta_1$) parameter space for both hole- and electron-doped regimes, at moderate Coulomb strength $\epsilon_{\text{r}} = 20$ and no Hund's coupling ($J_{\text{H}} = 0$). Different phases are denoted by their color and hatching (see \cref{tab:symmetry-orders-legends} for legends). \textbf{(b)} Corresponding DOS at Fermi level     using a $\SI{500}{\micro\electronvolt}$ broadening of energy levels. Bottom panels show Fermi surfaces at numbered points in \textbf{(a)}; colors denote the number of mean-field valence or conduction bands occupied by carriers.}
	\label{fig:results-wo-hunds-wo-soc}
\end{figure*}

We present in \cref{fig:results-wo-hunds-wo-soc-phasediag} the phase diagram of RTG without Hund's coupling, $J_{\text{H}} = 0$, determined through self-consistent Hartree-Fock calculations as a function of electronic density $n_{\text{e}}$ and interlayer potential $\Delta_1$. [All phases in the figure are degenerate with those related by the unphysical $\text{SU}(2) \times \text{SU}(2)$ symmetry present at $J_{\text{H}} = 0$.]

A variety of correlated phases emerge in both hole- and electron-doped regimes. At first glance, the phase diagram resembles that expected from a generalized Stoner ferromagnet model~\cite{zhou2021half}: as either the electron or hole density is increased from charge neutrality, the system undergoes successive transitions to a quarter-metal, a half-metal and a three-quarter-metal phase, wherein $1$, $2$, or $3$ of the underlying spin and valley flavors are respectively occupied. We also find cases where the Stoner polarization is incomplete---namely, where a subset of spin-valley flavors is predominantly occupied, but where minority Fermi surfaces also exist; see below for further discussion. Due to the $\text{SU}(4)$ symmetry of the long-range Coulomb interactions, Stoner ferromagnets with the same number $g$ of majority-occupied flavors are degenerate---\eg~the spin-polarized (SP), valley-polarized (VP), and spin-valley-locked (SVL) states.\footnote{This observation relates phases, such as the valley-polarized and spin-polarized states, that are not connected by the non-generic $\text{SU}(2) \times \text{SU}(2)$ symmetry noted earlier.} 

Beyond Stoner-type ferromagnets, we find IVC orders---where again the two graphene valleys hybridize spontaneously---consistent with recent theoretical studies~\cite{Zhumagulov2023, Chatterjee2022, Wang2023, Huang2023}. The energetic advantage of IVC states arises from the fact that the valley pseudospin $\vb*{\tau} = (\tau^x, \tau^y, \tau^z)$ can rotate as a function of momentum $\vb{k}$ (see \cref{fig:results-ivc-textures}) to exploit the trigonally warped Fermi surfaces of RTG. When the energy difference between the two valleys $E_+(\vb{k}) - E_-(\vb{k})$ is small, $\vb*{\tau}$ points in the plane (thus hybridizing the two valleys). In contrast, when the energy difference is large it is favorable to rotate $\bm{\tau}$ out of the plane to benefit from the lower kinetic energy associated with populating a single valley. The in-plane components of $\vb*{\tau}$ wind six times when encircling the origin of the Brillouin zone~\cite{Chatterjee2022}---a consequence of the Berry phase of $3 \pi$ per valley in the low-energy bands of RTG.

\begin{figure*}
    \centering
    \includegraphics[width = 1\linewidth]{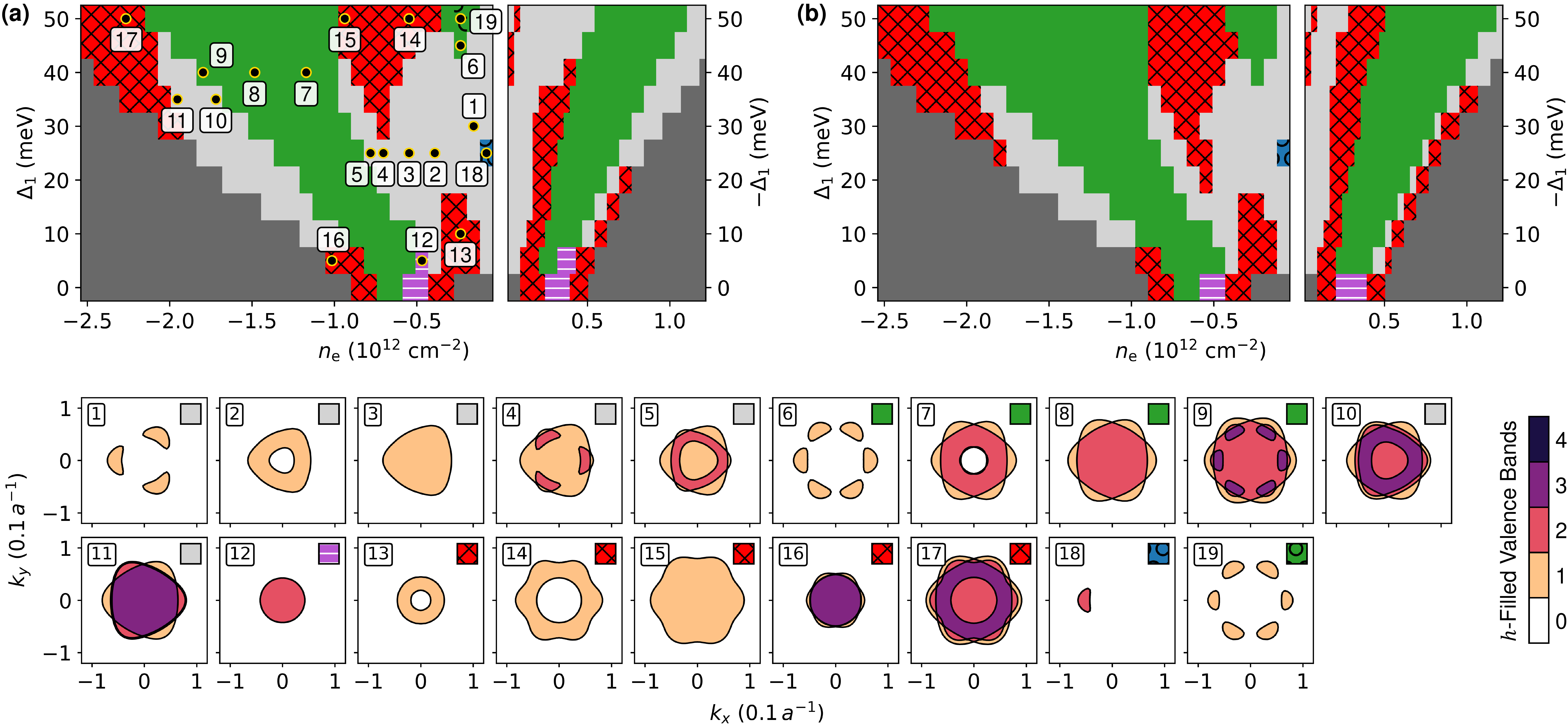}
    \phantomsubfloat{\label{fig:results-w-hunds-wo-soc-Jh-04}}
    \phantomsubfloat{\label{fig:results-w-hunds-wo-soc-Jh-08}}
    \vspace{-1.8\baselineskip}
    \caption{\textbf{RTG phases without SOC in the presence of Hund's coupling.} \textbf{(a)} Phase diagram of RTG in the electron density--interlayer potential ($n_{\text{e}}$--$\Delta_1$) parameter space in hole- and electron-doped regimes, at moderate Coulomb strength $\epsilon_{\text{r}} = 20$ and $J_{\text{H}} = \SI{4}{\electronvolt\ucdot\unitcellarea}$. Different phases are denoted by their color and hatching (see \cref{tab:symmetry-orders-legends} for legends). \textbf{(b)} Phase diagram at increased $J_{\text{H}} = \SI{8}{\electronvolt\ucdot\unitcellarea}$. Hund's coupling breaks the degeneracy between the $g = 2$ Stoner ferromagnets in favor of the spin-polarized ($\text{SP}$) phase, suppresses the SVP phases, and promotes SP-IVC and $\text{IVC}_{\text{z}}$ phases. Bottom panels show Fermi surfaces at numbered points in \textbf{(a)}; colors denote the number of mean-field valence bands occupied by carriers.}
    \label{fig:results-w-hunds-wo-soc}
\end{figure*}

We obtain two types of IVC orders. The $\text{IVC}_0$ state \mbox{(\icontext{icon-HM-IVC-0-nom})} preserves spin-rotation and time-reversal symmetries, and comprises a doubly degenerate Fermi surface where each spin projection develops identical intervalley coherent order (\cref{fig:results-ivc-textures-[IVC-0]}). In contrast, the non-degenerate SP-IVC state \mbox{(\icontext{icon-QM-IVC-0+SP-Z-nom})} is obtained starting from a spin-polarized state and lifting its valley degeneracy through the development of intervalley coherence (\cref{fig:results-ivc-textures-[SP-IVC]}). 

Representative Fermi surfaces in the bottom panels of \cref{fig:results-wo-hunds-wo-soc} illustrate the large variety of metallic phases that are stabilized. The Stoner-type symmetry-breaking cascade, wherein spin and valley flavors are successively filled, is clearly visible (subpanels 1--12 and 21--27), alongside the IVC phases (subpanels 13--18 and 28--30). Some states are partially polarized, featuring minority Fermi surfaces not expected from a pure half- or quarter-metal picture, \eg~subpanels 5, 8, 16, 17, 20 and 29. In addition to symmetry-breaking transitions precipitated by Coulomb interactions, Lifshitz-type phase transitions, where the Fermi surface changes topology, can also occur within a given phase and are associated with local extrema in the Fermi-level density of states shown in \cref{fig:results-wo-hunds-wo-soc-dos}; see also the fixed interlayer potential ($\Delta_1$) slice in \cref{fig:energy-slice}. Nematic ordering, characterized by the spontaneous polarization of low-density Fermi pockets in momentum space~\cite{Jung2015, Dong2021, Huang2023}, is also observed in subpanels 19 and 20. The spontaneous reorganization of low-density Fermi pockets into a reduced number of larger, $\text{C}_3$-breaking pockets lowers exchange energy but increases kinetic energy, and can be advantageous in certain regions of both Stoner-type and IVC phases. 

\subsection{Nonzero Hund's coupling}
\label{sec:phase-diagram-wo-soc/w-hunds}

We next study the effects of short-range Hund's coupling on the phase diagram of RTG. Experimental constraints including the observation of a spin-polarized half-metal phase~\cite{zhou2021half} indicate that Hund's coupling should be ferromagnetic ($J_{\text{H}} > 0$), such that the degeneracy between the $g = 2$ Stoner states (VP, SVL and SP) is broken in favor of the SP state \mbox{(\icontext{icon-HM-SP-Z-nom})}. In \cref{fig:results-w-hunds-wo-soc} we consider $J_{\text{H}} = \SI{4}{\electronvolt\ucdot\unitcellarea} \approx \SI{210}{\milli\electronvolt\nano\meter\squared}$ (\cref{fig:results-w-hunds-wo-soc-Jh-04}) and $J_{\text{H}} = \SI{8}{\electronvolt\ucdot\unitcellarea} \approx \SI{420}{\milli\electronvolt\nano\meter\squared}$ (\cref{fig:results-w-hunds-wo-soc-Jh-08}), and show representative Fermi surfaces in the bottom panels.

\begin{figure*}
    \centering
    \includegraphics[width = 1\linewidth]{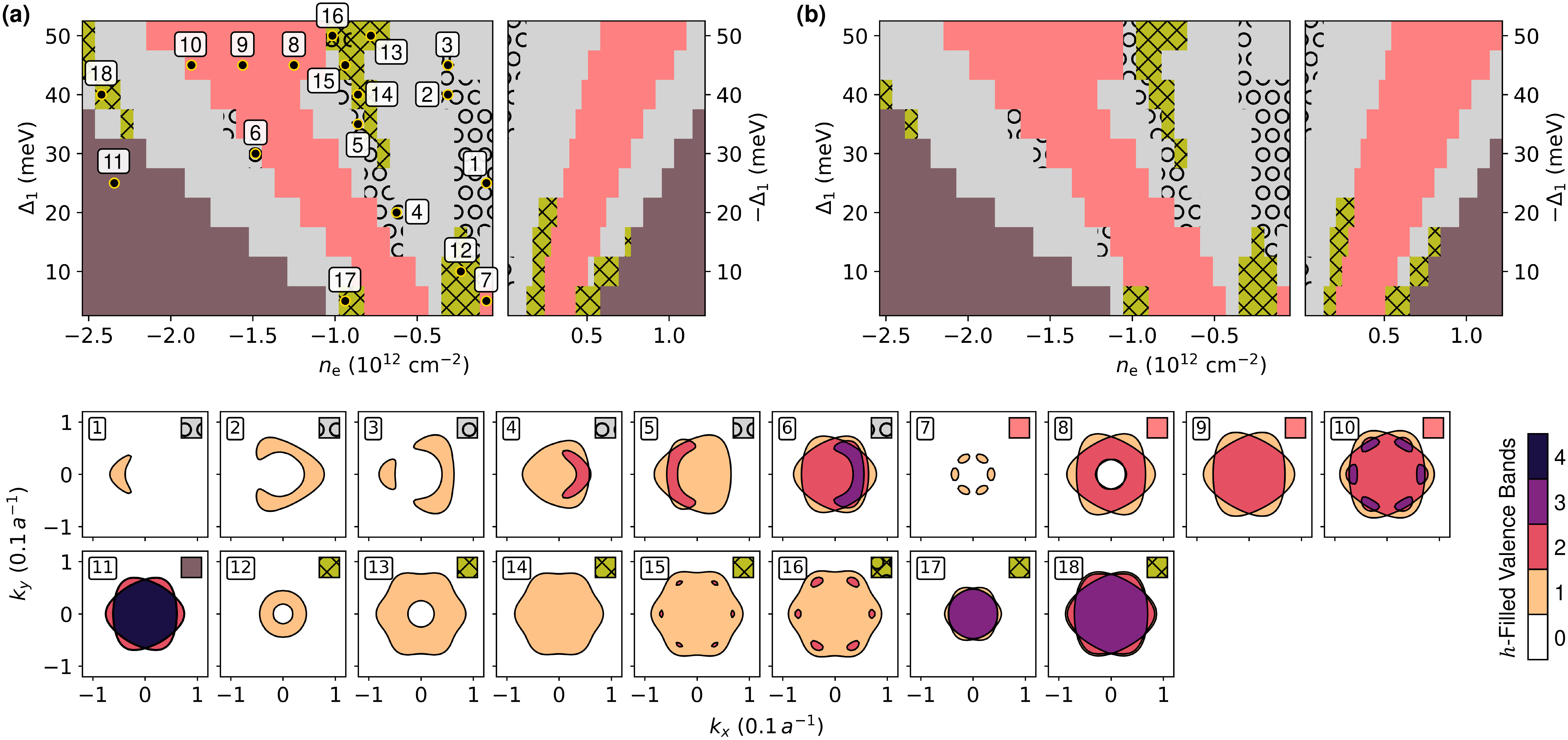}
    \phantomsubfloat{\label{fig:results-wo-hunds-w-soc-lambdaI-10}}
    \phantomsubfloat{\label{fig:results-wo-hunds-w-soc-lambdaI-30}}
    \vspace{-1.8\baselineskip}
    \caption{\textbf{Spin-orbit-coupled RTG phases in the absence of Hund's coupling.} \textbf{(a)} Phase diagram of RTG in the electron density--interlayer potential ($n_{\text{e}}$--$\Delta_1$) parameter space in hole- and electron-doped regimes, at moderate Coulomb strength $\epsilon_{\text{r}} = 20$ and Ising SOC splitting $\lambda_{\text{I}} = \SI{1}{\milli\electronvolt}$, without Hund's coupling ($J_{\text{H}} = 0$). Different phases are denoted by their color and hatching (see \cref{tab:symmetry-orders-legends} for legends). \textbf{(b)} Phase diagram at increased $\lambda_{\text{I}} = \SI{3}{\milli\electronvolt}$. Ising SOC breaks the degeneracy between the $g = 2$ Stoner ferromagnets in favor of the spin-valley-locked ($\text{SVL}$) phase, promotes nematicity at low carrier density, and converts SP-IVC phases to SVL-IVC phases. The fully degenerate ($g = 4$) phases without SOC acquire marginal spin-valley-locked polarization (\hspace{-3pt}\smash{\raisebox{-4.5pt}{\includegraphics[height = 13pt]{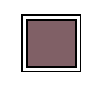}}}\hspace{-1pt}) mostly due to band structure (non-interacting) effects. Bottom panels show Fermi surfaces at numbered points in \textbf{(a)}; colors denote the number of mean-field valence bands occupied by carriers.}
    \label{fig:results-wo-hunds-w-soc}
\end{figure*}

As discussed above, the ferromagnetic Hund's coupling breaks the degeneracy between the $g = 2$ Stoner ferromagnets in favor of the spin-polarized phase. The energy advantage conferred can be sufficiently large to favor the spin-polarized phase deep into regions previously occupied by SVP phases ($g = 1$) in the absence of Hund's coupling, \eg~around point 6 of \cref{fig:results-w-hunds-wo-soc-Jh-04} and the analogous region in \cref{fig:results-w-hunds-wo-soc-Jh-08}. Because spin-unpolarized IVC states do not benefit from Hund's coupling, they are replaced by either the Stoner-type spin-polarized phase or its IVC counterpart (SP-IVC), both of which can take advantage of Hund's coupling for a reduction in energy. For the same reason, the SP-IVC phase grows at the expanse of the neighboring SVP phases when introducing Hund's coupling. Furthermore, at low interlayer potential we find a spin-triplet IVC state ($\text{IVC}_{\text{z}}$) characterized by the order parameter ${\sim} \tau^x s^z$ (\icontext{icon-HM-IVC-Z-nom} regions). Such a state is depicted in \cref{fig:results-ivc-textures-[IVC-Z]} and is characterized by an IVC order that spontaneously breaks the $\text{SU}(2)_{\text{s}}$ spin rotation symmetry: the two spin components each exhibit intervalley coherence, but with a relative sign difference between their respective IVC order parameters. As a result, the IVC character of this order will not manifest in charge density modulations but as a spin density wave.

Nematicity is also observed in certain regions of the phase diagram, for example near points 18 and 19 of \cref{fig:results-w-hunds-wo-soc-Jh-04}, with the latter exhibiting a partial polarization of the low-density Fermi pockets---\ie~the pockets deform in a $\text{C}_3$-breaking manner, although no pocket is entirely removed.

\section{Phase diagram of spin-orbit-coupled RTG}
\label{sec:phase-diagram-w-soc}

In this Section we address our key motivating question, namely the role of induced Ising SOC on the interacting phase diagram of RTG.

\subsection{Zero Hund's coupling}
\label{sec:phase-diagram-w-soc/wo-hunds}

\begin{figure*}
	\centering
	\includegraphics[width = 1\linewidth]{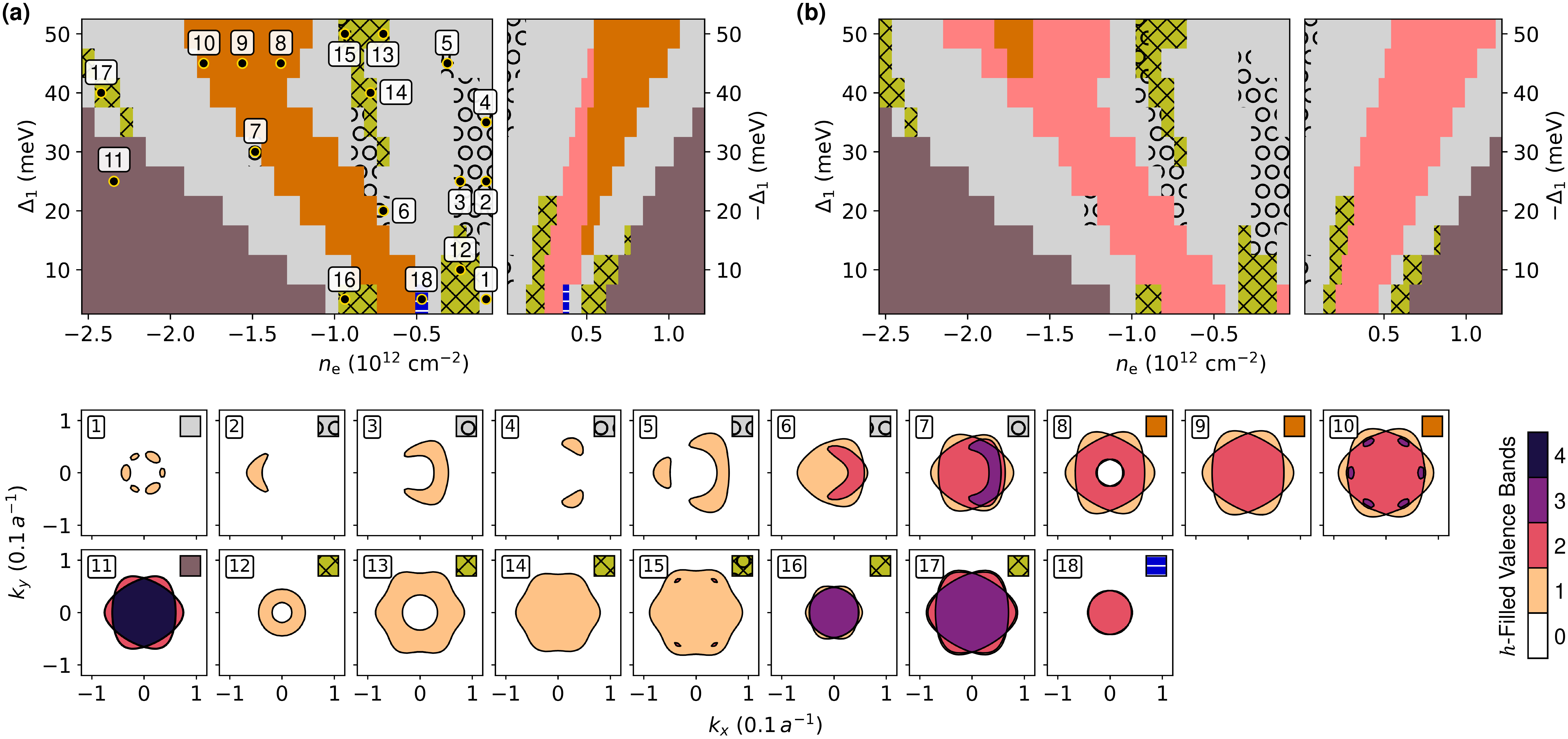}
    \phantomsubfloat{\label{fig:results-w-hunds-w-soc-lambdaI-10}}
    \phantomsubfloat{\label{fig:results-w-hunds-w-soc-lambdaI-30}}
    \vspace{-1.8\baselineskip}
	\caption{\textbf{Spin-orbit-coupled RTG phases in the presence of Hund's coupling.} \textbf{(a)} Phase diagram of RTG in the electron density--interlayer potential ($n_{\text{e}}$--$\Delta_1$) parameter space in hole- and electron-doped regimes, at moderate Coulomb strength $\epsilon_{\text{r}} = 20$, $J_{\text{H}} = \SI{4}{\electronvolt\ucdot\unitcellarea}$, and Ising SOC splitting $\lambda_{\text{I}} = \SI{1}{\milli\electronvolt}$. Different phases are denoted by their color and hatching (see \cref{tab:symmetry-orders-legends} for legends). \textbf{(b)} Phase diagram at increased $\lambda_{\text{I}} = \SI{3}{\milli\electronvolt}$. The simultaneous presence of Ising SOC and Hund's coupling promotes the composite
    $\text{SVL} + \text{SP}_\pi$ phase over the plain spin-valley-locked ($\text{SVL}$) phase; for larger Ising SOC, a larger Hund's coupling is required for this transition. The dominant IVC instability is towards the SVL-IVC state, except for a small region of SVL-$\text{IVC}_{\text{z}}$ phase for weak Ising SOC and interlayer potential. Bottom panels show Fermi surfaces at numbered points in \textbf{(a)}; colors denote the number of mean-field valence bands occupied by carriers.}
	\label{fig:results-w-hunds-w-soc}
\end{figure*}

We first consider the case without Hund's coupling, and compare Hartree-Fock phase diagrams with $\lambda_{\text{I}} = \SI{1}{\milli\electronvolt}$ and $\lambda_{\text{I}} = \SI{3}{\milli\electronvolt}$ in \cref{fig:results-wo-hunds-w-soc-lambdaI-10} and \cref{fig:results-wo-hunds-w-soc-lambdaI-30}, respectively. [Ising SOC reduces the non-generic $\text{SU}(2) \times \text{SU}(2)$ symmetry present without Hund's coupling down to $\text{U}(1) \times \text{U}(1)$, corresponding to spin rotations about the Ising axis that can be carried out independently for each valley. All states related by this $\text{U}(1) \times \text{U}(1)$ symmetry are degenerate.] 

The degeneracy between the SP, VP and SVL states observed in the limit with $\text{SU}(4)$-symmetric Coulomb interactions (\cref{fig:results-wo-hunds-wo-soc}) is now lifted to favor the spin-valley locked (SVL) phase \mbox{(\icontext{icon-HM-VSP-nom})}. Note that regions previously occupied by the four-fold degenerate metal also acquire a slight spin-valley polarization, due primarily to a non-interacting band structure effect. In these regions \mbox{(\icontext{icon-VSP-NI-nom})}, the extent of Ising polarization ($\tau^z s^z$) is an order of magnitude smaller than in the SVL phase and roughly matches non-interacting expectations.

Interestingly, a new type of intervalley coherent phase emerges: the spin-valley-locked SVL-IVC order~\cite{Zhumagulov2023} occupying hatched yellow regions \mbox{(\icontext{icon-QM-SVP-P-nom})}. Similarly to the SVL state, SVL-IVC exhibits a large and uniform Ising polarization ${\sim} \tau^z s^z$, as shown in \cref{fig:results-ivc-textures-[SVL-IVC]}, while additionally developing intervalley coherence within the relevant subspaces $(K^+ {\uparrow}, K^- {\downarrow})$ and/or $(K^+ {\downarrow}, K^- {\uparrow})$, depending on the sign of $\lambda_{\text{I}}$ as well as the electronic density. Interestingly, SVL-IVC states preserve an effective anti-unitary symmetry $\mathcal{T}' = \tau^y s^y \mathcal{K}$ that corresponds to the electronic time-reversal symmetry $\mathcal{T}$ followed by a $\pi$ valley rotation around the $\tau^z$ axis. Representative Fermi surfaces of SVL-IVC states appear in subpanels 12--18 of \cref{fig:results-wo-hunds-w-soc}. 

We also find that nematic tendencies are greatly enhanced in the presence of induced SOC. In particular, large regions of spin- and valley-polarized Stoner states exhibiting nematicity are observed in \cref{fig:results-wo-hunds-w-soc-lambdaI-10}---comprising one or two deformed Fermi pockets, in some cases superimposed on larger trigonally warped Fermi surfaces (subpanels 1--6). This observation contrasts with the SOC-free problem, \cref{fig:results-wo-hunds-wo-soc}, where only small (mostly low-density) regions exhibit nematicity. At large interlayer potential, a region of nematic SVL-IVC phase is also observed (subpanel 16), with partial momentum polarization of the $6$ small Fermi pockets lying atop a valley-hybridized hexagonal Fermi surface. The extended nematic regions in \cref{fig:results-wo-hunds-w-soc-lambdaI-10} and \cref{fig:results-wo-hunds-w-soc-lambdaI-30} are quite similar, suggesting that the enhancement of nematicity saturates rapidly upon increasing Ising SOC.

\subsection{Nonzero Hund's coupling}
\label{sec:phase-diagram-w-soc/w-hunds}

The phase competition in RTG becomes most complex when both Hund's interaction and Ising SOC are present, as shown in \cref{fig:results-w-hunds-w-soc}. In this regime, the doubly degenerate Stoner phase that is selected depends on the dominant perturbation to the long-range Coulomb interaction. When $\lambda_{\text{I}}$ dominates, as in \cref{fig:results-w-hunds-w-soc-lambdaI-30}, the spin-valley-locked (SVL) state dominates. However, when both $J_{\text{H}}$ and $\lambda_{\text{I}}$ compete, a compromise solution is found in the form of a state \mbox{(\icontext{icon-HM-VSP+SP-X-nom})} that combines spin-valley locking ($\tau^z s^z$) and in-plane spin-polarization (${\sim} \tau^0 s^{x/y}$). Spin polarization is favored by Hund's coupling---but due to the energy cost of polarizing along the Ising axis, an in-plane spin polarization is preferred over the out-of-plane alternative (${\sim} \tau^0 s^z$). The resulting Fermi surfaces remain doubly degenerate as the two order parameters anti-commute. 

The preferred IVC state also depends on the competition between Hund's and Ising terms. We find that the SVL-IVC order is preferred almost everywhere, with the exception of small regions of parameter space for weak Ising SOC and interlayer potential (on both electron- and hole-dope sides) that host a new state dubbed SVL-$\text{IVC}_{\text{z}}$. This state is similar to the $\text{IVC}_{\text{z}}$ order stabilized in a similar parameter regime in the absence of SOC (\cref{fig:results-w-hunds-wo-soc}), but now with an additional ${\sim} \tau^z s^z$ polarization induced by Ising SOC (see \cref{fig:results-ivc-textures-[SVL-IVC-Z]}). The SVL-$\text{IVC}_{\text{z}}$ order is also invariant under the anti-unitary $\mathcal{T}'$, and its Fermi surfaces remain doubly degenerate due to the anti-commutation of its constituent order parameters (see \cref{tab:symmetry-orders-legends}). The SVL-$\text{IVC}_{\text{z}}$ and $\text{IVC}_{\text{z}}$ states are distinguished by a subtle symmetry feature: while both states break $\text{SU}(2)_{\text{s}}$ and $\text{U}(1)_{\text{v}}$, $\text{IVC}_{\text{z}}$ preserves the product $\tau^z s^x$ of a $\pi$ valley rotation and a $\pi$ spin rotation, while SVL-$\text{IVC}_{\text{z}}$ does not. 
 
\section{Benchmarking With Experiments}
\label{sec:benchmarking}

Thus far we have treated the interaction parameters $\epsilon_{\text{r}}$ and $J_{\text{H}}$ largely from an exploratory standpoint, which has afforded a discussion of RTG ground states in successively more complicated scenarios (\cref{sec:phase-diagram-wo-soc,sec:phase-diagram-w-soc}). To close our discussion, we attempt a benchmarking of our Hartree-Fock phase diagrams to available experimental data on Stoner-type symmetry breaking in RTG~\cite{zhou2021half, zhou2021superconductivity}. 

\begin{figure}
    \centering
    \includegraphics[width = 1\linewidth]{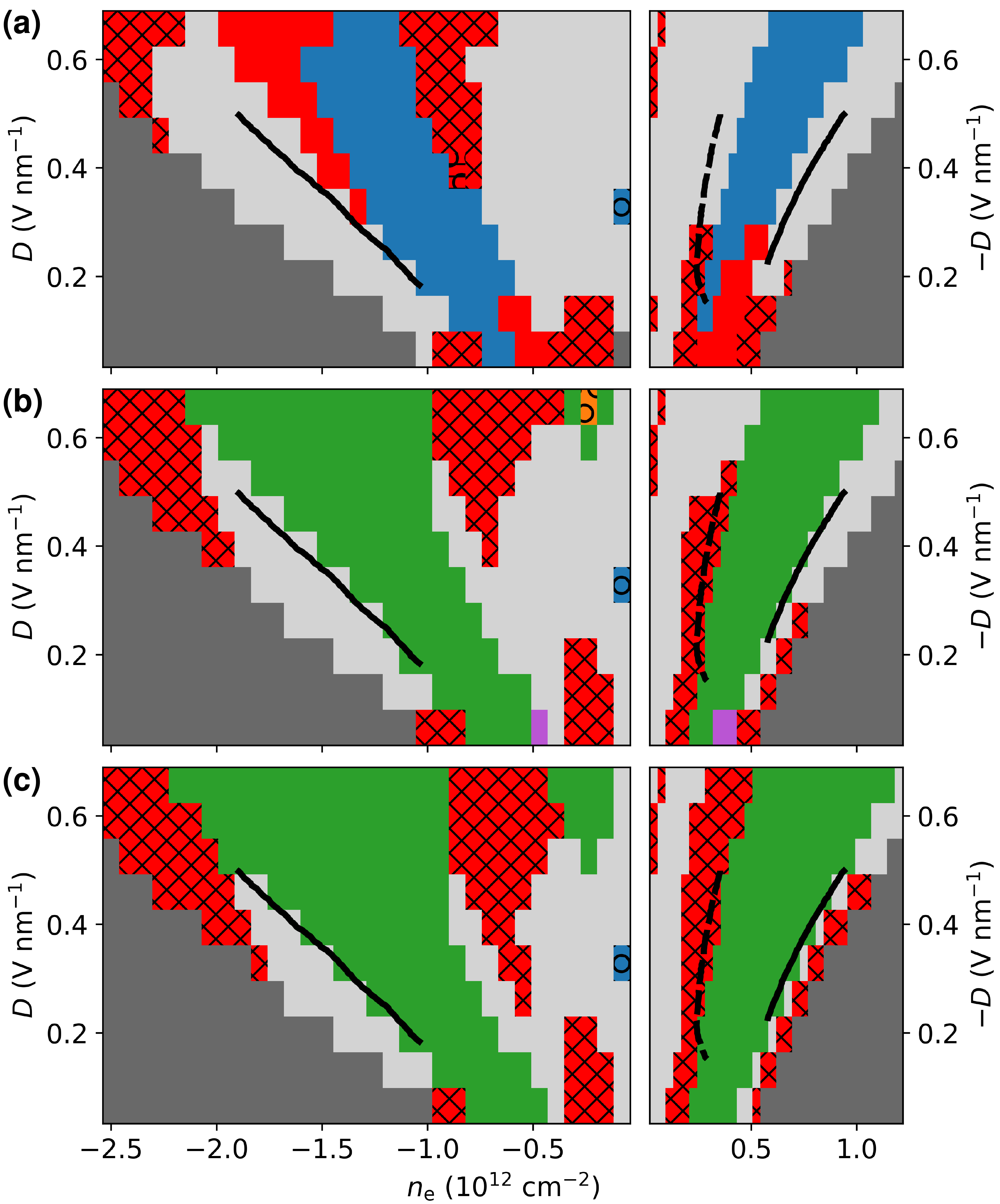}
    \phantomsubfloat{\label{fig:results-benchmarking-Jh-00}}
    \phantomsubfloat{\label{fig:results-benchmarking-Jh-04}}
    \phantomsubfloat{\label{fig:results-benchmarking-Jh-08}}
    \vspace{-1.8\baselineskip}
    \caption{\textbf{Comparison of self-consistent Hartree-Fock phase diagrams to prior experiments.} RTG phase diagrams in the electron density--displacement field ($n_{\text{e}}$--$D$) parameter space in hole- and electron-doped regimes without induced SOC, at $\epsilon_{\text{r}} = 20$ and \textbf{(a)} $J_{\text{H}} = 0$, \textbf{(b)} $J_{\text{H}} = \SI{4}{\electronvolt\ucdot\unitcellarea}$, \textbf{(c)} $J_{\text{H}} = \SI{8}{\electronvolt\ucdot\unitcellarea}$. In the hole-doped regime, solid lines denote experimental phase boundaries between a fully symmetric and a partially polarized $g = 2$ phase. In the electron-doped regime, dashed lines denote phase boundaries between a quarter-metal and a spin-polarized half-metal phase, and solid lines denote phase boundaries into the fully symmetric metal. Experimental data reproduced from Ref.~\cite{zhou2021half}. An out-of-plane $\smash{\epsilon_{\text{r}}^\perp = 4.4}$ is used to convert between interlayer potential $\Delta_1$ and displacement field $D$ (see \cref{eq:Delta1-def}).}
    \label{fig:results-benchmarking}
\end{figure}

To date, experiments on RTG have reported results without induced SOC from an adjacent WSe$_2$ monolayer. Therefore we present in \cref{fig:results-benchmarking} phase diagrams computed with $\lambda_{\text{I}} = 0$ as well as $J_{\text{H}} = 0$, $\SI{4}{\electronvolt\ucdot\unitcellarea}$, and $\SI{8}{\electronvolt\ucdot\unitcellarea}$. Experimental phase boundaries~\cite{zhou2021half} separating the fully symmetric metal and half-metal phases (solid lines), and a spin-polarized half-metal and a quarter-metal phase (dashed lines) are drawn for comparison.  We continue to assume an out-of-plane dielectric constant $\epsilon_{\text{r}}^\perp = 4.4$, typical of \hBN-encapsulated devices, to convert between displacement field $D$ and interlayer potential $\Delta_1$ via Eq.~\eqref{eq:Delta1-def}. As mentioned in \cref{sec:phase-diagram-wo-soc/w-hunds}, $J_{\text{H}} > 0$ is needed to account for the spin-polarized half-metal in experiments, as opposed to other Stoner-type half-metals. 

Notably, \cref{fig:results-benchmarking-Jh-04} suggests a reasonable agreement between the numerical and experimental phase boundaries at $\epsilon_{\text{r}} \sim 20$ and $J_{\text{H}} \sim \SI{4}{\electronvolt\ucdot\unitcellarea} \approx \SI{210}{\milli\electronvolt\nano\meter\squared}$, which explains our choice of parameters in the preceding figures (\cref{fig:results-wo-hunds-wo-soc,fig:results-w-hunds-wo-soc,fig:results-wo-hunds-w-soc,fig:results-w-hunds-w-soc}). A larger $J_{\text{H}} \sim \SI{8}{\electronvolt\ucdot\unitcellarea} \approx \SI{420}{\milli\electronvolt\nano\meter\squared}$ produces also a plausible agreement to experiment phase boundaries (see \cref{fig:results-benchmarking-Jh-08}). The above energy scales for Hund's coupling represent a significant perturbation on the long-range Coulomb interaction: for reference, at $\epsilon_{\text{r}} = 20$ and typical Fermi momentum $q \approx 0.1 a^{-1}$, the screened Coulomb potential $V_{\text{C}}(\vb{q}) \approx \SI{1}{\electronvolt\nano\meter\squared}$. Because local interactions beyond Coulomb, such as electron-phonon coupling, can also contribute to the Hund's interaction, a direct estimation of $J_{\text{H}}$ is not straightforward. Additionally, our Hartree-Fock treatment features weak residual three-quarter-metal phases that are absent from the experimental observations; beyond-Hartree-Fock effects~\cite{Wolf2023} may destabilize such phases and thus change the $J_{\text{H}}$ values that give best agreement with the data. 

We also show comparisons with phase diagrams obtained with $\epsilon_{\text{r}} = 15$ and $\epsilon_{\text{r}} = 30$ in \cref{app-sec:results/benchmarking}, which do not agree comparably well with experiments. In all, we estimate $18 \lesssim \epsilon_{\text{r}} \lesssim 25$ and $\SI{3}{\electronvolt\ucdot\unitcellarea} \lesssim J_{\text{H}} \lesssim \SI{10}{\electronvolt\ucdot\unitcellarea}$ to be consistent (at the mean-field level) with currently available data for RTG. The value $\epsilon_{\text{r}} = 20$ used in this work represents a weaker Coulomb interaction than in Ref.~\cite{Huang2023}, but is considerably stronger (especially near the van Hove singularities) than RPA treatments of electronic screening used in Refs.~\cite{Zhumagulov2023, Chatterjee2022}.

\section{Conclusion and Outlook}
\label{sec:outlook}

In this work we performed extensive self-consistent Hartree-Fock simulations to investigate the phase diagram of RTG in the presence of long-range Coulomb interactions, short-range Hund's coupling, and Ising SOC induced by proximity to a neighboring WSe$_2$ monolayer. Our main conclusions can be summarized as follows. 

In the absence of SOC, there is a competition between Stoner-type ferromagnets (where a subset of the spin-valley flavors are uniformly polarized along the Fermi surface), and IVC states which exploit the momentum-space structure of the trigonally warped Fermi surfaces of RTG to lower their kinetic energy.
As such, IVC states tend to be stabilized in regimes of `intermediate' correlations, which are likely relevant in RTG for experimentally accessible displacement fields $D$ of order ${\sim} \SI{0.5}{\volt\per\nano\meter}$. The addition of a ferromagnetic Hund's coupling, motivated by empirical observations~\cite{zhou2021half}, breaks the degeneracy of the $g = 2$ Stoner ferromagnets in favor of spin polarization (SP), which dominates the corresponding part of the phase diagram with the exception of a small region of ($g = 2$) $\text{IVC}_{\text{z}}$ phase for weak interlayer potential. In the parameter regime where $g = 1$ phases are preferred, Hund's coupling enlarges the regions occupied by SP-IVC states at the expense of Stoner SVP states.

Among the different ground states we obtained in the SOC-free limit, three (SP, SP-IVC and $\text{IVC}_{\text{z}}$) may be conducive to zero-momentum Cooper pairing at low temperature, as their Fermi surfaces preserve the $\vb{k} \leftrightarrow -\vb{k}$ resonance condition. The SP and SP-IVC states exhibit spin-polarized Fermi surfaces that naturally lead to (intra-band) superconductivity of spin-triplet character, and may therefore be relevant for the SC2 phase identified in Ref.~\cite{zhou2021superconductivity}. For these two states, the resonance condition appears as a consequence of the  spinless time-reversal symmetry $\mathcal{T}_{\text{spinless}} = \tau^x \mathcal{K}$. We thus expect perturbations that break $\mathcal{T}_{\text{spinless}}$  to be detrimental to pairing in the SP or SP-IVC state. Such perturbations include spin-orbit scattering from impurities but also the native SOC in graphene. The energy scale of native SOC is estimated to be of order ${\sim} \SI{10}{\micro\electronvolt}$~\cite{Avsar2020} and is usually neglected---including in our Hartree-Fock treatment where typical energy differences between competing ground states are of order $\SI{100}{\micro\electronvolt}$ (see \cref{fig:energy-slice}). However, Cooper pairing in RTG is characterized by energy scales $\Delta_{\text{BCS}} = 1.76 \, k_{\text{B}} T_{\text{c}} \approx 15$ and $\SI{4}{\micro\electronvolt}$ for the SC1 and SC2 phases, respectively---sufficiently low to be adversely affected by native SOC. Similarly, the $\text{IVC}_{\text{z}}$ state is invariant under the anti-unitary $\mathcal{T}'$, which is also an approximate symmetry as it relies on emergent $\text{U}(1)_{\text{v}}$ valley conservation at low energies. Short-range potential disorder and edge terminations will therefore be pair breaking, but such effects are presumably small as superconductivity resides deep in the clean limit. Unfortunately, in our simulations $\text{IVC}_{\text{z}}$ is only stabilized for low interlayer potential, where superconductivity is not observed in RTG.

The addition of Ising SOC of order $\lambda_{\text{I}} \sim \SI{1}{\milli\electronvolt}$, the relevant scale for BLG/WSe$_2$~\cite{Island2019, Zhang2023, Holleis2023}, significantly tilts the energetic balance between the various candidate phases. In the electronic density and $D$ field region where the Stoner picture predicts a $g = 2$ state, a sufficiently large $\lambda_{\text{I}}$ (compared to the Hund's energy scale) favors the spin-valley-locked (SVL) state. This phase is naturally conducive to Cooper pairing as it preserves time-reversal symmetry $\mathcal{T}$---and further admits a non-zero projection of a spin-singlet $s$-wave pairing interaction on its Fermi surface. When Ising and Hund's coupling terms are comparable, a linear combination of their respective (SVL and SP) preferred order parameters is selected. If the spin polarization points \emph{in the plane}, this state can benefit energetically from Hund's coupling while avoiding paying the penalty associated with polarizing along the Ising quantization axis. The corresponding Fermi surfaces are doubly degenerate ($g = 2$) because the order parameters $\tau^z s^z$ and $\tau^0 s^x$ anti-commute, and are also $\vb{k} \leftrightarrow -\vb{k}$ symmetric. However, this resonance condition is \emph{not} symmetry-enforced and can be understood as an artifact of neglecting symmetry-allowed terms (in the presence of the WSe$_2$ substrate) such as Rashba SOC, which would deform the SVL+SP$_\pi$ Fermi surfaces in a pair-breaking manner~\cite{Yuan2021}.

The most robust intervalley coherent order in the presence of Ising SOC is the SVL-IVC state, which hybridizes both valley and spin degrees of freedom. Such a state has non-degenerate Fermi surfaces ($g = 1$) and arises from developing intervalley coherence within the subset of electronic states favored by the spin-valley-locked order. Crucially, the SVL-IVC state respects the effective anti-unitary $\mathcal{T}'$ that guarantees the $\vb{k} \leftrightarrow - \vb{k}$ symmetry of its Fermi surfaces (again provided that $\text{U}(1)_{\text{v}}$ valley rotations remain a good symmetry), and therefore represents a promising candidate to host zero-momentum superconductivity at low temperature. If realized experimentally, the SVL-IVC state could be used as a resource to engineer topological superconductivity in gate-defined Josephson junctions, following ideas in Ref.~\cite{Xie2023}, due to its unique combination of non-degenerate Fermi surfaces protected by an anti-unitary symmetry. 

Interestingly, we find that
Ising SOC promotes nematic ordering tendencies among the small Fermi pockets of RTG, a phenomenon for which experimental evidence is mounting in the closely related BLG/WSe$_2$ platform~\cite{Zhang2023, Holleis2023}. Provided that nematic ordering preferentially selects pairs of pockets that are related by $\vb{k} \leftrightarrow -\vb{k}$, superconductivity may naturally coexist with nematicity.

Our study neglects Rashba SOC, which is symmetry-allowed in experiments due to the breaking of vertical mirror symmetry at the graphene/TMD interace. While the impact of Rashba SOC is expected to be suppressed by wavefunction effects at large $D$ field~\cite{Zaletel2019}, it may have a stronger effect in the weak $D$ field regime, where two of our IVC ground states ($\text{IVC}_{\text{z}}$ and SVL-$\text{IVC}_{\text{z}}$) are stabilized. Moreover, even weak Rashba SOC could have important effects on potential pairing instabilities, especially for phases that develop non-zero in-plane spin components. As mentioned above, in the SVL+SP$_{\pi}$ state the introduction of Rashba SOC is expected to be detrimental to pairing. In contrast, in the SVL-IVC state the $\mathcal{T}'$ symmetry does not rely on preserving the $\text{U}(1)_{\text{s}}$ rotations along the Ising axis; the Fermi surface resonance condition is therefore a robust feature. In fact, Rashba SOC may even favor the SVL-IVC state at the expense of competing Stoner SVP states, due to its non-trivial in-plane spin texture. Exploring this interplay represents an interesting avenue of future work. Conversely, it should be possible to minimize Rashba effects experimentally by constructing devices encapsulated with WSe$_2$ on both sides~\cite{Island2019, Wang2023}, which may furnish a more robust platform for stabilizing superconductivity.

Intriguingly, RTG is to date the only member of the rhombohedral graphene multilayer family known to superconduct in the absence of external perturbations (other than the applied perpendicular displacement field $D$). Disentangling the physical mechanisms underlying this observation, and understanding the perturbations that may favor Cooper pairing in the other members of the family, represent promising opportunities for future work.

\emph{Note added.--} While finalizing this work, we became aware of a parallel study~\cite{Zhumagulov2023} investigating the interplay of (long-range) Coulomb interactions and induced Ising SOC in RTG. The reported phase diagrams are qualitatively similar to ours (when setting Hund's coupling $J_{\text{H}} = 0$). The authors also uncover a quarter-metal IVC order that takes advantage of spin-valley-locked polarization, equivalent to our SVL-IVC state (which they name ``spin-valley coherent'' order).

\section*{Acknowledgments}

We are grateful to Yiran Zhang, Alex Thomson, Cyprian Lewandowski, and Stevan Nadj-Perge for insightful discussions and collaborations on related projects. J.~M.~K. acknowledges support from the SURF programme at Caltech. \'E.~L.-H. was supported by the Gordon and Betty Moore Foundation’s EPiQS Initiative, Grant GBMF8682. The U.S. Department of Energy, Office of Science, National Quantum Information Science Research Centers, Quantum Science Center supported the high-performance computing as well as the symmetry analysis component of this work. Additional support was provided by the Caltech Institute for Quantum Information and Matter, an NSF Physics Frontiers Center with support of the Gordon and Betty Moore Foundation through Grant GBMF1250, and the Walter Burke Institute for Theoretical Physics at Caltech.

\bibliography{references}

\begin{thebibliography}{70}%
\makeatletter
\providecommand \@ifxundefined [1]{%
 \@ifx{#1\undefined}
}%
\providecommand \@ifnum [1]{%
 \ifnum #1\expandafter \@firstoftwo
 \else \expandafter \@secondoftwo
 \fi
}%
\providecommand \@ifx [1]{%
 \ifx #1\expandafter \@firstoftwo
 \else \expandafter \@secondoftwo
 \fi
}%
\providecommand \natexlab [1]{#1}%
\providecommand \enquote  [1]{``#1''}%
\providecommand \bibnamefont  [1]{#1}%
\providecommand \bibfnamefont [1]{#1}%
\providecommand \citenamefont [1]{#1}%
\providecommand \href@noop [0]{\@secondoftwo}%
\providecommand \href [0]{\begingroup \@sanitize@url \@href}%
\providecommand \@href[1]{\@@startlink{#1}\@@href}%
\providecommand \@@href[1]{\endgroup#1\@@endlink}%
\providecommand \@sanitize@url [0]{\catcode `\\12\catcode `\$12\catcode
  `\&12\catcode `\#12\catcode `\^12\catcode `\_12\catcode `\%12\relax}%
\providecommand \@@startlink[1]{}%
\providecommand \@@endlink[0]{}%
\providecommand \url  [0]{\begingroup\@sanitize@url \@url }%
\providecommand \@url [1]{\endgroup\@href {#1}{\urlprefix }}%
\providecommand \urlprefix  [0]{URL }%
\providecommand \Eprint [0]{\href }%
\providecommand \doibase [0]{https://doi.org/}%
\providecommand \selectlanguage [0]{\@gobble}%
\providecommand \bibinfo  [0]{\@secondoftwo}%
\providecommand \bibfield  [0]{\@secondoftwo}%
\providecommand \translation [1]{[#1]}%
\providecommand \BibitemOpen [0]{}%
\providecommand \bibitemStop [0]{}%
\providecommand \bibitemNoStop [0]{.\EOS\space}%
\providecommand \EOS [0]{\spacefactor3000\relax}%
\providecommand \BibitemShut  [1]{\csname bibitem#1\endcsname}%
\let\auto@bib@innerbib\@empty
\bibitem [{\citenamefont {Weitz}\ \emph {et~al.}(2010)\citenamefont {Weitz},
  \citenamefont {Allen}, \citenamefont {Feldman}, \citenamefont {Martin},\ and\
  \citenamefont {Yacoby}}]{Weitz2010}%
  \BibitemOpen
  \bibfield  {author} {\bibinfo {author} {\bibfnamefont {R.~T.}\ \bibnamefont
  {Weitz}}, \bibinfo {author} {\bibfnamefont {M.~T.}\ \bibnamefont {Allen}},
  \bibinfo {author} {\bibfnamefont {B.~E.}\ \bibnamefont {Feldman}}, \bibinfo
  {author} {\bibfnamefont {J.}~\bibnamefont {Martin}},\ and\ \bibinfo {author}
  {\bibfnamefont {A.}~\bibnamefont {Yacoby}},\ }\bibfield  {title} {\bibinfo
  {title} {Broken-symmetry states in doubly gated suspended bilayer graphene},\
  }\href {https://doi.org/10.1126/science.1194988} {\bibfield  {journal}
  {\bibinfo  {journal} {Science}\ }\textbf {\bibinfo {volume} {330}},\ \bibinfo
  {pages} {812} (\bibinfo {year} {2010})}\BibitemShut {NoStop}%
\bibitem [{\citenamefont {Shi}\ \emph {et~al.}(2020)\citenamefont {Shi},
  \citenamefont {Xu}, \citenamefont {Yang}, \citenamefont {Slizovskiy},
  \citenamefont {Morozov}, \citenamefont {Son}, \citenamefont {Ozdemir},
  \citenamefont {Mullan}, \citenamefont {Barrier}, \citenamefont {Yin},
  \citenamefont {Berdyugin}, \citenamefont {Piot}, \citenamefont {Taniguchi},
  \citenamefont {Watanabe}, \citenamefont {Fal'ko}, \citenamefont {Novoselov},
  \citenamefont {Geim},\ and\ \citenamefont {Mishchenko}}]{Shi2020}%
  \BibitemOpen
  \bibfield  {author} {\bibinfo {author} {\bibfnamefont {Y.}~\bibnamefont
  {Shi}}, \bibinfo {author} {\bibfnamefont {S.}~\bibnamefont {Xu}}, \bibinfo
  {author} {\bibfnamefont {Y.}~\bibnamefont {Yang}}, \bibinfo {author}
  {\bibfnamefont {S.}~\bibnamefont {Slizovskiy}}, \bibinfo {author}
  {\bibfnamefont {S.~V.}\ \bibnamefont {Morozov}}, \bibinfo {author}
  {\bibfnamefont {S.-K.}\ \bibnamefont {Son}}, \bibinfo {author} {\bibfnamefont
  {S.}~\bibnamefont {Ozdemir}}, \bibinfo {author} {\bibfnamefont
  {C.}~\bibnamefont {Mullan}}, \bibinfo {author} {\bibfnamefont
  {J.}~\bibnamefont {Barrier}}, \bibinfo {author} {\bibfnamefont
  {J.}~\bibnamefont {Yin}}, \bibinfo {author} {\bibfnamefont {A.~I.}\
  \bibnamefont {Berdyugin}}, \bibinfo {author} {\bibfnamefont {B.~A.}\
  \bibnamefont {Piot}}, \bibinfo {author} {\bibfnamefont {T.}~\bibnamefont
  {Taniguchi}}, \bibinfo {author} {\bibfnamefont {K.}~\bibnamefont {Watanabe}},
  \bibinfo {author} {\bibfnamefont {V.~I.}\ \bibnamefont {Fal'ko}}, \bibinfo
  {author} {\bibfnamefont {K.~S.}\ \bibnamefont {Novoselov}}, \bibinfo {author}
  {\bibfnamefont {A.~K.}\ \bibnamefont {Geim}},\ and\ \bibinfo {author}
  {\bibfnamefont {A.}~\bibnamefont {Mishchenko}},\ }\bibfield  {title}
  {\bibinfo {title} {Electronic phase separation in multilayer rhombohedral
  graphite},\ }\href {https://doi.org/10.1038/s41586-020-2568-2} {\bibfield
  {journal} {\bibinfo  {journal} {Nature}\ }\textbf {\bibinfo {volume} {584}},\
  \bibinfo {pages} {210} (\bibinfo {year} {2020})}\BibitemShut {NoStop}%
\bibitem [{\citenamefont {Zhou}\ \emph
  {et~al.}(2021{\natexlab{a}})\citenamefont {Zhou}, \citenamefont {Xie},
  \citenamefont {Ghazaryan}, \citenamefont {Holder}, \citenamefont {Ehrets},
  \citenamefont {Spanton}, \citenamefont {Taniguchi}, \citenamefont {Watanabe},
  \citenamefont {Berg}, \citenamefont {Serbyn},\ and\ \citenamefont
  {Young}}]{zhou2021half}%
  \BibitemOpen
  \bibfield  {author} {\bibinfo {author} {\bibfnamefont {H.}~\bibnamefont
  {Zhou}}, \bibinfo {author} {\bibfnamefont {T.}~\bibnamefont {Xie}}, \bibinfo
  {author} {\bibfnamefont {A.}~\bibnamefont {Ghazaryan}}, \bibinfo {author}
  {\bibfnamefont {T.}~\bibnamefont {Holder}}, \bibinfo {author} {\bibfnamefont
  {J.~R.}\ \bibnamefont {Ehrets}}, \bibinfo {author} {\bibfnamefont {E.~M.}\
  \bibnamefont {Spanton}}, \bibinfo {author} {\bibfnamefont {T.}~\bibnamefont
  {Taniguchi}}, \bibinfo {author} {\bibfnamefont {K.}~\bibnamefont {Watanabe}},
  \bibinfo {author} {\bibfnamefont {E.}~\bibnamefont {Berg}}, \bibinfo {author}
  {\bibfnamefont {M.}~\bibnamefont {Serbyn}},\ and\ \bibinfo {author}
  {\bibfnamefont {A.~F.}\ \bibnamefont {Young}},\ }\bibfield  {title} {\bibinfo
  {title} {Half- and quarter-metals in rhombohedral trilayer graphene},\ }\href
  {https://doi.org/10.1038/s41586-021-03938-w} {\bibfield  {journal} {\bibinfo
  {journal} {Nature}\ }\textbf {\bibinfo {volume} {598}},\ \bibinfo {pages}
  {429} (\bibinfo {year} {2021}{\natexlab{a}})}\BibitemShut {NoStop}%
\bibitem [{\citenamefont {Zhou}\ \emph
  {et~al.}(2021{\natexlab{b}})\citenamefont {Zhou}, \citenamefont {Xie},
  \citenamefont {Taniguchi}, \citenamefont {Watanabe},\ and\ \citenamefont
  {Young}}]{zhou2021superconductivity}%
  \BibitemOpen
  \bibfield  {author} {\bibinfo {author} {\bibfnamefont {H.}~\bibnamefont
  {Zhou}}, \bibinfo {author} {\bibfnamefont {T.}~\bibnamefont {Xie}}, \bibinfo
  {author} {\bibfnamefont {T.}~\bibnamefont {Taniguchi}}, \bibinfo {author}
  {\bibfnamefont {K.}~\bibnamefont {Watanabe}},\ and\ \bibinfo {author}
  {\bibfnamefont {A.~F.}\ \bibnamefont {Young}},\ }\bibfield  {title} {\bibinfo
  {title} {Superconductivity in rhombohedral trilayer graphene},\ }\href
  {https://doi.org/10.1038/s41586-021-03926-0} {\bibfield  {journal} {\bibinfo
  {journal} {Nature}\ }\textbf {\bibinfo {volume} {598}},\ \bibinfo {pages}
  {434} (\bibinfo {year} {2021}{\natexlab{b}})}\BibitemShut {NoStop}%
\bibitem [{\citenamefont {Kerelsky}\ \emph {et~al.}(2021)\citenamefont
  {Kerelsky}, \citenamefont {Rubio-Verd{\'{u}}}, \citenamefont {Xian},
  \citenamefont {Kennes}, \citenamefont {Halbertal}, \citenamefont {Finney},
  \citenamefont {Song}, \citenamefont {Turkel}, \citenamefont {Wang},
  \citenamefont {Watanabe}, \citenamefont {Taniguchi}, \citenamefont {Hone},
  \citenamefont {Dean}, \citenamefont {Basov}, \citenamefont {Rubio},\ and\
  \citenamefont {Pasupathy}}]{Kerelsky2021}%
  \BibitemOpen
  \bibfield  {author} {\bibinfo {author} {\bibfnamefont {A.}~\bibnamefont
  {Kerelsky}}, \bibinfo {author} {\bibfnamefont {C.}~\bibnamefont
  {Rubio-Verd{\'{u}}}}, \bibinfo {author} {\bibfnamefont {L.}~\bibnamefont
  {Xian}}, \bibinfo {author} {\bibfnamefont {D.~M.}\ \bibnamefont {Kennes}},
  \bibinfo {author} {\bibfnamefont {D.}~\bibnamefont {Halbertal}}, \bibinfo
  {author} {\bibfnamefont {N.}~\bibnamefont {Finney}}, \bibinfo {author}
  {\bibfnamefont {L.}~\bibnamefont {Song}}, \bibinfo {author} {\bibfnamefont
  {S.}~\bibnamefont {Turkel}}, \bibinfo {author} {\bibfnamefont
  {L.}~\bibnamefont {Wang}}, \bibinfo {author} {\bibfnamefont {K.}~\bibnamefont
  {Watanabe}}, \bibinfo {author} {\bibfnamefont {T.}~\bibnamefont {Taniguchi}},
  \bibinfo {author} {\bibfnamefont {J.}~\bibnamefont {Hone}}, \bibinfo {author}
  {\bibfnamefont {C.}~\bibnamefont {Dean}}, \bibinfo {author} {\bibfnamefont
  {D.~N.}\ \bibnamefont {Basov}}, \bibinfo {author} {\bibfnamefont
  {A.}~\bibnamefont {Rubio}},\ and\ \bibinfo {author} {\bibfnamefont {A.~N.}\
  \bibnamefont {Pasupathy}},\ }\bibfield  {title} {\bibinfo {title}
  {Moir{\'{e}}less correlations in {ABCA} graphene},\ }\bibfield  {journal}
  {\bibinfo  {journal} {Proceedings of the National Academy of Sciences}\
  }\textbf {\bibinfo {volume} {118}},\ \href
  {https://doi.org/10.1073/pnas.2017366118} {10.1073/pnas.2017366118} (\bibinfo
  {year} {2021})\BibitemShut {NoStop}%
\bibitem [{\citenamefont {Zhou}\ \emph {et~al.}(2022)\citenamefont {Zhou},
  \citenamefont {Holleis}, \citenamefont {Saito}, \citenamefont {Cohen},
  \citenamefont {Huynh}, \citenamefont {Patterson}, \citenamefont {Yang},
  \citenamefont {Taniguchi}, \citenamefont {Watanabe},\ and\ \citenamefont
  {Young}}]{zhou2022isospin}%
  \BibitemOpen
  \bibfield  {author} {\bibinfo {author} {\bibfnamefont {H.}~\bibnamefont
  {Zhou}}, \bibinfo {author} {\bibfnamefont {L.}~\bibnamefont {Holleis}},
  \bibinfo {author} {\bibfnamefont {Y.}~\bibnamefont {Saito}}, \bibinfo
  {author} {\bibfnamefont {L.}~\bibnamefont {Cohen}}, \bibinfo {author}
  {\bibfnamefont {W.}~\bibnamefont {Huynh}}, \bibinfo {author} {\bibfnamefont
  {C.~L.}\ \bibnamefont {Patterson}}, \bibinfo {author} {\bibfnamefont
  {F.}~\bibnamefont {Yang}}, \bibinfo {author} {\bibfnamefont {T.}~\bibnamefont
  {Taniguchi}}, \bibinfo {author} {\bibfnamefont {K.}~\bibnamefont
  {Watanabe}},\ and\ \bibinfo {author} {\bibfnamefont {A.~F.}\ \bibnamefont
  {Young}},\ }\bibfield  {title} {\bibinfo {title} {Isospin magnetism and
  spin-polarized superconductivity in bernal bilayer graphene},\ }\href
  {https://doi.org/10.1126/science.abm8386} {\bibfield  {journal} {\bibinfo
  {journal} {Science}\ }\textbf {\bibinfo {volume} {375}},\ \bibinfo {pages}
  {774} (\bibinfo {year} {2022})}\BibitemShut {NoStop}%
\bibitem [{\citenamefont {Seiler}\ \emph {et~al.}(2022)\citenamefont {Seiler},
  \citenamefont {Geisenhof}, \citenamefont {Winterer}, \citenamefont
  {Watanabe}, \citenamefont {Taniguchi}, \citenamefont {Xu}, \citenamefont
  {Zhang},\ and\ \citenamefont {Weitz}}]{Seiler2022}%
  \BibitemOpen
  \bibfield  {author} {\bibinfo {author} {\bibfnamefont {A.~M.}\ \bibnamefont
  {Seiler}}, \bibinfo {author} {\bibfnamefont {F.~R.}\ \bibnamefont
  {Geisenhof}}, \bibinfo {author} {\bibfnamefont {F.}~\bibnamefont {Winterer}},
  \bibinfo {author} {\bibfnamefont {K.}~\bibnamefont {Watanabe}}, \bibinfo
  {author} {\bibfnamefont {T.}~\bibnamefont {Taniguchi}}, \bibinfo {author}
  {\bibfnamefont {T.}~\bibnamefont {Xu}}, \bibinfo {author} {\bibfnamefont
  {F.}~\bibnamefont {Zhang}},\ and\ \bibinfo {author} {\bibfnamefont {R.~T.}\
  \bibnamefont {Weitz}},\ }\bibfield  {title} {\bibinfo {title} {Quantum
  cascade of correlated phases in trigonally warped bilayer graphene},\ }\href
  {https://doi.org/10.1038/s41586-022-04937-1} {\bibfield  {journal} {\bibinfo
  {journal} {Nature}\ }\textbf {\bibinfo {volume} {608}},\ \bibinfo {pages}
  {298} (\bibinfo {year} {2022})}\BibitemShut {NoStop}%
\bibitem [{\citenamefont {de~la Barrera}\ \emph {et~al.}(2022)\citenamefont
  {de~la Barrera}, \citenamefont {Aronson}, \citenamefont {Zheng},
  \citenamefont {Watanabe}, \citenamefont {Taniguchi}, \citenamefont {Ma},
  \citenamefont {Jarillo-Herrero},\ and\ \citenamefont
  {Ashoori}}]{delaBarrera2022}%
  \BibitemOpen
  \bibfield  {author} {\bibinfo {author} {\bibfnamefont {S.~C.}\ \bibnamefont
  {de~la Barrera}}, \bibinfo {author} {\bibfnamefont {S.}~\bibnamefont
  {Aronson}}, \bibinfo {author} {\bibfnamefont {Z.}~\bibnamefont {Zheng}},
  \bibinfo {author} {\bibfnamefont {K.}~\bibnamefont {Watanabe}}, \bibinfo
  {author} {\bibfnamefont {T.}~\bibnamefont {Taniguchi}}, \bibinfo {author}
  {\bibfnamefont {Q.}~\bibnamefont {Ma}}, \bibinfo {author} {\bibfnamefont
  {P.}~\bibnamefont {Jarillo-Herrero}},\ and\ \bibinfo {author} {\bibfnamefont
  {R.}~\bibnamefont {Ashoori}},\ }\bibfield  {title} {\bibinfo {title} {Cascade
  of isospin phase transitions in bernal-stacked bilayer graphene at zero
  magnetic field},\ }\href {https://doi.org/10.1038/s41567-022-01616-w}
  {\bibfield  {journal} {\bibinfo  {journal} {Nature Physics}\ }\textbf
  {\bibinfo {volume} {18}},\ \bibinfo {pages} {771} (\bibinfo {year}
  {2022})}\BibitemShut {NoStop}%
\bibitem [{\citenamefont {Zhang}\ \emph {et~al.}(2023)\citenamefont {Zhang},
  \citenamefont {Polski}, \citenamefont {Thomson}, \citenamefont
  {Lantagne-Hurtubise}, \citenamefont {Lewandowski}, \citenamefont {Zhou},
  \citenamefont {Watanabe}, \citenamefont {Taniguchi}, \citenamefont {Alicea},\
  and\ \citenamefont {Nadj-Perge}}]{Zhang2023}%
  \BibitemOpen
  \bibfield  {author} {\bibinfo {author} {\bibfnamefont {Y.}~\bibnamefont
  {Zhang}}, \bibinfo {author} {\bibfnamefont {R.}~\bibnamefont {Polski}},
  \bibinfo {author} {\bibfnamefont {A.}~\bibnamefont {Thomson}}, \bibinfo
  {author} {\bibfnamefont {{\'E}.}~\bibnamefont {Lantagne-Hurtubise}}, \bibinfo
  {author} {\bibfnamefont {C.}~\bibnamefont {Lewandowski}}, \bibinfo {author}
  {\bibfnamefont {H.}~\bibnamefont {Zhou}}, \bibinfo {author} {\bibfnamefont
  {K.}~\bibnamefont {Watanabe}}, \bibinfo {author} {\bibfnamefont
  {T.}~\bibnamefont {Taniguchi}}, \bibinfo {author} {\bibfnamefont
  {J.}~\bibnamefont {Alicea}},\ and\ \bibinfo {author} {\bibfnamefont
  {S.}~\bibnamefont {Nadj-Perge}},\ }\bibfield  {title} {\bibinfo {title}
  {Enhanced superconductivity in spin--orbit proximitized bilayer graphene},\
  }\href {https://doi.org/10.1038/s41586-022-05446-x} {\bibfield  {journal}
  {\bibinfo  {journal} {Nature}\ }\textbf {\bibinfo {volume} {613}},\ \bibinfo
  {pages} {268} (\bibinfo {year} {2023})}\BibitemShut {NoStop}%
\bibitem [{\citenamefont {Holleis}\ \emph {et~al.}(2023)\citenamefont
  {Holleis}, \citenamefont {Patterson}, \citenamefont {Zhang}, \citenamefont
  {Yoo}, \citenamefont {Zhou}, \citenamefont {Taniguchi}, \citenamefont
  {Watanabe}, \citenamefont {Nadj-Perge},\ and\ \citenamefont
  {Young}}]{Holleis2023}%
  \BibitemOpen
  \bibfield  {author} {\bibinfo {author} {\bibfnamefont {L.}~\bibnamefont
  {Holleis}}, \bibinfo {author} {\bibfnamefont {C.~L.}\ \bibnamefont
  {Patterson}}, \bibinfo {author} {\bibfnamefont {Y.}~\bibnamefont {Zhang}},
  \bibinfo {author} {\bibfnamefont {H.~M.}\ \bibnamefont {Yoo}}, \bibinfo
  {author} {\bibfnamefont {H.}~\bibnamefont {Zhou}}, \bibinfo {author}
  {\bibfnamefont {T.}~\bibnamefont {Taniguchi}}, \bibinfo {author}
  {\bibfnamefont {K.}~\bibnamefont {Watanabe}}, \bibinfo {author}
  {\bibfnamefont {S.}~\bibnamefont {Nadj-Perge}},\ and\ \bibinfo {author}
  {\bibfnamefont {A.~F.}\ \bibnamefont {Young}},\ }\href@noop {} {\bibinfo
  {title} {Ising superconductivity and nematicity in bernal bilayer graphene
  with strong spin orbit coupling}} (\bibinfo {year} {2023}),\ \Eprint
  {https://arxiv.org/abs/2303.00742} {arXiv:2303.00742 [cond-mat.supr-con]}
  \BibitemShut {NoStop}%
\bibitem [{\citenamefont {Han}\ \emph {et~al.}(2023{\natexlab{a}})\citenamefont
  {Han}, \citenamefont {Lu}, \citenamefont {Scuri}, \citenamefont {Sung},
  \citenamefont {Wang}, \citenamefont {Han}, \citenamefont {Watanabe},
  \citenamefont {Taniguchi}, \citenamefont {Park},\ and\ \citenamefont
  {Ju}}]{Han2023}%
  \BibitemOpen
  \bibfield  {author} {\bibinfo {author} {\bibfnamefont {T.}~\bibnamefont
  {Han}}, \bibinfo {author} {\bibfnamefont {Z.}~\bibnamefont {Lu}}, \bibinfo
  {author} {\bibfnamefont {G.}~\bibnamefont {Scuri}}, \bibinfo {author}
  {\bibfnamefont {J.}~\bibnamefont {Sung}}, \bibinfo {author} {\bibfnamefont
  {J.}~\bibnamefont {Wang}}, \bibinfo {author} {\bibfnamefont {T.}~\bibnamefont
  {Han}}, \bibinfo {author} {\bibfnamefont {K.}~\bibnamefont {Watanabe}},
  \bibinfo {author} {\bibfnamefont {T.}~\bibnamefont {Taniguchi}}, \bibinfo
  {author} {\bibfnamefont {H.}~\bibnamefont {Park}},\ and\ \bibinfo {author}
  {\bibfnamefont {L.}~\bibnamefont {Ju}},\ }\href@noop {} {\bibinfo {title}
  {Correlated insulator and chern insulators in pentalayer rhombohedral stacked
  graphene}} (\bibinfo {year} {2023}{\natexlab{a}}),\ \Eprint
  {https://arxiv.org/abs/2305.03151} {arXiv:2305.03151 [cond-mat.mes-hall]}
  \BibitemShut {NoStop}%
\bibitem [{\citenamefont {Liu}\ \emph {et~al.}(2023)\citenamefont {Liu},
  \citenamefont {Zheng}, \citenamefont {Sha}, \citenamefont {Lyu},
  \citenamefont {Li}, \citenamefont {Park}, \citenamefont {Ren}, \citenamefont
  {Watanabe}, \citenamefont {Taniguchi}, \citenamefont {Jia}, \citenamefont
  {Luo}, \citenamefont {Shi}, \citenamefont {Jung},\ and\ \citenamefont
  {Chen}}]{liu2023interactiondriven}%
  \BibitemOpen
  \bibfield  {author} {\bibinfo {author} {\bibfnamefont {K.}~\bibnamefont
  {Liu}}, \bibinfo {author} {\bibfnamefont {J.}~\bibnamefont {Zheng}}, \bibinfo
  {author} {\bibfnamefont {Y.}~\bibnamefont {Sha}}, \bibinfo {author}
  {\bibfnamefont {B.}~\bibnamefont {Lyu}}, \bibinfo {author} {\bibfnamefont
  {F.}~\bibnamefont {Li}}, \bibinfo {author} {\bibfnamefont {Y.}~\bibnamefont
  {Park}}, \bibinfo {author} {\bibfnamefont {Y.}~\bibnamefont {Ren}}, \bibinfo
  {author} {\bibfnamefont {K.}~\bibnamefont {Watanabe}}, \bibinfo {author}
  {\bibfnamefont {T.}~\bibnamefont {Taniguchi}}, \bibinfo {author}
  {\bibfnamefont {J.}~\bibnamefont {Jia}}, \bibinfo {author} {\bibfnamefont
  {W.}~\bibnamefont {Luo}}, \bibinfo {author} {\bibfnamefont {Z.}~\bibnamefont
  {Shi}}, \bibinfo {author} {\bibfnamefont {J.}~\bibnamefont {Jung}},\ and\
  \bibinfo {author} {\bibfnamefont {G.}~\bibnamefont {Chen}},\ }\href@noop {}
  {\bibinfo {title} {Interaction-driven spontaneous broken-symmetry insulator
  and metals in abca tetralayer graphene}} (\bibinfo {year} {2023}),\ \Eprint
  {https://arxiv.org/abs/2306.11042} {arXiv:2306.11042 [cond-mat.mes-hall]}
  \BibitemShut {NoStop}%
\bibitem [{\citenamefont {Han}\ \emph {et~al.}(2023{\natexlab{b}})\citenamefont
  {Han}, \citenamefont {Lu}, \citenamefont {Scuri}, \citenamefont {Sung},
  \citenamefont {Wang}, \citenamefont {Han}, \citenamefont {Watanabe},
  \citenamefont {Taniguchi}, \citenamefont {Fu}, \citenamefont {Park},\ and\
  \citenamefont {Ju}}]{han2023orbital}%
  \BibitemOpen
  \bibfield  {author} {\bibinfo {author} {\bibfnamefont {T.}~\bibnamefont
  {Han}}, \bibinfo {author} {\bibfnamefont {Z.}~\bibnamefont {Lu}}, \bibinfo
  {author} {\bibfnamefont {G.}~\bibnamefont {Scuri}}, \bibinfo {author}
  {\bibfnamefont {J.}~\bibnamefont {Sung}}, \bibinfo {author} {\bibfnamefont
  {J.}~\bibnamefont {Wang}}, \bibinfo {author} {\bibfnamefont {T.}~\bibnamefont
  {Han}}, \bibinfo {author} {\bibfnamefont {K.}~\bibnamefont {Watanabe}},
  \bibinfo {author} {\bibfnamefont {T.}~\bibnamefont {Taniguchi}}, \bibinfo
  {author} {\bibfnamefont {L.}~\bibnamefont {Fu}}, \bibinfo {author}
  {\bibfnamefont {H.}~\bibnamefont {Park}},\ and\ \bibinfo {author}
  {\bibfnamefont {L.}~\bibnamefont {Ju}},\ }\href@noop {} {\bibinfo {title}
  {Orbital multiferroicity in pentalayer rhombohedral graphene}} (\bibinfo
  {year} {2023}{\natexlab{b}}),\ \Eprint {https://arxiv.org/abs/2308.08837}
  {arXiv:2308.08837 [cond-mat.mes-hall]} \BibitemShut {NoStop}%
\bibitem [{\citenamefont {Lu}\ \emph {et~al.}(2023)\citenamefont {Lu},
  \citenamefont {Han}, \citenamefont {Yao}, \citenamefont {Reddy},
  \citenamefont {Yang}, \citenamefont {Seo}, \citenamefont {Watanabe},
  \citenamefont {Taniguchi}, \citenamefont {Fu},\ and\ \citenamefont
  {Ju}}]{lu2023fractional}%
  \BibitemOpen
  \bibfield  {author} {\bibinfo {author} {\bibfnamefont {Z.}~\bibnamefont
  {Lu}}, \bibinfo {author} {\bibfnamefont {T.}~\bibnamefont {Han}}, \bibinfo
  {author} {\bibfnamefont {Y.}~\bibnamefont {Yao}}, \bibinfo {author}
  {\bibfnamefont {A.~P.}\ \bibnamefont {Reddy}}, \bibinfo {author}
  {\bibfnamefont {J.}~\bibnamefont {Yang}}, \bibinfo {author} {\bibfnamefont
  {J.}~\bibnamefont {Seo}}, \bibinfo {author} {\bibfnamefont {K.}~\bibnamefont
  {Watanabe}}, \bibinfo {author} {\bibfnamefont {T.}~\bibnamefont {Taniguchi}},
  \bibinfo {author} {\bibfnamefont {L.}~\bibnamefont {Fu}},\ and\ \bibinfo
  {author} {\bibfnamefont {L.}~\bibnamefont {Ju}},\ }\href@noop {} {\bibinfo
  {title} {Fractional quantum anomalous hall effect in a graphene moire
  superlattice}} (\bibinfo {year} {2023}),\ \Eprint
  {https://arxiv.org/abs/2309.17436} {arXiv:2309.17436 [cond-mat.mes-hall]}
  \BibitemShut {NoStop}%
\bibitem [{\citenamefont {Chou}\ \emph
  {et~al.}(2022{\natexlab{a}})\citenamefont {Chou}, \citenamefont {Wu},
  \citenamefont {Sau},\ and\ \citenamefont {Das~Sarma}}]{Chou2022}%
  \BibitemOpen
  \bibfield  {author} {\bibinfo {author} {\bibfnamefont {Y.-Z.}\ \bibnamefont
  {Chou}}, \bibinfo {author} {\bibfnamefont {F.}~\bibnamefont {Wu}}, \bibinfo
  {author} {\bibfnamefont {J.~D.}\ \bibnamefont {Sau}},\ and\ \bibinfo {author}
  {\bibfnamefont {S.}~\bibnamefont {Das~Sarma}},\ }\bibfield  {title} {\bibinfo
  {title} {Acoustic-phonon-mediated superconductivity in bernal bilayer
  graphene},\ }\href {https://doi.org/10.1103/PhysRevB.105.L100503} {\bibfield
  {journal} {\bibinfo  {journal} {Phys. Rev. B}\ }\textbf {\bibinfo {volume}
  {105}},\ \bibinfo {pages} {L100503} (\bibinfo {year}
  {2022}{\natexlab{a}})}\BibitemShut {NoStop}%
\bibitem [{\citenamefont {Dong}\ \emph
  {et~al.}(2023{\natexlab{a}})\citenamefont {Dong}, \citenamefont {Chubukov},\
  and\ \citenamefont {Levitov}}]{Dong2023}%
  \BibitemOpen
  \bibfield  {author} {\bibinfo {author} {\bibfnamefont {Z.}~\bibnamefont
  {Dong}}, \bibinfo {author} {\bibfnamefont {A.~V.}\ \bibnamefont {Chubukov}},\
  and\ \bibinfo {author} {\bibfnamefont {L.}~\bibnamefont {Levitov}},\
  }\bibfield  {title} {\bibinfo {title} {Transformer spin-triplet
  superconductivity at the onset of isospin order in bilayer graphene},\ }\href
  {https://doi.org/10.1103/PhysRevB.107.174512} {\bibfield  {journal} {\bibinfo
   {journal} {Phys. Rev. B}\ }\textbf {\bibinfo {volume} {107}},\ \bibinfo
  {pages} {174512} (\bibinfo {year} {2023}{\natexlab{a}})}\BibitemShut
  {NoStop}%
\bibitem [{\citenamefont {Dong}\ \emph
  {et~al.}(2023{\natexlab{b}})\citenamefont {Dong}, \citenamefont {Lee},\ and\
  \citenamefont {Levitov}}]{dong2023signatures}%
  \BibitemOpen
  \bibfield  {author} {\bibinfo {author} {\bibfnamefont {Z.}~\bibnamefont
  {Dong}}, \bibinfo {author} {\bibfnamefont {P.~A.}\ \bibnamefont {Lee}},\ and\
  \bibinfo {author} {\bibfnamefont {L.~S.}\ \bibnamefont {Levitov}},\
  }\href@noop {} {\bibinfo {title} {Signatures of cooper pair dynamics and
  quantum-critical superconductivity in tunable carrier bands}} (\bibinfo
  {year} {2023}{\natexlab{b}}),\ \Eprint {https://arxiv.org/abs/2304.09812}
  {arXiv:2304.09812 [cond-mat.supr-con]} \BibitemShut {NoStop}%
\bibitem [{\citenamefont {Shavit}\ and\ \citenamefont
  {Oreg}(2023)}]{Shavit2023}%
  \BibitemOpen
  \bibfield  {author} {\bibinfo {author} {\bibfnamefont {G.}~\bibnamefont
  {Shavit}}\ and\ \bibinfo {author} {\bibfnamefont {Y.}~\bibnamefont {Oreg}},\
  }\href@noop {} {\bibinfo {title} {Inducing superconductivity in bilayer
  graphene by alleviation of the stoner blockade}} (\bibinfo {year} {2023}),\
  \Eprint {https://arxiv.org/abs/2303.04176} {arXiv:2303.04176
  [cond-mat.supr-con]} \BibitemShut {NoStop}%
\bibitem [{\citenamefont {Wagner}\ \emph {et~al.}(2023)\citenamefont {Wagner},
  \citenamefont {Kwan}, \citenamefont {Bultinck}, \citenamefont {Simon},\ and\
  \citenamefont {Parameswaran}}]{Wagner2023}%
  \BibitemOpen
  \bibfield  {author} {\bibinfo {author} {\bibfnamefont {G.}~\bibnamefont
  {Wagner}}, \bibinfo {author} {\bibfnamefont {Y.~H.}\ \bibnamefont {Kwan}},
  \bibinfo {author} {\bibfnamefont {N.}~\bibnamefont {Bultinck}}, \bibinfo
  {author} {\bibfnamefont {S.~H.}\ \bibnamefont {Simon}},\ and\ \bibinfo
  {author} {\bibfnamefont {S.~A.}\ \bibnamefont {Parameswaran}},\ }\href@noop
  {} {\bibinfo {title} {Superconductivity from repulsive interactions in
  bernal-stacked bilayer graphene}} (\bibinfo {year} {2023}),\ \Eprint
  {https://arxiv.org/abs/2302.00682} {arXiv:2302.00682 [cond-mat.supr-con]}
  \BibitemShut {NoStop}%
\bibitem [{\citenamefont {Jimeno-Pozo}\ \emph {et~al.}(2023)\citenamefont
  {Jimeno-Pozo}, \citenamefont {Sainz-Cruz}, \citenamefont {Cea}, \citenamefont
  {Pantale\'on},\ and\ \citenamefont {Guinea}}]{Jimeno-Pozo2022}%
  \BibitemOpen
  \bibfield  {author} {\bibinfo {author} {\bibfnamefont {A.}~\bibnamefont
  {Jimeno-Pozo}}, \bibinfo {author} {\bibfnamefont {H.}~\bibnamefont
  {Sainz-Cruz}}, \bibinfo {author} {\bibfnamefont {T.}~\bibnamefont {Cea}},
  \bibinfo {author} {\bibfnamefont {P.~A.}\ \bibnamefont {Pantale\'on}},\ and\
  \bibinfo {author} {\bibfnamefont {F.}~\bibnamefont {Guinea}},\ }\bibfield
  {title} {\bibinfo {title} {Superconductivity from electronic interactions and
  spin-orbit enhancement in bilayer and trilayer graphene},\ }\href
  {https://doi.org/10.1103/PhysRevB.107.L161106} {\bibfield  {journal}
  {\bibinfo  {journal} {Phys. Rev. B}\ }\textbf {\bibinfo {volume} {107}},\
  \bibinfo {pages} {L161106} (\bibinfo {year} {2023})}\BibitemShut {NoStop}%
\bibitem [{\citenamefont {Li}\ \emph {et~al.}(2023)\citenamefont {Li},
  \citenamefont {Kuang}, \citenamefont {Jimeno-Pozo}, \citenamefont
  {Sainz-Cruz}, \citenamefont {Zhan}, \citenamefont {Yuan},\ and\ \citenamefont
  {Guinea}}]{li2023charge}%
  \BibitemOpen
  \bibfield  {author} {\bibinfo {author} {\bibfnamefont {Z.}~\bibnamefont
  {Li}}, \bibinfo {author} {\bibfnamefont {X.}~\bibnamefont {Kuang}}, \bibinfo
  {author} {\bibfnamefont {A.}~\bibnamefont {Jimeno-Pozo}}, \bibinfo {author}
  {\bibfnamefont {H.}~\bibnamefont {Sainz-Cruz}}, \bibinfo {author}
  {\bibfnamefont {Z.}~\bibnamefont {Zhan}}, \bibinfo {author} {\bibfnamefont
  {S.}~\bibnamefont {Yuan}},\ and\ \bibinfo {author} {\bibfnamefont
  {F.}~\bibnamefont {Guinea}},\ }\href@noop {} {\bibinfo {title} {Charge
  fluctuations, phonons and superconductivity in multilayer graphene}}
  (\bibinfo {year} {2023}),\ \Eprint {https://arxiv.org/abs/2303.17286}
  {arXiv:2303.17286 [cond-mat.mes-hall]} \BibitemShut {NoStop}%
\bibitem [{\citenamefont {Dong}\ \emph
  {et~al.}(2023{\natexlab{c}})\citenamefont {Dong}, \citenamefont {Levitov},\
  and\ \citenamefont {Chubukov}}]{Dong2023superconductivity}%
  \BibitemOpen
  \bibfield  {author} {\bibinfo {author} {\bibfnamefont {Z.}~\bibnamefont
  {Dong}}, \bibinfo {author} {\bibfnamefont {L.}~\bibnamefont {Levitov}},\ and\
  \bibinfo {author} {\bibfnamefont {A.~V.}\ \bibnamefont {Chubukov}},\
  }\href@noop {} {\bibinfo {title} {Superconductivity near spin and valley
  orders in graphene multilayers: a systematic study}} (\bibinfo {year}
  {2023}{\natexlab{c}}),\ \Eprint {https://arxiv.org/abs/2306.11005}
  {arXiv:2306.11005 [cond-mat.supr-con]} \BibitemShut {NoStop}%
\bibitem [{\citenamefont {Szab\'o}\ and\ \citenamefont
  {Roy}(2022{\natexlab{a}})}]{szabo2022competing}%
  \BibitemOpen
  \bibfield  {author} {\bibinfo {author} {\bibfnamefont {A.~L.}\ \bibnamefont
  {Szab\'o}}\ and\ \bibinfo {author} {\bibfnamefont {B.}~\bibnamefont {Roy}},\
  }\bibfield  {title} {\bibinfo {title} {Competing orders and cascade of
  degeneracy lifting in doped bernal bilayer graphene},\ }\href
  {https://doi.org/10.1103/PhysRevB.105.L201107} {\bibfield  {journal}
  {\bibinfo  {journal} {Phys. Rev. B}\ }\textbf {\bibinfo {volume} {105}},\
  \bibinfo {pages} {L201107} (\bibinfo {year}
  {2022}{\natexlab{a}})}\BibitemShut {NoStop}%
\bibitem [{\citenamefont {Curtis}\ \emph {et~al.}(2023)\citenamefont {Curtis},
  \citenamefont {Poniatowski}, \citenamefont {Xie}, \citenamefont {Yacoby},
  \citenamefont {Demler},\ and\ \citenamefont {Narang}}]{Curtis2023}%
  \BibitemOpen
  \bibfield  {author} {\bibinfo {author} {\bibfnamefont {J.~B.}\ \bibnamefont
  {Curtis}}, \bibinfo {author} {\bibfnamefont {N.~R.}\ \bibnamefont
  {Poniatowski}}, \bibinfo {author} {\bibfnamefont {Y.}~\bibnamefont {Xie}},
  \bibinfo {author} {\bibfnamefont {A.}~\bibnamefont {Yacoby}}, \bibinfo
  {author} {\bibfnamefont {E.}~\bibnamefont {Demler}},\ and\ \bibinfo {author}
  {\bibfnamefont {P.}~\bibnamefont {Narang}},\ }\bibfield  {title} {\bibinfo
  {title} {Stabilizing fluctuating spin-triplet superconductivity in graphene
  via induced spin-orbit coupling},\ }\href
  {https://doi.org/10.1103/PhysRevLett.130.196001} {\bibfield  {journal}
  {\bibinfo  {journal} {Phys. Rev. Lett.}\ }\textbf {\bibinfo {volume} {130}},\
  \bibinfo {pages} {196001} (\bibinfo {year} {2023})}\BibitemShut {NoStop}%
\bibitem [{\citenamefont {Xie}\ and\ \citenamefont
  {Das~Sarma}(2023)}]{Ming2023}%
  \BibitemOpen
  \bibfield  {author} {\bibinfo {author} {\bibfnamefont {M.}~\bibnamefont
  {Xie}}\ and\ \bibinfo {author} {\bibfnamefont {S.}~\bibnamefont
  {Das~Sarma}},\ }\bibfield  {title} {\bibinfo {title} {Flavor symmetry
  breaking in spin-orbit coupled bilayer graphene},\ }\href
  {https://doi.org/10.1103/PhysRevB.107.L201119} {\bibfield  {journal}
  {\bibinfo  {journal} {Phys. Rev. B}\ }\textbf {\bibinfo {volume} {107}},\
  \bibinfo {pages} {L201119} (\bibinfo {year} {2023})}\BibitemShut {NoStop}%
\bibitem [{\citenamefont {Chou}\ \emph
  {et~al.}(2022{\natexlab{b}})\citenamefont {Chou}, \citenamefont {Wu},\ and\
  \citenamefont {Das~Sarma}}]{Yangzhi2022}%
  \BibitemOpen
  \bibfield  {author} {\bibinfo {author} {\bibfnamefont {Y.-Z.}\ \bibnamefont
  {Chou}}, \bibinfo {author} {\bibfnamefont {F.}~\bibnamefont {Wu}},\ and\
  \bibinfo {author} {\bibfnamefont {S.}~\bibnamefont {Das~Sarma}},\ }\bibfield
  {title} {\bibinfo {title} {Enhanced superconductivity through virtual
  tunneling in bernal bilayer graphene coupled to ${\mathrm{wse}}_{2}$},\
  }\href {https://doi.org/10.1103/PhysRevB.106.L180502} {\bibfield  {journal}
  {\bibinfo  {journal} {Phys. Rev. B}\ }\textbf {\bibinfo {volume} {106}},\
  \bibinfo {pages} {L180502} (\bibinfo {year}
  {2022}{\natexlab{b}})}\BibitemShut {NoStop}%
\bibitem [{\citenamefont {Chou}\ \emph {et~al.}(2021)\citenamefont {Chou},
  \citenamefont {Wu}, \citenamefont {Sau},\ and\ \citenamefont
  {Das~Sarma}}]{chou2021acoustic}%
  \BibitemOpen
  \bibfield  {author} {\bibinfo {author} {\bibfnamefont {Y.-Z.}\ \bibnamefont
  {Chou}}, \bibinfo {author} {\bibfnamefont {F.}~\bibnamefont {Wu}}, \bibinfo
  {author} {\bibfnamefont {J.~D.}\ \bibnamefont {Sau}},\ and\ \bibinfo {author}
  {\bibfnamefont {S.}~\bibnamefont {Das~Sarma}},\ }\bibfield  {title} {\bibinfo
  {title} {Acoustic-phonon-mediated superconductivity in rhombohedral trilayer
  graphene},\ }\href {https://doi.org/10.1103/PhysRevLett.127.187001}
  {\bibfield  {journal} {\bibinfo  {journal} {Phys. Rev. Lett.}\ }\textbf
  {\bibinfo {volume} {127}},\ \bibinfo {pages} {187001} (\bibinfo {year}
  {2021})}\BibitemShut {NoStop}%
\bibitem [{\citenamefont {Cea}\ \emph {et~al.}(2022)\citenamefont {Cea},
  \citenamefont {Pantale\'on}, \citenamefont {Phong},\ and\ \citenamefont
  {Guinea}}]{Cea2022}%
  \BibitemOpen
  \bibfield  {author} {\bibinfo {author} {\bibfnamefont {T.}~\bibnamefont
  {Cea}}, \bibinfo {author} {\bibfnamefont {P.~A.}\ \bibnamefont
  {Pantale\'on}}, \bibinfo {author} {\bibfnamefont {V.~o.~T.}\ \bibnamefont
  {Phong}},\ and\ \bibinfo {author} {\bibfnamefont {F.}~\bibnamefont
  {Guinea}},\ }\bibfield  {title} {\bibinfo {title} {Superconductivity from
  repulsive interactions in rhombohedral trilayer graphene: A
  kohn-luttinger-like mechanism},\ }\href
  {https://doi.org/10.1103/PhysRevB.105.075432} {\bibfield  {journal} {\bibinfo
   {journal} {Phys. Rev. B}\ }\textbf {\bibinfo {volume} {105}},\ \bibinfo
  {pages} {075432} (\bibinfo {year} {2022})}\BibitemShut {NoStop}%
\bibitem [{\citenamefont {You}\ and\ \citenamefont
  {Vishwanath}(2022)}]{You2022}%
  \BibitemOpen
  \bibfield  {author} {\bibinfo {author} {\bibfnamefont {Y.-Z.}\ \bibnamefont
  {You}}\ and\ \bibinfo {author} {\bibfnamefont {A.}~\bibnamefont
  {Vishwanath}},\ }\bibfield  {title} {\bibinfo {title} {Kohn-luttinger
  superconductivity and intervalley coherence in rhombohedral trilayer
  graphene},\ }\href {https://doi.org/10.1103/PhysRevB.105.134524} {\bibfield
  {journal} {\bibinfo  {journal} {Phys. Rev. B}\ }\textbf {\bibinfo {volume}
  {105}},\ \bibinfo {pages} {134524} (\bibinfo {year} {2022})}\BibitemShut
  {NoStop}%
\bibitem [{\citenamefont {Ghazaryan}\ \emph {et~al.}(2021)\citenamefont
  {Ghazaryan}, \citenamefont {Holder}, \citenamefont {Serbyn},\ and\
  \citenamefont {Berg}}]{Berg2021}%
  \BibitemOpen
  \bibfield  {author} {\bibinfo {author} {\bibfnamefont {A.}~\bibnamefont
  {Ghazaryan}}, \bibinfo {author} {\bibfnamefont {T.}~\bibnamefont {Holder}},
  \bibinfo {author} {\bibfnamefont {M.}~\bibnamefont {Serbyn}},\ and\ \bibinfo
  {author} {\bibfnamefont {E.}~\bibnamefont {Berg}},\ }\bibfield  {title}
  {\bibinfo {title} {Unconventional superconductivity in systems with annular
  fermi surfaces: Application to rhombohedral trilayer graphene},\ }\href
  {https://doi.org/10.1103/PhysRevLett.127.247001} {\bibfield  {journal}
  {\bibinfo  {journal} {Phys. Rev. Lett.}\ }\textbf {\bibinfo {volume} {127}},\
  \bibinfo {pages} {247001} (\bibinfo {year} {2021})}\BibitemShut {NoStop}%
\bibitem [{\citenamefont {Dong}\ and\ \citenamefont
  {Levitov}(2021)}]{dong2021superconductivity}%
  \BibitemOpen
  \bibfield  {author} {\bibinfo {author} {\bibfnamefont {Z.}~\bibnamefont
  {Dong}}\ and\ \bibinfo {author} {\bibfnamefont {L.}~\bibnamefont {Levitov}},\
  }\href@noop {} {\bibinfo {title} {Superconductivity in the vicinity of an
  isospin-polarized state in a cubic dirac band}} (\bibinfo {year} {2021}),\
  \Eprint {https://arxiv.org/abs/2109.01133} {arXiv:2109.01133
  [cond-mat.supr-con]} \BibitemShut {NoStop}%
\bibitem [{\citenamefont {Chatterjee}\ \emph {et~al.}(2022)\citenamefont
  {Chatterjee}, \citenamefont {Wang}, \citenamefont {Berg},\ and\ \citenamefont
  {Zaletel}}]{Chatterjee2022}%
  \BibitemOpen
  \bibfield  {author} {\bibinfo {author} {\bibfnamefont {S.}~\bibnamefont
  {Chatterjee}}, \bibinfo {author} {\bibfnamefont {T.}~\bibnamefont {Wang}},
  \bibinfo {author} {\bibfnamefont {E.}~\bibnamefont {Berg}},\ and\ \bibinfo
  {author} {\bibfnamefont {M.~P.}\ \bibnamefont {Zaletel}},\ }\bibfield
  {title} {\bibinfo {title} {Inter-valley coherent order and isospin
  fluctuation mediated superconductivity in rhombohedral trilayer graphene},\
  }\href {https://doi.org/10.1038/s41467-022-33561-w} {\bibfield  {journal}
  {\bibinfo  {journal} {Nature Communications}\ }\textbf {\bibinfo {volume}
  {13}},\ \bibinfo {pages} {6013} (\bibinfo {year} {2022})}\BibitemShut
  {NoStop}%
\bibitem [{\citenamefont {Szab\'o}\ and\ \citenamefont
  {Roy}(2022{\natexlab{b}})}]{Szab2022}%
  \BibitemOpen
  \bibfield  {author} {\bibinfo {author} {\bibfnamefont {A.~L.}\ \bibnamefont
  {Szab\'o}}\ and\ \bibinfo {author} {\bibfnamefont {B.}~\bibnamefont {Roy}},\
  }\bibfield  {title} {\bibinfo {title} {Metals, fractional metals, and
  superconductivity in rhombohedral trilayer graphene},\ }\href
  {https://doi.org/10.1103/PhysRevB.105.L081407} {\bibfield  {journal}
  {\bibinfo  {journal} {Phys. Rev. B}\ }\textbf {\bibinfo {volume} {105}},\
  \bibinfo {pages} {L081407} (\bibinfo {year}
  {2022}{\natexlab{b}})}\BibitemShut {NoStop}%
\bibitem [{\citenamefont {Qin}\ \emph {et~al.}(2023)\citenamefont {Qin},
  \citenamefont {Huang}, \citenamefont {Wolf}, \citenamefont {Wei},
  \citenamefont {Blinov},\ and\ \citenamefont {MacDonald}}]{Qin2023}%
  \BibitemOpen
  \bibfield  {author} {\bibinfo {author} {\bibfnamefont {W.}~\bibnamefont
  {Qin}}, \bibinfo {author} {\bibfnamefont {C.}~\bibnamefont {Huang}}, \bibinfo
  {author} {\bibfnamefont {T.}~\bibnamefont {Wolf}}, \bibinfo {author}
  {\bibfnamefont {N.}~\bibnamefont {Wei}}, \bibinfo {author} {\bibfnamefont
  {I.}~\bibnamefont {Blinov}},\ and\ \bibinfo {author} {\bibfnamefont {A.~H.}\
  \bibnamefont {MacDonald}},\ }\bibfield  {title} {\bibinfo {title} {Functional
  renormalization group study of superconductivity in rhombohedral trilayer
  graphene},\ }\href {https://doi.org/10.1103/PhysRevLett.130.146001}
  {\bibfield  {journal} {\bibinfo  {journal} {Phys. Rev. Lett.}\ }\textbf
  {\bibinfo {volume} {130}},\ \bibinfo {pages} {146001} (\bibinfo {year}
  {2023})}\BibitemShut {NoStop}%
\bibitem [{\citenamefont {Lu}\ \emph {et~al.}(2022)\citenamefont {Lu},
  \citenamefont {Wang}, \citenamefont {Chatterjee},\ and\ \citenamefont
  {You}}]{Lu2022}%
  \BibitemOpen
  \bibfield  {author} {\bibinfo {author} {\bibfnamefont {D.-C.}\ \bibnamefont
  {Lu}}, \bibinfo {author} {\bibfnamefont {T.}~\bibnamefont {Wang}}, \bibinfo
  {author} {\bibfnamefont {S.}~\bibnamefont {Chatterjee}},\ and\ \bibinfo
  {author} {\bibfnamefont {Y.-Z.}\ \bibnamefont {You}},\ }\bibfield  {title}
  {\bibinfo {title} {Correlated metals and unconventional superconductivity in
  rhombohedral trilayer graphene: A renormalization group analysis},\ }\href
  {https://doi.org/10.1103/PhysRevB.106.155115} {\bibfield  {journal} {\bibinfo
   {journal} {Phys. Rev. B}\ }\textbf {\bibinfo {volume} {106}},\ \bibinfo
  {pages} {155115} (\bibinfo {year} {2022})}\BibitemShut {NoStop}%
\bibitem [{\citenamefont {Dai}\ \emph {et~al.}(2023)\citenamefont {Dai},
  \citenamefont {Ma}, \citenamefont {Zhang}, \citenamefont {Guo},\ and\
  \citenamefont {Ma}}]{Dai2023}%
  \BibitemOpen
  \bibfield  {author} {\bibinfo {author} {\bibfnamefont {H.}~\bibnamefont
  {Dai}}, \bibinfo {author} {\bibfnamefont {R.}~\bibnamefont {Ma}}, \bibinfo
  {author} {\bibfnamefont {X.}~\bibnamefont {Zhang}}, \bibinfo {author}
  {\bibfnamefont {T.}~\bibnamefont {Guo}},\ and\ \bibinfo {author}
  {\bibfnamefont {T.}~\bibnamefont {Ma}},\ }\bibfield  {title} {\bibinfo
  {title} {Quantum monte carlo study of superconductivity in rhombohedral
  trilayer graphene under an electric field},\ }\href
  {https://doi.org/10.1103/PhysRevB.107.245106} {\bibfield  {journal} {\bibinfo
   {journal} {Phys. Rev. B}\ }\textbf {\bibinfo {volume} {107}},\ \bibinfo
  {pages} {245106} (\bibinfo {year} {2023})}\BibitemShut {NoStop}%
\bibitem [{\citenamefont {Liu}\ \emph {et~al.}(2022)\citenamefont {Liu},
  \citenamefont {Farahi}, \citenamefont {Chiu}, \citenamefont {Papic},
  \citenamefont {Watanabe}, \citenamefont {Taniguchi}, \citenamefont
  {Zaletel},\ and\ \citenamefont {Yazdani}}]{Liu2022}%
  \BibitemOpen
  \bibfield  {author} {\bibinfo {author} {\bibfnamefont {X.}~\bibnamefont
  {Liu}}, \bibinfo {author} {\bibfnamefont {G.}~\bibnamefont {Farahi}},
  \bibinfo {author} {\bibfnamefont {C.-L.}\ \bibnamefont {Chiu}}, \bibinfo
  {author} {\bibfnamefont {Z.}~\bibnamefont {Papic}}, \bibinfo {author}
  {\bibfnamefont {K.}~\bibnamefont {Watanabe}}, \bibinfo {author}
  {\bibfnamefont {T.}~\bibnamefont {Taniguchi}}, \bibinfo {author}
  {\bibfnamefont {M.~P.}\ \bibnamefont {Zaletel}},\ and\ \bibinfo {author}
  {\bibfnamefont {A.}~\bibnamefont {Yazdani}},\ }\bibfield  {title} {\bibinfo
  {title} {Visualizing broken symmetry and topological defects in a quantum
  hall ferromagnet},\ }\href {https://doi.org/10.1126/science.abm3770}
  {\bibfield  {journal} {\bibinfo  {journal} {Science}\ }\textbf {\bibinfo
  {volume} {375}},\ \bibinfo {pages} {321} (\bibinfo {year}
  {2022})}\BibitemShut {NoStop}%
\bibitem [{\citenamefont {Coissard}\ \emph {et~al.}(2022)\citenamefont
  {Coissard}, \citenamefont {Wander}, \citenamefont {Vignaud}, \citenamefont
  {Grushin}, \citenamefont {Repellin}, \citenamefont {Watanabe}, \citenamefont
  {Taniguchi}, \citenamefont {Gay}, \citenamefont {Winkelmann}, \citenamefont
  {Courtois}, \citenamefont {Sellier},\ and\ \citenamefont
  {Sac{\'{e}}p{\'{e}}}}]{Coissard2022}%
  \BibitemOpen
  \bibfield  {author} {\bibinfo {author} {\bibfnamefont {A.}~\bibnamefont
  {Coissard}}, \bibinfo {author} {\bibfnamefont {D.}~\bibnamefont {Wander}},
  \bibinfo {author} {\bibfnamefont {H.}~\bibnamefont {Vignaud}}, \bibinfo
  {author} {\bibfnamefont {A.~G.}\ \bibnamefont {Grushin}}, \bibinfo {author}
  {\bibfnamefont {C.}~\bibnamefont {Repellin}}, \bibinfo {author}
  {\bibfnamefont {K.}~\bibnamefont {Watanabe}}, \bibinfo {author}
  {\bibfnamefont {T.}~\bibnamefont {Taniguchi}}, \bibinfo {author}
  {\bibfnamefont {F.}~\bibnamefont {Gay}}, \bibinfo {author} {\bibfnamefont
  {C.~B.}\ \bibnamefont {Winkelmann}}, \bibinfo {author} {\bibfnamefont
  {H.}~\bibnamefont {Courtois}}, \bibinfo {author} {\bibfnamefont
  {H.}~\bibnamefont {Sellier}},\ and\ \bibinfo {author} {\bibfnamefont
  {B.}~\bibnamefont {Sac{\'{e}}p{\'{e}}}},\ }\bibfield  {title} {\bibinfo
  {title} {Imaging tunable quantum hall broken-symmetry orders in graphene},\
  }\href {https://doi.org/10.1038/s41586-022-04513-7} {\bibfield  {journal}
  {\bibinfo  {journal} {Nature}\ }\textbf {\bibinfo {volume} {605}},\ \bibinfo
  {pages} {51} (\bibinfo {year} {2022})}\BibitemShut {NoStop}%
\bibitem [{\citenamefont {Nuckolls}\ \emph {et~al.}(2023)\citenamefont
  {Nuckolls}, \citenamefont {Lee}, \citenamefont {Oh}, \citenamefont {Wong},
  \citenamefont {Soejima}, \citenamefont {Hong}, \citenamefont {Călugăru},
  \citenamefont {Herzog-Arbeitman}, \citenamefont {Bernevig}, \citenamefont
  {Watanabe}, \citenamefont {Taniguchi}, \citenamefont {Regnault},
  \citenamefont {Zaletel},\ and\ \citenamefont {Yazdani}}]{Nuckolls2023}%
  \BibitemOpen
  \bibfield  {author} {\bibinfo {author} {\bibfnamefont {K.~P.}\ \bibnamefont
  {Nuckolls}}, \bibinfo {author} {\bibfnamefont {R.~L.}\ \bibnamefont {Lee}},
  \bibinfo {author} {\bibfnamefont {M.}~\bibnamefont {Oh}}, \bibinfo {author}
  {\bibfnamefont {D.}~\bibnamefont {Wong}}, \bibinfo {author} {\bibfnamefont
  {T.}~\bibnamefont {Soejima}}, \bibinfo {author} {\bibfnamefont {J.~P.}\
  \bibnamefont {Hong}}, \bibinfo {author} {\bibfnamefont {D.}~\bibnamefont
  {Călugăru}}, \bibinfo {author} {\bibfnamefont {J.}~\bibnamefont
  {Herzog-Arbeitman}}, \bibinfo {author} {\bibfnamefont {B.~A.}\ \bibnamefont
  {Bernevig}}, \bibinfo {author} {\bibfnamefont {K.}~\bibnamefont {Watanabe}},
  \bibinfo {author} {\bibfnamefont {T.}~\bibnamefont {Taniguchi}}, \bibinfo
  {author} {\bibfnamefont {N.}~\bibnamefont {Regnault}}, \bibinfo {author}
  {\bibfnamefont {M.~P.}\ \bibnamefont {Zaletel}},\ and\ \bibinfo {author}
  {\bibfnamefont {A.}~\bibnamefont {Yazdani}},\ }\bibfield  {title} {\bibinfo
  {title} {Quantum textures of the many-body wavefunctions in magic-angle
  graphene},\ }\href {https://doi.org/10.1038/s41586-023-06226-x} {\bibfield
  {journal} {\bibinfo  {journal} {Nature}\ }\textbf {\bibinfo {volume} {620}},\
  \bibinfo {pages} {525–532} (\bibinfo {year} {2023})}\BibitemShut {NoStop}%
\bibitem [{\citenamefont {Kim}\ \emph {et~al.}(2023)\citenamefont {Kim},
  \citenamefont {Choi}, \citenamefont {Lantagne-Hurtubise}, \citenamefont
  {Lewandowski}, \citenamefont {Thomson}, \citenamefont {Kong}, \citenamefont
  {Zhou}, \citenamefont {Baum}, \citenamefont {Zhang}, \citenamefont {Holleis}
  \emph {et~al.}}]{Kim2023}%
  \BibitemOpen
  \bibfield  {author} {\bibinfo {author} {\bibfnamefont {H.}~\bibnamefont
  {Kim}}, \bibinfo {author} {\bibfnamefont {Y.}~\bibnamefont {Choi}}, \bibinfo
  {author} {\bibfnamefont {{\'E}.}~\bibnamefont {Lantagne-Hurtubise}}, \bibinfo
  {author} {\bibfnamefont {C.}~\bibnamefont {Lewandowski}}, \bibinfo {author}
  {\bibfnamefont {A.}~\bibnamefont {Thomson}}, \bibinfo {author} {\bibfnamefont
  {L.}~\bibnamefont {Kong}}, \bibinfo {author} {\bibfnamefont {H.}~\bibnamefont
  {Zhou}}, \bibinfo {author} {\bibfnamefont {E.}~\bibnamefont {Baum}}, \bibinfo
  {author} {\bibfnamefont {Y.}~\bibnamefont {Zhang}}, \bibinfo {author}
  {\bibfnamefont {L.}~\bibnamefont {Holleis}}, \emph {et~al.},\ }\bibfield
  {title} {\bibinfo {title} {Imaging inter-valley coherent order in magic-angle
  twisted trilayer graphene},\ }\href
  {https://doi.org/10.1038/s41586-023-06663-8} {\bibfield  {journal} {\bibinfo
  {journal} {Nature}\ }\textbf {\bibinfo {volume} {623}},\ \bibinfo {pages}
  {942} (\bibinfo {year} {2023})}\BibitemShut {NoStop}%
\bibitem [{\citenamefont {Jung}\ \emph {et~al.}(2015)\citenamefont {Jung},
  \citenamefont {Polini},\ and\ \citenamefont {MacDonald}}]{Jung2015}%
  \BibitemOpen
  \bibfield  {author} {\bibinfo {author} {\bibfnamefont {J.}~\bibnamefont
  {Jung}}, \bibinfo {author} {\bibfnamefont {M.}~\bibnamefont {Polini}},\ and\
  \bibinfo {author} {\bibfnamefont {A.~H.}\ \bibnamefont {MacDonald}},\
  }\bibfield  {title} {\bibinfo {title} {Persistent current states in bilayer
  graphene},\ }\href {https://doi.org/10.1103/PhysRevB.91.155423} {\bibfield
  {journal} {\bibinfo  {journal} {Phys. Rev. B}\ }\textbf {\bibinfo {volume}
  {91}},\ \bibinfo {pages} {155423} (\bibinfo {year} {2015})}\BibitemShut
  {NoStop}%
\bibitem [{\citenamefont {Dong}\ \emph
  {et~al.}(2023{\natexlab{d}})\citenamefont {Dong}, \citenamefont {Davydova},
  \citenamefont {Ogunnaike},\ and\ \citenamefont {Levitov}}]{Dong2021}%
  \BibitemOpen
  \bibfield  {author} {\bibinfo {author} {\bibfnamefont {Z.}~\bibnamefont
  {Dong}}, \bibinfo {author} {\bibfnamefont {M.}~\bibnamefont {Davydova}},
  \bibinfo {author} {\bibfnamefont {O.}~\bibnamefont {Ogunnaike}},\ and\
  \bibinfo {author} {\bibfnamefont {L.}~\bibnamefont {Levitov}},\ }\bibfield
  {title} {\bibinfo {title} {Isospin- and momentum-polarized orders in bilayer
  graphene},\ }\href {https://doi.org/10.1103/PhysRevB.107.075108} {\bibfield
  {journal} {\bibinfo  {journal} {Phys. Rev. B}\ }\textbf {\bibinfo {volume}
  {107}},\ \bibinfo {pages} {075108} (\bibinfo {year}
  {2023}{\natexlab{d}})}\BibitemShut {NoStop}%
\bibitem [{\citenamefont {Huang}\ \emph {et~al.}(2023)\citenamefont {Huang},
  \citenamefont {Wolf}, \citenamefont {Qin}, \citenamefont {Wei}, \citenamefont
  {Blinov},\ and\ \citenamefont {MacDonald}}]{Huang2023}%
  \BibitemOpen
  \bibfield  {author} {\bibinfo {author} {\bibfnamefont {C.}~\bibnamefont
  {Huang}}, \bibinfo {author} {\bibfnamefont {T.~M.~R.}\ \bibnamefont {Wolf}},
  \bibinfo {author} {\bibfnamefont {W.}~\bibnamefont {Qin}}, \bibinfo {author}
  {\bibfnamefont {N.}~\bibnamefont {Wei}}, \bibinfo {author} {\bibfnamefont
  {I.~V.}\ \bibnamefont {Blinov}},\ and\ \bibinfo {author} {\bibfnamefont
  {A.~H.}\ \bibnamefont {MacDonald}},\ }\bibfield  {title} {\bibinfo {title}
  {Spin and orbital metallic magnetism in rhombohedral trilayer graphene},\
  }\href {https://doi.org/10.1103/PhysRevB.107.L121405} {\bibfield  {journal}
  {\bibinfo  {journal} {Phys. Rev. B}\ }\textbf {\bibinfo {volume} {107}},\
  \bibinfo {pages} {L121405} (\bibinfo {year} {2023})}\BibitemShut {NoStop}%
\bibitem [{\citenamefont {Lin}\ \emph {et~al.}(2023)\citenamefont {Lin},
  \citenamefont {Wang}, \citenamefont {Zhang}, \citenamefont {Watanabe},
  \citenamefont {Taniguchi}, \citenamefont {Fu},\ and\ \citenamefont
  {Li}}]{Lin2023}%
  \BibitemOpen
  \bibfield  {author} {\bibinfo {author} {\bibfnamefont {J.-X.}\ \bibnamefont
  {Lin}}, \bibinfo {author} {\bibfnamefont {Y.}~\bibnamefont {Wang}}, \bibinfo
  {author} {\bibfnamefont {N.~J.}\ \bibnamefont {Zhang}}, \bibinfo {author}
  {\bibfnamefont {K.}~\bibnamefont {Watanabe}}, \bibinfo {author}
  {\bibfnamefont {T.}~\bibnamefont {Taniguchi}}, \bibinfo {author}
  {\bibfnamefont {L.}~\bibnamefont {Fu}},\ and\ \bibinfo {author}
  {\bibfnamefont {J.~I.~A.}\ \bibnamefont {Li}},\ }\href@noop {} {\bibinfo
  {title} {Spontaneous momentum polarization and diodicity in bernal bilayer
  graphene}} (\bibinfo {year} {2023}),\ \Eprint
  {https://arxiv.org/abs/2302.04261} {arXiv:2302.04261 [cond-mat.mes-hall]}
  \BibitemShut {NoStop}%
\bibitem [{\citenamefont {Wang}\ \emph {et~al.}(2023)\citenamefont {Wang},
  \citenamefont {Vila}, \citenamefont {Zaletel},\ and\ \citenamefont
  {Chatterjee}}]{Wang2023}%
  \BibitemOpen
  \bibfield  {author} {\bibinfo {author} {\bibfnamefont {T.}~\bibnamefont
  {Wang}}, \bibinfo {author} {\bibfnamefont {M.}~\bibnamefont {Vila}}, \bibinfo
  {author} {\bibfnamefont {M.~P.}\ \bibnamefont {Zaletel}},\ and\ \bibinfo
  {author} {\bibfnamefont {S.}~\bibnamefont {Chatterjee}},\ }\href@noop {}
  {\bibinfo {title} {Electrical control of magnetism in spin-orbit coupled
  graphene multilayers}} (\bibinfo {year} {2023}),\ \Eprint
  {https://arxiv.org/abs/2303.04855} {arXiv:2303.04855 [cond-mat.str-el]}
  \BibitemShut {NoStop}%
\bibitem [{\citenamefont {Xie}\ \emph {et~al.}(2023)\citenamefont {Xie},
  \citenamefont {Étienne Lantagne-Hurtubise}, \citenamefont {Young},
  \citenamefont {Nadj-Perge},\ and\ \citenamefont {Alicea}}]{Xie2023}%
  \BibitemOpen
  \bibfield  {author} {\bibinfo {author} {\bibfnamefont {Y.-M.}\ \bibnamefont
  {Xie}}, \bibinfo {author} {\bibnamefont {Étienne Lantagne-Hurtubise}},
  \bibinfo {author} {\bibfnamefont {A.~F.}\ \bibnamefont {Young}}, \bibinfo
  {author} {\bibfnamefont {S.}~\bibnamefont {Nadj-Perge}},\ and\ \bibinfo
  {author} {\bibfnamefont {J.}~\bibnamefont {Alicea}},\ }\href@noop {}
  {\bibinfo {title} {Gate-defined topological josephson junctions in bernal
  bilayer graphene}} (\bibinfo {year} {2023}),\ \Eprint
  {https://arxiv.org/abs/2304.11807} {arXiv:2304.11807 [cond-mat.mes-hall]}
  \BibitemShut {NoStop}%
\bibitem [{\citenamefont {Zhang}\ \emph {et~al.}(2010)\citenamefont {Zhang},
  \citenamefont {Sahu}, \citenamefont {Min},\ and\ \citenamefont
  {MacDonald}}]{zhang2010band}%
  \BibitemOpen
  \bibfield  {author} {\bibinfo {author} {\bibfnamefont {F.}~\bibnamefont
  {Zhang}}, \bibinfo {author} {\bibfnamefont {B.}~\bibnamefont {Sahu}},
  \bibinfo {author} {\bibfnamefont {H.}~\bibnamefont {Min}},\ and\ \bibinfo
  {author} {\bibfnamefont {A.~H.}\ \bibnamefont {MacDonald}},\ }\bibfield
  {title} {\bibinfo {title} {Band structure of $abc$-stacked graphene
  trilayers},\ }\href {https://doi.org/10.1103/PhysRevB.82.035409} {\bibfield
  {journal} {\bibinfo  {journal} {Phys. Rev. B}\ }\textbf {\bibinfo {volume}
  {82}},\ \bibinfo {pages} {035409} (\bibinfo {year} {2010})}\BibitemShut
  {NoStop}%
\bibitem [{\citenamefont {Jung}\ and\ \citenamefont
  {MacDonald}(2013)}]{jung2013gapped}%
  \BibitemOpen
  \bibfield  {author} {\bibinfo {author} {\bibfnamefont {J.}~\bibnamefont
  {Jung}}\ and\ \bibinfo {author} {\bibfnamefont {A.~H.}\ \bibnamefont
  {MacDonald}},\ }\bibfield  {title} {\bibinfo {title} {Gapped broken symmetry
  states in abc-stacked trilayer graphene},\ }\href
  {https://doi.org/10.1103/PhysRevB.88.075408} {\bibfield  {journal} {\bibinfo
  {journal} {Phys. Rev. B}\ }\textbf {\bibinfo {volume} {88}},\ \bibinfo
  {pages} {075408} (\bibinfo {year} {2013})}\BibitemShut {NoStop}%
\bibitem [{\citenamefont {Koshino}\ and\ \citenamefont
  {McCann}(2009)}]{Koshino2009}%
  \BibitemOpen
  \bibfield  {author} {\bibinfo {author} {\bibfnamefont {M.}~\bibnamefont
  {Koshino}}\ and\ \bibinfo {author} {\bibfnamefont {E.}~\bibnamefont
  {McCann}},\ }\bibfield  {title} {\bibinfo {title} {Trigonal warping and
  berry's phase $n\ensuremath{\pi}$ in abc-stacked multilayer graphene},\
  }\href {https://doi.org/10.1103/PhysRevB.80.165409} {\bibfield  {journal}
  {\bibinfo  {journal} {Phys. Rev. B}\ }\textbf {\bibinfo {volume} {80}},\
  \bibinfo {pages} {165409} (\bibinfo {year} {2009})}\BibitemShut {NoStop}%
\bibitem [{\citenamefont {Gmitra}\ \emph {et~al.}(2016)\citenamefont {Gmitra},
  \citenamefont {Kochan}, \citenamefont {H\"ogl},\ and\ \citenamefont
  {Fabian}}]{Gmitra2016}%
  \BibitemOpen
  \bibfield  {author} {\bibinfo {author} {\bibfnamefont {M.}~\bibnamefont
  {Gmitra}}, \bibinfo {author} {\bibfnamefont {D.}~\bibnamefont {Kochan}},
  \bibinfo {author} {\bibfnamefont {P.}~\bibnamefont {H\"ogl}},\ and\ \bibinfo
  {author} {\bibfnamefont {J.}~\bibnamefont {Fabian}},\ }\bibfield  {title}
  {\bibinfo {title} {Trivial and inverted dirac bands and the emergence of
  quantum spin hall states in graphene on transition-metal dichalcogenides},\
  }\href {https://doi.org/10.1103/PhysRevB.93.155104} {\bibfield  {journal}
  {\bibinfo  {journal} {Phys. Rev. B}\ }\textbf {\bibinfo {volume} {93}},\
  \bibinfo {pages} {155104} (\bibinfo {year} {2016})}\BibitemShut {NoStop}%
\bibitem [{\citenamefont {Yang}\ \emph {et~al.}(2017)\citenamefont {Yang},
  \citenamefont {Lohmann}, \citenamefont {Barroso}, \citenamefont {Liao},
  \citenamefont {Lin}, \citenamefont {Liu}, \citenamefont {Bartels},
  \citenamefont {Watanabe}, \citenamefont {Taniguchi},\ and\ \citenamefont
  {Shi}}]{Yang2017}%
  \BibitemOpen
  \bibfield  {author} {\bibinfo {author} {\bibfnamefont {B.}~\bibnamefont
  {Yang}}, \bibinfo {author} {\bibfnamefont {M.}~\bibnamefont {Lohmann}},
  \bibinfo {author} {\bibfnamefont {D.}~\bibnamefont {Barroso}}, \bibinfo
  {author} {\bibfnamefont {I.}~\bibnamefont {Liao}}, \bibinfo {author}
  {\bibfnamefont {Z.}~\bibnamefont {Lin}}, \bibinfo {author} {\bibfnamefont
  {Y.}~\bibnamefont {Liu}}, \bibinfo {author} {\bibfnamefont {L.}~\bibnamefont
  {Bartels}}, \bibinfo {author} {\bibfnamefont {K.}~\bibnamefont {Watanabe}},
  \bibinfo {author} {\bibfnamefont {T.}~\bibnamefont {Taniguchi}},\ and\
  \bibinfo {author} {\bibfnamefont {J.}~\bibnamefont {Shi}},\ }\bibfield
  {title} {\bibinfo {title} {Strong electron-hole symmetric rashba spin-orbit
  coupling in graphene/monolayer transition metal dichalcogenide
  heterostructures},\ }\href {https://doi.org/10.1103/PhysRevB.96.041409}
  {\bibfield  {journal} {\bibinfo  {journal} {Phys. Rev. B}\ }\textbf {\bibinfo
  {volume} {96}},\ \bibinfo {pages} {041409(R)} (\bibinfo {year}
  {2017})}\BibitemShut {NoStop}%
\bibitem [{\citenamefont {Zihlmann}\ \emph {et~al.}(2018)\citenamefont
  {Zihlmann}, \citenamefont {Cummings}, \citenamefont {Garcia}, \citenamefont
  {Kedves}, \citenamefont {Watanabe}, \citenamefont {Taniguchi}, \citenamefont
  {Sch\"onenberger},\ and\ \citenamefont {Makk}}]{Zihlmann2018}%
  \BibitemOpen
  \bibfield  {author} {\bibinfo {author} {\bibfnamefont {S.}~\bibnamefont
  {Zihlmann}}, \bibinfo {author} {\bibfnamefont {A.~W.}\ \bibnamefont
  {Cummings}}, \bibinfo {author} {\bibfnamefont {J.~H.}\ \bibnamefont
  {Garcia}}, \bibinfo {author} {\bibfnamefont {M.}~\bibnamefont {Kedves}},
  \bibinfo {author} {\bibfnamefont {K.}~\bibnamefont {Watanabe}}, \bibinfo
  {author} {\bibfnamefont {T.}~\bibnamefont {Taniguchi}}, \bibinfo {author}
  {\bibfnamefont {C.}~\bibnamefont {Sch\"onenberger}},\ and\ \bibinfo {author}
  {\bibfnamefont {P.}~\bibnamefont {Makk}},\ }\bibfield  {title} {\bibinfo
  {title} {Large spin relaxation anisotropy and valley-zeeman spin-orbit
  coupling in ${\mathrm{wse}}_{2}$/graphene/$h$-bn heterostructures},\ }\href
  {https://doi.org/10.1103/PhysRevB.97.075434} {\bibfield  {journal} {\bibinfo
  {journal} {Phys. Rev. B}\ }\textbf {\bibinfo {volume} {97}},\ \bibinfo
  {pages} {075434} (\bibinfo {year} {2018})}\BibitemShut {NoStop}%
\bibitem [{\citenamefont {Wang}\ \emph {et~al.}(2019)\citenamefont {Wang},
  \citenamefont {Che}, \citenamefont {Cao}, \citenamefont {Lyu}, \citenamefont
  {Watanabe}, \citenamefont {Taniguchi}, \citenamefont {Lau},\ and\
  \citenamefont {Bockrath}}]{Wang2019}%
  \BibitemOpen
  \bibfield  {author} {\bibinfo {author} {\bibfnamefont {D.}~\bibnamefont
  {Wang}}, \bibinfo {author} {\bibfnamefont {S.}~\bibnamefont {Che}}, \bibinfo
  {author} {\bibfnamefont {G.}~\bibnamefont {Cao}}, \bibinfo {author}
  {\bibfnamefont {R.}~\bibnamefont {Lyu}}, \bibinfo {author} {\bibfnamefont
  {K.}~\bibnamefont {Watanabe}}, \bibinfo {author} {\bibfnamefont
  {T.}~\bibnamefont {Taniguchi}}, \bibinfo {author} {\bibfnamefont {C.~N.}\
  \bibnamefont {Lau}},\ and\ \bibinfo {author} {\bibfnamefont {M.}~\bibnamefont
  {Bockrath}},\ }\bibfield  {title} {\bibinfo {title} {Quantum hall effect
  measurement of spin{\textendash}orbit coupling strengths in ultraclean
  bilayer graphene/${\mathrm{wse}}_{2}$ heterostructures},\ }\href
  {https://doi.org/10.1021/acs.nanolett.9b02445} {\bibfield  {journal}
  {\bibinfo  {journal} {Nano Letters}\ }\textbf {\bibinfo {volume} {19}},\
  \bibinfo {pages} {7028} (\bibinfo {year} {2019})}\BibitemShut {NoStop}%
\bibitem [{\citenamefont {Island}\ \emph {et~al.}(2019)\citenamefont {Island},
  \citenamefont {Cui}, \citenamefont {Lewandowski}, \citenamefont {Khoo},
  \citenamefont {Spanton}, \citenamefont {Zhou}, \citenamefont {Rhodes},
  \citenamefont {Hone}, \citenamefont {Taniguchi}, \citenamefont {Watanabe},
  \citenamefont {Levitov}, \citenamefont {Zaletel},\ and\ \citenamefont
  {Young}}]{Island2019}%
  \BibitemOpen
  \bibfield  {author} {\bibinfo {author} {\bibfnamefont {J.~O.}\ \bibnamefont
  {Island}}, \bibinfo {author} {\bibfnamefont {X.}~\bibnamefont {Cui}},
  \bibinfo {author} {\bibfnamefont {C.}~\bibnamefont {Lewandowski}}, \bibinfo
  {author} {\bibfnamefont {J.~Y.}\ \bibnamefont {Khoo}}, \bibinfo {author}
  {\bibfnamefont {E.~M.}\ \bibnamefont {Spanton}}, \bibinfo {author}
  {\bibfnamefont {H.}~\bibnamefont {Zhou}}, \bibinfo {author} {\bibfnamefont
  {D.}~\bibnamefont {Rhodes}}, \bibinfo {author} {\bibfnamefont {J.~C.}\
  \bibnamefont {Hone}}, \bibinfo {author} {\bibfnamefont {T.}~\bibnamefont
  {Taniguchi}}, \bibinfo {author} {\bibfnamefont {K.}~\bibnamefont {Watanabe}},
  \bibinfo {author} {\bibfnamefont {L.~S.}\ \bibnamefont {Levitov}}, \bibinfo
  {author} {\bibfnamefont {M.~P.}\ \bibnamefont {Zaletel}},\ and\ \bibinfo
  {author} {\bibfnamefont {A.~F.}\ \bibnamefont {Young}},\ }\bibfield  {title}
  {\bibinfo {title} {Spin{\textendash}orbit-driven band inversion in bilayer
  graphene by the van der waals proximity effect},\ }\href
  {https://doi.org/10.1038/s41586-019-1304-2} {\bibfield  {journal} {\bibinfo
  {journal} {Nature}\ }\textbf {\bibinfo {volume} {571}},\ \bibinfo {pages}
  {85} (\bibinfo {year} {2019})}\BibitemShut {NoStop}%
\bibitem [{\citenamefont {Wang}\ \emph {et~al.}(2016)\citenamefont {Wang},
  \citenamefont {Ki}, \citenamefont {Khoo}, \citenamefont {Mauro},
  \citenamefont {Berger}, \citenamefont {Levitov},\ and\ \citenamefont
  {Morpurgo}}]{Wang2020}%
  \BibitemOpen
  \bibfield  {author} {\bibinfo {author} {\bibfnamefont {Z.}~\bibnamefont
  {Wang}}, \bibinfo {author} {\bibfnamefont {D.-K.}\ \bibnamefont {Ki}},
  \bibinfo {author} {\bibfnamefont {J.~Y.}\ \bibnamefont {Khoo}}, \bibinfo
  {author} {\bibfnamefont {D.}~\bibnamefont {Mauro}}, \bibinfo {author}
  {\bibfnamefont {H.}~\bibnamefont {Berger}}, \bibinfo {author} {\bibfnamefont
  {L.~S.}\ \bibnamefont {Levitov}},\ and\ \bibinfo {author} {\bibfnamefont
  {A.~F.}\ \bibnamefont {Morpurgo}},\ }\bibfield  {title} {\bibinfo {title}
  {Origin and magnitude of `designer' spin-orbit interaction in graphene on
  semiconducting transition metal dichalcogenides},\ }\href
  {https://doi.org/10.1103/PhysRevX.6.041020} {\bibfield  {journal} {\bibinfo
  {journal} {Phys. Rev. X}\ }\textbf {\bibinfo {volume} {6}},\ \bibinfo {pages}
  {041020} (\bibinfo {year} {2016})}\BibitemShut {NoStop}%
\bibitem [{\citenamefont {Amann}\ \emph {et~al.}(2022)\citenamefont {Amann},
  \citenamefont {V\"olkl}, \citenamefont {Rockinger}, \citenamefont {Kochan},
  \citenamefont {Watanabe}, \citenamefont {Taniguchi}, \citenamefont {Fabian},
  \citenamefont {Weiss},\ and\ \citenamefont {Eroms}}]{Amann2022}%
  \BibitemOpen
  \bibfield  {author} {\bibinfo {author} {\bibfnamefont {J.}~\bibnamefont
  {Amann}}, \bibinfo {author} {\bibfnamefont {T.}~\bibnamefont {V\"olkl}},
  \bibinfo {author} {\bibfnamefont {T.}~\bibnamefont {Rockinger}}, \bibinfo
  {author} {\bibfnamefont {D.}~\bibnamefont {Kochan}}, \bibinfo {author}
  {\bibfnamefont {K.}~\bibnamefont {Watanabe}}, \bibinfo {author}
  {\bibfnamefont {T.}~\bibnamefont {Taniguchi}}, \bibinfo {author}
  {\bibfnamefont {J.}~\bibnamefont {Fabian}}, \bibinfo {author} {\bibfnamefont
  {D.}~\bibnamefont {Weiss}},\ and\ \bibinfo {author} {\bibfnamefont
  {J.}~\bibnamefont {Eroms}},\ }\bibfield  {title} {\bibinfo {title}
  {Counterintuitive gate dependence of weak antilocalization in bilayer
  $\mathrm{graphene}/{\mathrm{wse}}_{2}$ heterostructures},\ }\href
  {https://doi.org/10.1103/PhysRevB.105.115425} {\bibfield  {journal} {\bibinfo
   {journal} {Phys. Rev. B}\ }\textbf {\bibinfo {volume} {105}},\ \bibinfo
  {pages} {115425} (\bibinfo {year} {2022})}\BibitemShut {NoStop}%
\bibitem [{\citenamefont {Sun}\ \emph {et~al.}(2022)\citenamefont {Sun},
  \citenamefont {Rademaker}, \citenamefont {Mauro}, \citenamefont {Scarfato},
  \citenamefont {Árpád Pásztor}, \citenamefont {Gutiérrez-Lezama},
  \citenamefont {Wang}, \citenamefont {Martinez-Castro}, \citenamefont
  {Morpurgo},\ and\ \citenamefont {Renner}}]{Sun2022determining}%
  \BibitemOpen
  \bibfield  {author} {\bibinfo {author} {\bibfnamefont {L.}~\bibnamefont
  {Sun}}, \bibinfo {author} {\bibfnamefont {L.}~\bibnamefont {Rademaker}},
  \bibinfo {author} {\bibfnamefont {D.}~\bibnamefont {Mauro}}, \bibinfo
  {author} {\bibfnamefont {A.}~\bibnamefont {Scarfato}}, \bibinfo {author}
  {\bibnamefont {Árpád Pásztor}}, \bibinfo {author} {\bibfnamefont
  {I.}~\bibnamefont {Gutiérrez-Lezama}}, \bibinfo {author} {\bibfnamefont
  {Z.}~\bibnamefont {Wang}}, \bibinfo {author} {\bibfnamefont {J.}~\bibnamefont
  {Martinez-Castro}}, \bibinfo {author} {\bibfnamefont {A.~F.}\ \bibnamefont
  {Morpurgo}},\ and\ \bibinfo {author} {\bibfnamefont {C.}~\bibnamefont
  {Renner}},\ }\href@noop {} {\bibinfo {title} {Determining spin-orbit coupling
  in graphene by quasiparticle interference imaging}} (\bibinfo {year}
  {2022}),\ \Eprint {https://arxiv.org/abs/2212.04926} {arXiv:2212.04926
  [cond-mat.mes-hall]} \BibitemShut {NoStop}%
\bibitem [{\citenamefont {Gmitra}\ and\ \citenamefont
  {Fabian}(2017)}]{Gmitra2017}%
  \BibitemOpen
  \bibfield  {author} {\bibinfo {author} {\bibfnamefont {M.}~\bibnamefont
  {Gmitra}}\ and\ \bibinfo {author} {\bibfnamefont {J.}~\bibnamefont
  {Fabian}},\ }\bibfield  {title} {\bibinfo {title} {Proximity effects in
  bilayer graphene on monolayer ${\mathrm{wse}}_{2}$: Field-effect spin valley
  locking, spin-orbit valve, and spin transistor},\ }\href
  {https://doi.org/10.1103/PhysRevLett.119.146401} {\bibfield  {journal}
  {\bibinfo  {journal} {Phys. Rev. Lett.}\ }\textbf {\bibinfo {volume} {119}},\
  \bibinfo {pages} {146401} (\bibinfo {year} {2017})}\BibitemShut {NoStop}%
\bibitem [{\citenamefont {Khoo}\ \emph {et~al.}(2017)\citenamefont {Khoo},
  \citenamefont {Morpurgo},\ and\ \citenamefont {Levitov}}]{Khoo2017}%
  \BibitemOpen
  \bibfield  {author} {\bibinfo {author} {\bibfnamefont {J.~Y.}\ \bibnamefont
  {Khoo}}, \bibinfo {author} {\bibfnamefont {A.~F.}\ \bibnamefont {Morpurgo}},\
  and\ \bibinfo {author} {\bibfnamefont {L.}~\bibnamefont {Levitov}},\
  }\bibfield  {title} {\bibinfo {title} {On-demand spin{\textendash}orbit
  interaction from which-layer tunability in bilayer graphene},\ }\href
  {https://doi.org/10.1021/acs.nanolett.7b03604} {\bibfield  {journal}
  {\bibinfo  {journal} {Nano Letters}\ }\textbf {\bibinfo {volume} {17}},\
  \bibinfo {pages} {7003} (\bibinfo {year} {2017})}\BibitemShut {NoStop}%
\bibitem [{\citenamefont {Li}\ and\ \citenamefont {Koshino}(2019)}]{Li2019}%
  \BibitemOpen
  \bibfield  {author} {\bibinfo {author} {\bibfnamefont {Y.}~\bibnamefont
  {Li}}\ and\ \bibinfo {author} {\bibfnamefont {M.}~\bibnamefont {Koshino}},\
  }\bibfield  {title} {\bibinfo {title} {Twist-angle dependence of the
  proximity spin-orbit coupling in graphene on transition-metal
  dichalcogenides},\ }\href {https://doi.org/10.1103/PhysRevB.99.075438}
  {\bibfield  {journal} {\bibinfo  {journal} {Phys. Rev. B}\ }\textbf {\bibinfo
  {volume} {99}},\ \bibinfo {pages} {075438} (\bibinfo {year}
  {2019})}\BibitemShut {NoStop}%
\bibitem [{\citenamefont {David}\ \emph {et~al.}(2019)\citenamefont {David},
  \citenamefont {Rakyta}, \citenamefont {Korm\'anyos},\ and\ \citenamefont
  {Burkard}}]{David2019}%
  \BibitemOpen
  \bibfield  {author} {\bibinfo {author} {\bibfnamefont {A.}~\bibnamefont
  {David}}, \bibinfo {author} {\bibfnamefont {P.}~\bibnamefont {Rakyta}},
  \bibinfo {author} {\bibfnamefont {A.}~\bibnamefont {Korm\'anyos}},\ and\
  \bibinfo {author} {\bibfnamefont {G.}~\bibnamefont {Burkard}},\ }\bibfield
  {title} {\bibinfo {title} {Induced spin-orbit coupling in twisted
  graphene--transition metal dichalcogenide heterobilayers: Twistronics meets
  spintronics},\ }\href {https://doi.org/10.1103/PhysRevB.100.085412}
  {\bibfield  {journal} {\bibinfo  {journal} {Phys. Rev. B}\ }\textbf {\bibinfo
  {volume} {100}},\ \bibinfo {pages} {085412} (\bibinfo {year}
  {2019})}\BibitemShut {NoStop}%
\bibitem [{\citenamefont {Naimer}\ \emph {et~al.}(2021)\citenamefont {Naimer},
  \citenamefont {Zollner}, \citenamefont {Gmitra},\ and\ \citenamefont
  {Fabian}}]{Naimer2021}%
  \BibitemOpen
  \bibfield  {author} {\bibinfo {author} {\bibfnamefont {T.}~\bibnamefont
  {Naimer}}, \bibinfo {author} {\bibfnamefont {K.}~\bibnamefont {Zollner}},
  \bibinfo {author} {\bibfnamefont {M.}~\bibnamefont {Gmitra}},\ and\ \bibinfo
  {author} {\bibfnamefont {J.}~\bibnamefont {Fabian}},\ }\bibfield  {title}
  {\bibinfo {title} {Twist-angle dependent proximity induced spin-orbit
  coupling in graphene/transition metal dichalcogenide heterostructures},\
  }\href {https://doi.org/10.1103/PhysRevB.104.195156} {\bibfield  {journal}
  {\bibinfo  {journal} {Phys. Rev. B}\ }\textbf {\bibinfo {volume} {104}},\
  \bibinfo {pages} {195156} (\bibinfo {year} {2021})}\BibitemShut {NoStop}%
\bibitem [{\citenamefont {Zaletel}\ and\ \citenamefont
  {Khoo}(2019)}]{Zaletel2019}%
  \BibitemOpen
  \bibfield  {author} {\bibinfo {author} {\bibfnamefont {M.~P.}\ \bibnamefont
  {Zaletel}}\ and\ \bibinfo {author} {\bibfnamefont {J.~Y.}\ \bibnamefont
  {Khoo}},\ }\href@noop {} {\bibinfo {title} {The gate-tunable strong and
  fragile topology of multilayer-graphene on a transition metal
  dichalcogenide}} (\bibinfo {year} {2019}),\ \Eprint
  {https://arxiv.org/abs/1901.01294} {arXiv:1901.01294 [cond-mat.mes-hall]}
  \BibitemShut {NoStop}%
\bibitem [{\citenamefont {Zhumagulov}\ \emph {et~al.}(2023)\citenamefont
  {Zhumagulov}, \citenamefont {Kochan},\ and\ \citenamefont
  {Fabian}}]{Zhumagulov2023}%
  \BibitemOpen
  \bibfield  {author} {\bibinfo {author} {\bibfnamefont {Y.}~\bibnamefont
  {Zhumagulov}}, \bibinfo {author} {\bibfnamefont {D.}~\bibnamefont {Kochan}},\
  and\ \bibinfo {author} {\bibfnamefont {J.}~\bibnamefont {Fabian}},\
  }\href@noop {} {\bibinfo {title} {Emergent correlated phases in rhombohedral
  trilayer graphene induced by proximity spin-orbit and exchange coupling}}
  (\bibinfo {year} {2023}),\ \Eprint {https://arxiv.org/abs/2305.14277}
  {arXiv:2305.14277 [cond-mat.str-el]} \BibitemShut {NoStop}%
\bibitem [{\citenamefont {Wolf}\ \emph {et~al.}(2023)\citenamefont {Wolf},
  \citenamefont {Huang}, \citenamefont {Lin},\ and\ \citenamefont
  {MacDonald}}]{Wolf2023}%
  \BibitemOpen
  \bibfield  {author} {\bibinfo {author} {\bibfnamefont {T.}~\bibnamefont
  {Wolf}}, \bibinfo {author} {\bibfnamefont {C.}~\bibnamefont {Huang}},
  \bibinfo {author} {\bibfnamefont {S.-z.}\ \bibnamefont {Lin}},\ and\ \bibinfo
  {author} {\bibfnamefont {A.}~\bibnamefont {MacDonald}},\ }\bibfield  {title}
  {\bibinfo {title} {Correlation energy in bernal bilayer graphene under strong
  displacement field},\ }\href
  {https://meetings.aps.org/Meeting/MAR23/Session/Q38.10} {\bibfield  {journal}
  {\bibinfo  {journal} {Bulletin of the American Physical Society}\ } (\bibinfo
  {year} {2023})}\BibitemShut {NoStop}%
\bibitem [{\citenamefont {Avsar}\ \emph {et~al.}(2020)\citenamefont {Avsar},
  \citenamefont {Ochoa}, \citenamefont {Guinea}, \citenamefont {\"Ozyilmaz},
  \citenamefont {van Wees},\ and\ \citenamefont {Vera-Marun}}]{Avsar2020}%
  \BibitemOpen
  \bibfield  {author} {\bibinfo {author} {\bibfnamefont {A.}~\bibnamefont
  {Avsar}}, \bibinfo {author} {\bibfnamefont {H.}~\bibnamefont {Ochoa}},
  \bibinfo {author} {\bibfnamefont {F.}~\bibnamefont {Guinea}}, \bibinfo
  {author} {\bibfnamefont {B.}~\bibnamefont {\"Ozyilmaz}}, \bibinfo {author}
  {\bibfnamefont {B.~J.}\ \bibnamefont {van Wees}},\ and\ \bibinfo {author}
  {\bibfnamefont {I.~J.}\ \bibnamefont {Vera-Marun}},\ }\bibfield  {title}
  {\bibinfo {title} {Colloquium: Spintronics in graphene and other
  two-dimensional materials},\ }\href
  {https://doi.org/10.1103/RevModPhys.92.021003} {\bibfield  {journal}
  {\bibinfo  {journal} {Rev. Mod. Phys.}\ }\textbf {\bibinfo {volume} {92}},\
  \bibinfo {pages} {021003} (\bibinfo {year} {2020})}\BibitemShut {NoStop}%
\bibitem [{\citenamefont {Yuan}\ and\ \citenamefont {Fu}(2021)}]{Yuan2021}%
  \BibitemOpen
  \bibfield  {author} {\bibinfo {author} {\bibfnamefont {N.~F.~Q.}\
  \bibnamefont {Yuan}}\ and\ \bibinfo {author} {\bibfnamefont {L.}~\bibnamefont
  {Fu}},\ }\bibfield  {title} {\bibinfo {title} {Topological metals and
  finite-momentum superconductors},\ }\bibfield  {journal} {\bibinfo  {journal}
  {Proceedings of the National Academy of Sciences}\ }\textbf {\bibinfo
  {volume} {118}},\ \href {https://doi.org/10.1073/pnas.2019063118}
  {10.1073/pnas.2019063118} (\bibinfo {year} {2021})\BibitemShut {NoStop}%
\bibitem [{\citenamefont {McCann}\ and\ \citenamefont
  {Koshino}(2013)}]{McCann2013}%
  \BibitemOpen
  \bibfield  {author} {\bibinfo {author} {\bibfnamefont {E.}~\bibnamefont
  {McCann}}\ and\ \bibinfo {author} {\bibfnamefont {M.}~\bibnamefont
  {Koshino}},\ }\bibfield  {title} {\bibinfo {title} {The electronic properties
  of bilayer graphene},\ }\href {https://doi.org/10.1088/0034-4885/76/5/056503}
  {\bibfield  {journal} {\bibinfo  {journal} {Reports on Progress in Physics}\
  }\textbf {\bibinfo {volume} {76}},\ \bibinfo {pages} {056503} (\bibinfo
  {year} {2013})}\BibitemShut {NoStop}%
\bibitem [{\citenamefont {Bena}\ and\ \citenamefont
  {Montambaux}(2009)}]{bena2009remarks}%
  \BibitemOpen
  \bibfield  {author} {\bibinfo {author} {\bibfnamefont {C.}~\bibnamefont
  {Bena}}\ and\ \bibinfo {author} {\bibfnamefont {G.}~\bibnamefont
  {Montambaux}},\ }\bibfield  {title} {\bibinfo {title} {Remarks on the
  tight-binding model of graphene},\ }\href
  {https://doi.org/10.1088/1367-2630/11/9/095002} {\bibfield  {journal}
  {\bibinfo  {journal} {New Journal of Physics}\ }\textbf {\bibinfo {volume}
  {11}},\ \bibinfo {pages} {095003} (\bibinfo {year} {2009})}\BibitemShut
  {NoStop}%
\bibitem [{\citenamefont {Fabrizio}(2022)}]{fabrizio2022}%
  \BibitemOpen
  \bibfield  {author} {\bibinfo {author} {\bibfnamefont {M.}~\bibnamefont
  {Fabrizio}},\ }\href {https://doi.org/10.1007/978-3-031-16305-0} {\emph
  {\bibinfo {title} {A Course in Quantum Many-Body Theory}}}\ (\bibinfo
  {publisher} {Springer International Publishing},\ \bibinfo {year} {2022})\
  Chap.\ \bibinfo {chapter} {3 -- Hartree-Fock Approximation}\BibitemShut
  {NoStop}%
\end{thebibliography}%

\appendix

\clearpage 
\pagebreak

\makeatletter
\renewcommand\thetable{S\@arabic\c@table}
\renewcommand\thefigure{S\@arabic\c@figure}
\makeatother

\setcounter{figure}{0}
\setcounter{table}{0}

\begin{widetext}

\section{Hamiltonian}
\label{app-sec:hamiltonian}

\subsection{Tight-binding model and spin-orbit coupling}
\label{app-sec:hamiltonian/tight-binding-soc}

The full six-band (per spin) microscopic tight-binding Hamiltonian can be written in sublattice basis
\begin{equation}\begin{split}
    \hat{H}_{\text{B}} = \sum_{\vb{k}} \sum_{\tau s \sigma \sigma'} h(\tau \vb{K} + \vb{k})_{\sigma \sigma'} 
        c_{\tau s \sigma \vb{k}}^\dag c_{\tau s \sigma' \vb{k}},
\end{split}\end{equation}
where the fermion operator $c_{\tau s \sigma \vb{k}}$ annihilates an electron at momentum $\vb{k}$ for valley index $\tau \in \{\pm 1\}$, spin index $s \in \{\uparrow, \downarrow\}$ and sublattice index $\sigma$, with~\cite{zhang2010band}
\begin{equation}\begin{split}
    h(\vb{q})_{\sigma \sigma'} = \mqty[
        \Delta_1 + \Delta_2 + \delta & \gamma_2 / 2 & -\gamma_0 f_{\vb{q}} & -\gamma_4 f_{\vb{q}} & -\gamma_3 f_{\vb{q}}^\dag & 0 \\
        \gamma_2 / 2 & \Delta_2 - \Delta_1 + \delta & 0 & -\gamma_3 f_{\vb{q}} & -\gamma_4 f_{\vb{q}}^\dag & -\gamma_0 f_{\vb{q}}^\dag \\
        -\gamma_0 f_{\vb{q}}^\dag & 0 & \Delta_1 + \Delta_2 & \gamma_1 & -\gamma_4 f_{\vb{q}} & 0 \\
        -\gamma_4 f_{\vb{q}}^\dag & -\gamma_3 f_{\vb{q}}^\dag & \gamma_1 & -2 \Delta_2 & -\gamma_0 f_{\vb{q}} & -\gamma_4 f_{\vb{q}} \\
        -\gamma_3 f_{\vb{q}} & -\gamma_4 f_{\vb{q}} & -\gamma_4 f_{\vb{q}}^\dag & -\gamma_0 f_{\vb{q}}^\dag & -2 \Delta_2 & \gamma_1 \\
        0 & -\gamma_0 f_{\vb{q}} & 0 & -\gamma_4 f_{\vb{q}}^\dag & \gamma_1 & \Delta_2 - \Delta_1]_{\sigma \sigma'},
\end{split}\end{equation}
for microscopic momentum $\vb{q}$ such that $\vb{q} = \vb{0}$ at the $\Gamma$-point of the Brillouin zone. The sublattice basis is $(A_1, B_3, B_1, A_2, B_2, A_3)$ where $A, B$ label the sites and $l$ in $A_l, B_l$ labels the layer, and
\begin{equation*}\begin{split}
    f_{\vb{q}} = \smash{e^{i q_y a / \sqrt{3}} + 2 e^{-i q_y a / 2 \sqrt{3}} \cos{\left(q_x a / 2\right)}}
\end{split}\end{equation*}
describes the in-plane component of nearest-neighbor hopping centered at sublattices~\cite{McCann2013, bena2009remarks}. The valley point $\vb{K} = (4 \pi / 3 a) \vu{x}$ for graphene lattice constant $a = \SI{2.46}{\angstrom}$. Here $\Delta_1$ is a potential difference between adjacent layers due to an external perpendicular displacement field $D$, given by \cref{eq:Delta1-def} and of order ${\sim} 10$ to $\SI{50}{\milli\electronvolt}$ for experimentally relevant values of $D$. The values of all other parameters are fixed by fitting against ab-initio (DFT) calculations, as available in the literature~\cite{Chatterjee2022, Cea2022, chou2021acoustic, zhou2021half, zhang2010band}. In this work we use the values
\begin{table}[!h]
    \centering
    \begin{tabular}{c c c c c c c}
        \toprule 
        $\gamma_0$ & $\gamma_1$ & $\gamma_2$ & $\gamma_3$ & $\gamma_4$ & $\delta$ & $\Delta_2$ \\
        \midrule 
        $\quad \SI{3.1}{\electronvolt} \quad$ &
        $\quad \SI{380}{\milli\electronvolt} \quad$ &
        $\quad -\SI{15}{\milli\electronvolt} \quad$ &
        $\quad -\SI{290}{\milli\electronvolt} \quad$ &
        $\quad -\SI{141}{\milli\electronvolt} \quad$ &
        $\quad -\SI{10.5}{\milli\electronvolt} \quad$ &
        $\quad -\SI{2.3}{\milli\electronvolt} \quad$ \\
        \bottomrule
    \end{tabular}
\end{table}

The spin-orbit coupling (SOC) induced by the WSe$_2$ substrate is captured by Ising- and Rashba-type terms,
\begin{equation}\begin{split}
    \hat{H}_{\text{I}} = \frac{\lambda_{\text{I}}}{2} \sum_{\vb{k}} 
        \vb{c}^\dag_{\vb{k}} \left( \tau^z s^z \mathbb{P}_3 \right) \vb{c}_{\vb{k}}, \qquad
    \hat{H}_{\text{R}} = \frac{\lambda_{\text{R}}}{2} \sum_{\vb{k}} 
        \vb{c}_{\vb{k}}^\dag \left( \tau^z s^y \sigma^x - s^x \sigma^y \right) \mathbb{P}_3 \vb{c}_{\vb{k}},
\end{split}\end{equation}
where $\lambda_{\text{I}}$ and $\lambda_{\text{R}}$ are the SOC coefficients, $\tau^\mu$, $s^\mu$ and $\sigma^\mu$ are Pauli operators acting on the valley, spin and sublattice degrees of freedom respectively.
The operator $\mathbb{P}_3$ projects onto the top layer of RTG and $\trans{\vb{c}}_{\vb{k}} = \mqty[ c_{+ \uparrow A_1 \vb{k}} & \ldots & c_{- \downarrow A_3 \vb{k}} ]$ enumerates the fermion operators in valley, spin and sublattice basis. However, due to the suppression of Rashba SOC effects by the sublattice polarization of the low-energy wavefunctions of RTG at large $D$ fields~\cite{Zaletel2019}, we focus on Ising-type SOC in this work. Our starting point for the Hartree-Fock procedure is therefore the non-interacting Hamiltonian $\smash{\hat{H}_0} = \smash{\hat{H}_{\text{B}}} + \smash{\hat{H}_{\text{I}}}$. We detail the self-consistent Hartree-Fock method in \cref{app-sec:hartree-fock}.

\subsection{Screened Coulomb interactions}
\label{app-sec:hamiltonian/coulomb-hunds}

We consider gate-screened Coulomb interactions, which can be separated into long-range $\hat{H}_{\text{C}}$ and short-range $\hat{H}_{\text{V}}$ parts,
\begin{equation}\begin{split}
    \hat{H}_{\text{C}} &= \frac{1}{2A} \sum_{\vb{q}} V_{\text{C}}(\vb{q}) \normord{\rho(\vb{q}) \rho(-\vb{q})}, \\
    \hat{H}_{\text{V}} &= \frac{J_{\text{H}}}{2A} \sum_{\vb{k}, \vb{k}'} \sum_{\vb{q}} \sum_{\tau} \sum_{ss'} \sum_{\sigma \sigma'}
        \eta(\vb{q})_{\tau \sigma \sigma'}
        \normord{c_{(-\tau) s \sigma \vb{k}}^\dag c_{\tau s \sigma (\vb{k} + \vb{q})}
            c_{\tau s' \sigma' \vb{k}'}^\dag c_{(-\tau) s' \sigma' (\vb{k}' - \vb{q})} },
\end{split}\end{equation}
where $A$ is the sample area, $J_{\text{H}}$ the strength of short-range interactions, and $\normord{}$ denotes normal ordering. An approximate recasting of the short-range $\hat{H}_{\text{V}}$ in terms of spin operators in the two valley sectors is possible~\cite{Chatterjee2022}, which reveals a similarity in physical character to an intervalley Hund's coupling between electron spins---thus we refer to $\hat{H}_{\text{V}}$ also as the Hund's coupling Hamiltonian. Above in $\hat{H}_{\text{C}}$, $\rho(\vb{q})$ is the slowly varying component of the electron density operator involving only intra-valley terms,
\begin{equation}\begin{split}
    \rho(\vb{q}) = \sum_{\vb{k}} \sum_{\alpha} c_{\alpha \vb{k}}^\dag c_{\alpha (\vb{k} + \vb{q})},
\end{split}\end{equation}
where $\alpha = (\tau s \sigma)$ encompasses valley, spin and sublattice indices, and $V_{\text{C}}(\vb{q})$ is the repulsive dual gate-screened Coulomb interaction potential,
\begin{equation}
    V_{\text{C}}(\vb{q}) = \frac{q_{\text{e}}^2}{2 \epsilon_{\text{r}} \epsilon_0 q} 
        \tanh{(q d)},
\end{equation}
for $q_{\text{e}}$ the electron charge, screening length $d$ which can be taken as the distance from the graphene plane to the gates, $\epsilon_{\text{r}}$ the relative permittivity, and $\epsilon_0$ the permittivity of free space. We take $d \approx \SI{50}{\nano\meter}$ and phenomenologically model screening arising from the electron gas in the graphene plane by treating $\epsilon_{\text{r}}$ as a free parameter, which takes values larger than $\epsilon_{\text{r}} = 4.4$ expected simply from encapsulation in an \hBN dielectric environment.

Alternatively, to account for screening of the Coulomb interaction by mobile electrons, one may consider a random phase approximation (RPA) correction by itinerant electrons,
\begin{equation}\begin{split}
    V_{\text{C}}(\vb{q}) \leftarrow V_{\text{C}}^{\text{RPA}}(\vb{q}) = \frac{V_{\text{C}}(\vb{q})}{1 + \chi_{\rho\rho}(\vb{q}) V_{\text{C}}(\vb{q})},
\end{split}\end{equation}
for static Lindhard response function $\chi_{\rho\rho}(\vb{q})$. Disregarding the frequency dependence of the screening reduces $\chi_{\rho\rho}(\vb{q}) \approx \chi_0(1 - c q^2 / k_F^2 + \ldots) \approx \chi_0$ with the zero-temperature density of states at the Fermi surface $\smash{\chi_0 \sim \SI{0.16}{\per\electronvolt}}$ per unit cell, as was used in Ref.~\cite{Chatterjee2022}. However, we remark that the present RTG setting lies outside the conventionally accepted regime of validity of RPA. Fundamentally, RPA is an expansion in the small parameter $\chi_{\rho\rho}(\vb{q}) V_{\text{C}}(\vb{q}) \ll 1$, but in the present context near van Hove singularities, $\chi_{\rho\rho}(\vb{q}) V_{\text{C}}(\vb{q}) > 1$ and the expansion is not guaranteed to converge. For a ballpark estimate, at a characteristic Fermi momentum scale $q \approx 0.1 a^{-1}$ and $\epsilon_{\text{r}} \approx 4.4$, one finds $\chi_{\rho\rho}(\vb{q}) V_{\text{C}}(\vb{q}) \approx \chi_0 V_{\text{C}}(\vb{q}) \approx 16 \gg 1$. Qualitatively, large $\chi_0 V_{\text{C}}(\vb{q})$ produces an RPA potential $V_{\text{C}}^{\text{RPA}}(\vb{q}) \approx 1 / \chi_0$ that is momentum-independent, implying short-range Coulomb interactions. While not directly comparable due to the difference in momentum dependence, at a characteristic scale $q \approx 0.1 a^{-1}$ the RPA potential $V_{\text{C}}(\vb{q})^{\text{RPA}}$ matches $V_{\text{C}}(\vb{q})$ at corresponding $\epsilon_{\text{r}} \approx 70$. That is, the RPA interaction energy scale is considerably smaller than the screened Coulomb potential as benchmarked in our study (see \cref{fig:results-benchmarking,app-fig:results-benchmark-eps-15,app-fig:results-benchmark-eps-30}) which indicates $\epsilon_{\text{r}} \in [18, 25]$ approximately. The Coulomb energy scale in our study therefore lies in between the weak RPA interactions of Ref.~\cite{Chatterjee2022, Zhumagulov2023} and the stronger $\epsilon_{\text{r}} \approx 4.4$ interactions of Ref.~\cite{Huang2023}.

In the intervalley Hund's coupling $\smash{\hat{H}_{\text{V}}}$, the phase factors $\eta(\vb{q})_{\tau \sigma \sigma'}$ must be chosen to preserve $\text{C}_3$ rotation symmetry. The action of $\text{C}_3$ rotations on the fermionic operators is
\begin{equation}\begin{split}
    C_3 \left[ c^\dagger_{\tau s \sigma \vb{k}} \right] 
    = e^{i \tau \lambda_\sigma (2 \pi / 3)} 
        \left[ e^{i s^z \pi / 3} \right]_{ss} 
        c^\dagger_{\tau s \sigma (C_3 \vb{k})},
\end{split}\end{equation}
where $C_3 \vb{k}$ rotates $\vb{k}$ by $\SI{120}{\degree}$, and index $\lambda_\sigma = 0, +1, -1$ for the $A_1/B_3$, $B_1/A_2$ and $B_2/A_3$ sublattices $\sigma$ respectively. Applying these transformations to the operators in $\smash{\hat{H}_{\text{V}}}$ and demanding that $\smash{\hat{H}_{\text{V}}}$ is $\text{C}_3$-symmetric yields the following constraint on the phase factors,
\begin{equation}\begin{split}
    \eta(\vb{k})_{\tau \sigma \sigma'} = \eta(C_3 \vb{k})_{\tau \sigma \sigma'}
        e^{2 i \tau (\lambda_{\sigma'} - \lambda_\sigma) (2 \pi / 3)}.
    \label{app-eq:intervalley-hamiltonian-phase-factors-constraint}
\end{split}\end{equation}

We adopt a solution
\begin{equation}\begin{split}
    \eta(\vb{q})_{\tau \sigma \sigma'} = 
        \begin{dcases}
            e^{2 i \tau (\lambda_\sigma - \lambda_{\sigma'}) \theta_{\vb{q}}} & \vb{q} \neq \vb{0} \\
            \delta_{\sigma \sigma'} & \vb{q} = \vb{0},
        \end{dcases}
\end{split}\end{equation}
where $\theta_{\vb{k}} = \arg{(k_x + i k_y)}$, which satisfies the constraint (\cref{app-eq:intervalley-hamiltonian-phase-factors-constraint}) and fixes our gauge. Generically, while the short-range component of the Coulomb interaction is expected to give a ferromagnetic Hund's coupling ($J_{\text{H}} > 0$), which confers an energy advantage to aligned spins across valleys, lattice-scale effects as well as electron-phonon interactions may generate additional contributions; hence we treat $J_{\text{H}}$ as a phenomenological parameter. 

\clearpage 
\pagebreak

\section{Self-Consistent Hartree Fock methodology}
\label{app-sec:hartree-fock}

\subsection{Setup and overview}
\label{app-sec:hartree-fock/setup}

We focus on translation-symmetry preserving ground states in RTG, where the momentum $\vb{k}$ remains a good quantum number. The mean-field Hamiltonian is then diagonal in $\vb{k}$,
\begin{equation}\begin{split}
    \hat{H}_{\text{HF}} = \sum_{\vb{k}} \sum_{\alpha \alpha'} 
        H_{\text{HF}}(\vb{k})_{\alpha \alpha'} c^\dag_{\alpha \vb{k}} c_{\alpha' \vb{k}}.
\end{split}\end{equation}

The Hartree-Fock wavefunction $\ket{\Phi_{\text{HF}}}$ is accordingly identified as the (Slater determinant) ground-state of $\smash{\hat{H}_{\text{HF}}}$ at a mean-field energy $E$. 
We define the covariance matrix characterizing $\smash{\ket{\Phi_{\text{HF}}}}$ as
\begin{equation}\begin{split}
    \Delta(\vb{k})_{\alpha \alpha'}
    = \mel{\Phi_{\text{HF}}}{c^\dag_{\alpha \vb{k}} c_{\alpha' \vb{k}}}{\Phi_{\text{HF}}}.
\end{split}\end{equation}

Given a fermionic ground state $\ket{\Phi_{\text{HF}}}$, the covariance matrix $\Delta(\vb{k})$ is a Hermitian projector. The Hamiltonian we study is
\begin{equation}\begin{split}
    \hat{H} = \hat{H}_0 + \hat{H}_{\text{C}} + \hat{H}_{\text{V}} - \hat{H}_{\text{ref}},
    \label{app-eq:hamiltonian-overall}
\end{split}\end{equation}
where $\hat{H}_0 = \hat{H}_{\text{B}} + \hat{H}_{\text{I}}$ contains the non-interacting tight-binding and spin-orbit coupling components, $\smash{\hat{H}_{\text{C}}}$ and $\smash{\hat{H}_{\text{V}}}$ are the long- and short-range parts of the screened Coulomb interactions (see \cref{app-sec:hamiltonian}), and $\smash{\hat{H}_{\text{ref}}}$ is a reference Hamiltonian that is subtracted off to avoid a double-counting of interactions effects. Indeed, since $\smash{\hat{H}_{\text{B}}}$ is fitted to ab-initio calculations (see \cref{app-sec:hamiltonian/tight-binding-soc}), it already includes to an extent interaction effects at charge neutrality, and $\smash{\hat{H}_{\text{ref}}}$ is defined to cancel the double-counting of interactions by $\smash{\hat{H}_{\text{C}}}$ and $\smash{\hat{H}_{\text{V}}}$---we discuss the details shortly. Applying mean-field decoupling to $\smash{\hat{H}}$ yields a concrete form of $\smash{\hat{H}_{\text{HF}}}$, wherein $\smash{\hat{H}_{\text{C}}}$ and $\smash{\hat{H}_{\text{V}}}$ are replaced by effective single-body approximations dependent on the covariance matrix solution $\Delta$,
\begin{equation}\begin{split}
    \hat{H}_{\text{HF}}[\Delta] 
    &= \hat{H}_0 + \bar{H}_{\text{C}}[\Delta] + \bar{H}_{\text{V}}[\Delta] - \hat{H}_{\text{ref}}
    = \sum_{\vb{k}}
        \underbrace{\left[ \hat{H}_0(\vb{k})
            + \bar{H}_{\text{C}}[\Delta](\vb{k})
            + \bar{H}_{\text{V}}[\Delta](\vb{k}) 
            - \hat{H}_{\text{ref}}(\vb{k}) \right]}
            _{H_{\text{HF}}[\Delta](\vb{k})}.
\end{split}\end{equation}

The long-range part of Coulomb interactions $\smash{\hat{H}_{\text{C}}}$ splits into Hartree and Fock terms, arising from the diagonal and exchange decoupling channels respectively,
\begin{equation}\begin{split}
    \bar{H}_{\text{C}}[\Delta](\vb{k})
        &= \left(\bar{H}_{\text{C}}\right)^{\text{hart}}[\Delta](\vb{k}) +
            \left(\bar{H}_{\text{C}}\right)^{\text{fock}}[\Delta](\vb{k}), \\
    \left(\bar{H}_{\text{C}}\right)^{\text{hart}}[\Delta](\vb{k})
        &= \frac{N}{A} \sum_{\alpha} 
            V_{\text{C}}(\vb{0}) 
            c^\dag_{\alpha \vb{k}} c_{\alpha \vb{k}}, \\
    \left(\bar{H}_{\text{C}}\right)^{\text{fock}}[\Delta](\vb{k})
        &= - \frac{1}{A} \sum_{\vb{q}} \sum_{\alpha \alpha'} 
            V_{\text{C}}(\vb{q}) 
            \left[\trans{\Delta(\vb{k} + \vb{q})}\right]_{\alpha \alpha'} 
            c^\dag_{\alpha \vb{k}} c_{\alpha' \vb{k}},
\end{split}\end{equation}
where $N = \sum_{\vb{k}} \tr \Delta(\vb{k})$ is the total number of electrons in the system. The Hartree term represents a uniform background Coulomb potential arising from the average electron density, whereas the Fock term describes involves momentum transfer. Likewise, the intervalley Hund's coupling Hamiltonian $\smash{\hat{H}_{\text{V}}}$ splits into Hartree- and Fock-like contributions,
\begin{equation}\begin{split}
    \bar{H}_{\text{V}}[\Delta](\vb{k})
        &= \left(\bar{H}_{\text{V}}\right)^{\text{hart}}[\Delta](\vb{k}) +
            \left(\bar{H}_{\text{V}}\right)^{\text{fock}}[\Delta](\vb{k}), \\
    \left(\bar{H}_{\text{V}}\right)^{\text{hart}}[\Delta](\vb{k})
        &= \frac{J_{\text{H}}}{A} \sum_{\tau s \sigma}
            \left[ \tr_s \Delta^{\tau (-\tau)} \right]_{\sigma \sigma}
            \, c_{(-\tau) s \sigma \vb{k}}^\dag c_{\tau s \sigma \vb{k}}, \\
    \left(\bar{H}_{\text{C}}\right)^{\text{fock}}[\Delta](\vb{k})
        &= - \frac{J_{\text{H}}}{A} \sum_{\vb{q}} \sum_{\tau} \sum_{ss'} \sum_{\sigma \sigma'}
        \eta(\vb{q})_{\tau \sigma \sigma'}
        \left[\trans{\Delta^{\tau \tau}(\vb{k} + \vb{q})}\right]_{(s \sigma)(s' \sigma')} 
            \, c_{(-\tau) s \sigma \vb{k}}^\dag c_{(-\tau) s' \sigma' \vb{k}},
\end{split}\end{equation}
where, for convenience, we have defined $\smash{\Delta^{\tau \tau'}_{(s \sigma) (s' \sigma')} = \sum_{\vb{k}} \Delta(\vb{k})_{(\tau s \sigma) (\tau' s' \sigma')}}$ to be the $\tau \tau'$ valley sector of the covariance matrix traced over all $\vb{k}$-points, and $\tr_s$ denotes the partial trace over the spin degrees of freedom. As expressed above, the Hartree contribution is manifestly of an intervalley character. We then define the reference Hamiltonian
\begin{equation}\begin{split}
    \hat{H}_{\text{ref}}(\vb{k}) = 
        \frac{1}{2} \bar{H}_{\text{C}}[\Delta_{\text{ref}}](\vb{k}) +
        \frac{1}{2} \bar{H}_{\text{V}}[\Delta_{\text{ref}}](\vb{k}),
\end{split}\end{equation}
where the reference covariance matrix $\Delta_{\text{ref}}$ is the fully symmetric non-interacting ground state at charge neutrality.

The mean-field (Hartree-Fock) energy of a state $\ket{\Phi_{\text{HF}}}$ characterized by covariance matrix $\Delta$ is then
\begin{equation}\begin{split}
    E = \bra{\Phi_{\text{HF}}}{\hat{H}}\ket{\Phi_{\text{HF}}}
    = \sum_{\vb{k}} \tr\left[
        \trans{\Delta(\vb{k})}
        \left( \hat{H}_0(\vb{k})
            + \frac{1}{2} \bar{H}_{\text{C}}[\Delta](\vb{k})
            + \frac{1}{2} \bar{H}_{\text{V}}[\Delta](\vb{k})
            - \hat{H}_{\text{ref}}(\vb{k}) \right) \right],
    \label{app-eq:hartree-fock-energy}
\end{split}\end{equation}
where the factors of $1/2$ on the interacting Hamiltonians arise from Wick's theorem, that decouples two-body energy expectations into one-body contributions. Thus $\smash{\hat{H}_{\text{ref}}}$ indeed cancels energetic contributions from interactions at the charge-neutral reference point $\Delta = \Delta_{\text{ref}}$ as desired.

The self-consistent Hartree-Fock method computes a solution for $\ket{\Phi_{\text{HF}}}$ defining the covariance matrix $\Delta$, such that $\ket{\Phi_{\text{HF}}}$ is the ground-state of $\smash{\hat{H}_{\text{HF}}[\Delta]}$. The covariance matrix $\Delta$ characterizes the many-electron ground-state of the mean-field Hamiltonian $\smash{\hat{H}_{\text{HF}}[\Delta]}$; its construction therefore involves a projection onto the subspace spanned by filled electronic states. In a grand-canonical ensemble picture, electronic states are filled up to a pinned chemical potential $\mu$. In a canonical ensemble approach, electronic states are filled to match the desired carrier density; the dressed chemical potential, automatically including effects of renormalization due to the interactions, is thereby determined. We use the latter approach as it allows for convenient simulation sweeps across carrier densities.

\subsection{Fixed-point iteration and restricted symmetry-breaking}
\label{app-sec:hartree-fock/iteration}

We implement self-consistent Hartree-Fock through a variation of fixed-point iteration~\cite{fabrizio2022}. Conceptually, one begins with an initial $\Delta$ ansatz satisfying desired symmetry properties---\eg~fully symmetric metal, valley- or spin-polarized, intervalley coherent, \etc---constructs $\smash{\hat{H}_{\text{HF}}[\Delta]}$ and computes its ground state, revises $\Delta$ based on the updated ground state, and repeats until convergence. It is, however, non-trivial in practice to construct an initial guess $\Delta$ that satisfies all required properties of a fermionic covariance matrix, breaks only the desired symmetries, and is reasonably close to the true ground state within the symmetry sector of interest. Instead, it is generally easier to impose a transient perturbation $\smash{\hat{H}_\delta}$ to the Hamiltonian that encourages the breaking of desired symmetries; the symmetry-breaking is then inherited by the resultant ground state $\Delta$. In fact, for robustness and speed of convergence of the Hartree-Fock procedure, it is advantageous to apply a sequence of small perturbations $\smash{\hat{H}_\delta}$ at multiple scheduled time steps\footnote{The idea is similar to simulated annealing in numerical optimization.}, rather than a single larger perturbation $\smash{\hat{H}_\delta}$ at the beginning of the procedure. 

Concretely, let $t \in \mathbb{N}$ label time steps. We start with $\Delta_0$, the ground state of the non-interacting Hamiltonian $\smash{\hat{H}_0}$ at the desired carrier density. Then for $t \geq 0$, we consider the mean-field Hamiltonian
\begin{equation}\begin{split}
    (\hat{H}_{\text{HF}})_t = \hat{H}_{\text{HF}}[\Delta_t] + A_t \hat{H}_\delta,
    \label{app-eq:hartree-fock/setup/iteration}
\end{split}\end{equation}
and $\Delta_{t+1}$ is computed as the ground state of $\smash{(\hat{H}_{\text{HF}})_t}$ with associated mean-field energy $E_{t+1}$---see \cref{app-sec:hartree-fock/ground-state} for details on the computation of ground-state covariance matrices. Here $A_t$ are perturbation amplitudes, nonzero and decreasing in magnitude for a scheduled sequence of time steps $t \in T_\delta$, and zero for all $t > t_\delta$. The iteration continues---$\Delta_{t+1}$ yields $\smash{(\hat{H}_{\text{HF}})_{t+1}}$ by \cref{app-eq:hartree-fock/setup/iteration}, which produces $\Delta_{t+2}$ and $E_{t+2}$, henceforth---and terminates at time step $t > t_\delta$ when
\begin{equation}\begin{split}
    \max_{\alpha \alpha'} \abs{\left(\Delta_t - \Delta_{t-1}\right)_{\alpha \alpha'}} < \epsilon_\Delta 
    \qquad \text{and} \qquad 
    \abs{E_t - E_{t-1}} < \epsilon_{\text{e}} \abs{E_{t-1}},
\end{split}\end{equation}
for small convergence tolerances $\epsilon_\Delta$ and $\epsilon_{\text{e}}$. In our work we adopt
\begin{equation}\begin{split}
    A_t = \begin{dcases}
        1/4 & t = 0 \\
        1/8 & t = 6 \\
        1/16 & t = t_\delta = 12 \\
        0 & \text{all other} \,\, t,
    \end{dcases}
\end{split}\end{equation}
and tolerances $\epsilon_\Delta \approx 10^{-4}$ and $\epsilon_{\text{e}} \approx 10^{-8}$, more than sufficient to resolve the pertinent fine structure and degeneracy-breaking in our simulation sweeps. Note that standard machine precision presents ${\sim} 10^{-16}$ relative uncertainty per operation, which accumulates to ${\sim} 10^{-10}$ relative uncertainty per computed scalar on our RTG system ($24$ bands per $\vb{k}$-point across ${\sim} 10^3$ $\vb{k}$-points). For the vast majority of cases, convergence is achieved within $t \leq 200$. We summarize the symmetry-breaking Hamiltonian perturbations $\smash{\hat{H}_\delta}$ we used in \cref{app-table:hartree-fock/iteration/perturbations}.

Constructing $\smash{\hat{H}}_\delta$ with no $\text{C}_3$ symmetry requirement is straightforward---$\smash{\hat{H}}_\delta(\vb{k})$ at each $\vb{k}$-point can be initialized uniformly, as specified in \cref{app-table:hartree-fock/iteration/perturbations}. On the other hand, constructing $\smash{\hat{H}_\delta}$ that are $\text{C}_3$-symmetric requires one to impose the following gauge-fixing condition,
\begin{equation}\begin{split}
    \hat{H}_\delta(C_3 \vb{k}) = U_{C_3} \hat{H}_\delta(\vb{k}) U_{C_3}^\dagger, \qquad
    U_{C_3} = U_{C_3}^{\tau \sigma} U_{C_3}^s,
    \label{app-eq:hartree-fock/iteration/hamiltonian-C3-transform}
\end{split}\end{equation}
where $\smash{U_{C_3}^{\tau \sigma}}$ and $\smash{U_{C_3}^s}$ are rotation unitaries acting on valley-sublattice and spin degrees of freedom respectively,
\begin{equation}\begin{split}
    U_{C_3}^{\tau \sigma} = \exp[i \tau^z s^0 \left(\frac{2 \pi}{3}\right) \diag{\mqty(0 & 0 & 1 & 1 & -1 & -1)}], \qquad
    U_{C_3}^s = \exp[i s^z \left(\frac{\pi}{3}\right)].
\end{split}\end{equation}

\begin{table}[!t]
    \centering
    \begin{tabular}{p{7.5cm} p{4cm} p{2cm} p{2.5cm} p{2cm}}
        \toprule 
        Order Description & $\smash{\hat{H}_\delta}(\vb{k}) / \gamma_1$ & Degeneracy & $\text{C}_3$ \\
        \midrule 
        Fully symmetric & $0$ 
            & $4$ & \cmark \\
        Valley-polarized & $\tau^z$ 
            & $2$ & \cmark \\
        Spin-polarized (out-of-plane) & $s^z$ 
            & $2$ & \cmark \\
        Spin-polarized (in-plane) & $s^{x/y}$ 
            & $2$ & \xmark \\
        Spin-valley-locked & $\tau^z s^z$ 
            & $2$ & \cmark \\
        Intervalley-coherent spin-singlet & $\tau^x$ 
            & $2$ & \cmark \\
        Intervalley-coherent spin-triplet (out-of-plane) & $\tau^x s^z$ 
            & $2$ & \cmark \\
        Intervalley-coherent spin-triplet (in-plane) & $\tau^x s^{x/y}$ 
            & $2$ & \xmark \\
        Spin-valley-polarized & $(\tau^0 \pm \tau^z) (s^0 \pm s^z) / 4$ 
            & $1$ & \cmark \\
        Nematic ($y^2$) & $\tanh(k_y^2 / b^2)$ 
            & - & \xmark \\
        Nematic ($x$) & $\tanh(k_x / b)$ 
            & - & \xmark \\
        \bottomrule
    \end{tabular}
    \caption{Transient symmetry-breaking Hamiltonian perturbations $\smash{\hat{H}_\delta}$ used in self-consistent Hartree-Fock, specified up to phase factors (arising from transformation properties under $\text{C}_3$ rotations). The tight-binding interlayer hopping amplitude $\gamma_1$ (see \cref{app-sec:hamiltonian/tight-binding-soc}) sets the band gap between valence/conduction and lower occupied/higher unoccupied bands in hole-/electron-doped RTG, and is used as an energy scale for $\smash{\hat{H}_\delta}$ perturbations. For each order and associated $\smash{\hat{H}_\delta(\vb{k})}$, we present the band degeneracy within the four spin-valley sectors (with $g = 4$ representing a fully degenerate state), and whether the perturbation preserves $\text{C}_3$ rotation symmetry, which can be broken either due to an in-plane spin component or an orbital component in the case of nematicity. More complex composite orders (\eg~intervalley coherence with nematicity) are realized by superimposing multiple $\smash{\hat{H}_\delta}$ perturbations. For nematic perturbations, $b = 0.1 a^{-1}$ sets a momentum scale and corresponds to typical Fermi momenta.}
    \label{app-table:hartree-fock/iteration/perturbations}
\end{table}

\subsection{Computation of ground state covariance matrices}
\label{app-sec:hartree-fock/ground-state}

A single-body Hamiltonian $\widetilde{H}$ diagonal in $\vb{k}$ can be diagonalized as
\begin{equation}\begin{split}
    \widetilde{H}(\vb{k}) = \sum_\omega \varepsilon_{\omega \vb{k}} \phi^\dag_{\omega \vb{k}} \phi_{\omega \vb{k}}, \qquad 
    \phi^\dag_{\omega \vb{k}} = \sum_\alpha v(\vb{k})_{\alpha \omega} c^\dag_{\alpha \vb{k}}, \qquad 
    c^\dag_{\alpha \vb{k}} = \sum_{\omega} v(\vb{k})_{\alpha \omega}^* \phi^\dag_{\omega \vb{k}},
\end{split}\end{equation}
where $\omega$ enumerates the bands, $\phi_{\omega \vb{k}}$ are band fermionic operators, and $\varepsilon_{\omega \vb{k}}$ and $v(\vb{k})_{\alpha \omega}$ are respectively eigenenergies and normalized wavefunction coefficients. In our self-consistent Hartree-Fock procedure (\cref{app-sec:hartree-fock/iteration}), the Hamiltonian $\smash{\widetilde{H}}$ for which the ground state is of interest is either $\smash{\hat{H}_0}$ for initialization or $\smash{(\hat{H}_{\text{HF}})_t}$ as the iterations proceed. Consider sorting  the eigenenergies $\varepsilon_{\omega \vb{k}}$ to produce $\varepsilon_{\omega_1 \vb{k}_1} \leq \varepsilon_{\omega_2 \vb{k}_2} \leq \ldots$. Then, for an $N$-electron system, one straightforwardly identifies the Fermi energy $\varepsilon_{\text{F}} = \varepsilon_{\omega_N \vb{k}_N}$ and the occupied subspace of electron states $I = \{\omega_1 \vb{k}_1, \omega_2 \vb{k}_2, \ldots, \omega_N \vb{k}_N\}$. The $N$-electron Slater determinant ground state of $\smash{\widetilde{H}}$ is then
\begin{equation}\begin{split}
    \ket{\Phi_{\text{HF}}} = \prod_{\omega \vb{k} \in I} \phi^\dag_{\omega \vb{k}} \ket{0},
\end{split}\end{equation}
for electronic vacuum state $\ket{0}$. The ground state covariance matrix characterizing $\ket{\Phi_{\text{HF}}}$ is 
\begin{equation}\begin{split}
    \Delta(\vb{k})_{\alpha \alpha'}
    = \mel{\Phi_{\text{HF}}}{c^\dagger_{\alpha \vb{k}} c_{\alpha' \vb{k}}}{\Phi_{\text{HF}}}
    = \sum_{\omega: \, \omega \vb{k} \in I} v(\vb{k})_{\alpha \omega}^* v(\vb{k})_{\alpha' \omega}.
    \label{app-eq:hartree-fock/ground-state/covmat-solution-integer}
\end{split}\end{equation}

The above summarizes the standard construction of $\ket{\Phi_{\text{HF}}}$ and $\Delta(\vb{k})$ in a grand-canonical ensemble setting. However, a naïve application of this construction can lead to anomalous symmetry-breaking. The problem is that there may exist degeneracies at the Fermi level, and the selection of $\omega_N \vb{k}_N$ within this degenerate level need not preserve symmetries. More concretely, suppose $\smash{\varepsilon_{\omega_{m_-} \vb{k}_{m_-}} = \ldots = \varepsilon_{\omega_{m_+} \vb{k}_{m_+}}}$ is a degenerate level, and $m_- \leq N < m_+$ such that the level coincides with the Fermi energy. Then the selection of occupied electron states $\{\omega_{m_-} \vb{k}_{m_-}, \ldots, \omega_N \vb{k}_N\} \subset I$ can break symmetries---for example, the selection may comprise an unbalanced number of states in the two valleys, thus producing valley polarization. Note that $\omega_N \vb{k}_N$ can be arbitrarily chosen within the degenerate level up to reshuffling of states; thus the resulting symmetry-broken $\ket{\Phi_{\text{HF}}}$ and $\Delta(\vb{k})$ are non-unique and are all degenerate, none of which provides an energy advantage over a symmetry-unbroken solution. This kind of symmetry-breaking is unphysical.

A reasonable treatment is to identify the Hartree-Fock ground state $\Delta(\vb{k})$ as the ensemble average of the possible degenerate symmetry-broken solutions, each arising from a particular selection of occupied electron states at the Fermi level. This averaged $\Delta(\vb{k})$ then does not exhibit any anomalous symmetry-breaking. We write
\begin{equation}\begin{split}
    I_{\text{occ}} &= \left\{\omega_\ell \vb{k}_\ell : \, 
        \varepsilon_{\omega_\ell \vb{k}_\ell} < \varepsilon_{\omega_N \vb{k}_N}
        - \epsilon_{\text{g}} \Delta \varepsilon \right\}, \\ 
    I_{\text{avg}} &= \left\{\omega_\ell \vb{k}_\ell : \, 
        \varepsilon_{\omega_N \vb{k}_N}
        - \epsilon_{\text{g}} \Delta \varepsilon
        \leq \varepsilon_{\omega_\ell \vb{k}_\ell} \leq 
        \varepsilon_{\omega_N \vb{k}_N}
        + \epsilon_{\text{g}} \Delta \varepsilon\right\}, \\
    \Delta(\vb{k})_{\alpha \alpha'}
        &= \sum_{\omega: \, \omega \vb{k} \in I_{\text{occ}}} v(\vb{k})_{\alpha \omega}^* v(\vb{k})_{\alpha' \omega}
        + \frac{N - \abs{I_{\text{occ}}}}{\abs{I_{\text{avg}}}} 
        \sum_{\omega: \, \omega \vb{k} \in I_{\text{avg}}} v(\vb{k})_{\alpha \omega}^* v(\vb{k})_{\alpha' \omega},
    \label{app-eq:hartree-fock/ground-state/covmat-solution-fractional}
\end{split}\end{equation}
where $\epsilon_{\text{g}}$ is a small relative tolerance for detection of the degenerate level and $\Delta \varepsilon = \varepsilon_{\omega_N \vb{k}_N} - \varepsilon_{\omega_1 \vb{k}_1}$ sets the energy scale. Defined in this manner, $\Delta(\vb{k})$ is no longer a projector, in contrast to the solution in \cref{app-eq:hartree-fock/ground-state/covmat-solution-integer}, and does not correspond to any unique pure state $\ket{\Phi_{\text{HF}}}$; rather it is an ensemble-averaged mixed state. We refer to this construction also as the fractional filling scheme. We use $\epsilon_{\text{g}} \approx 10^{-10}$ in our work, which approaches numerical precision in the diagonalization of our Hamiltonians.

A slight further subtlety arises for self-consistent Hartree-Fock runs targeting the fully symmetric sector, wherein it is desired that the ground state $\Delta$ breaks no symmetries spontaneously. These calculations are ubiquitous in our study, as we assess all Hartree-Fock energies against fully symmetric solutions (see \eg~\cref{fig:energy-slice} of the main text). Throughout much of the parameter space explored---\ie~carrier densities, interaction and SOC strengths---spontaneous symmetry-broken ground states are energetically favorable, and thus the fully symmetric solution is unstable to perturbations. Numerical imprecision that minutely break degeneracies can lead to a proliferation of symmetry-breaking artifacts as the Hartree-Fock iterations proceed. Moreover, at $\lambda_{\text{I}} \neq 0$ the non-interacting $\smash{\hat{H}_0}$ already introduces a spin-valley splitting, such that spin-valley locking always confers an energy advantage and fully symmetric solutions cannot naturally arise through the iteration procedure. 

To enable and stabilize Hartree-Fock runs targeting fully symmetric states, we project out symmetry-breaking components of $\Delta$ that are possibly present before proceeding to the next iteration. Given a set of operators $P = \smash{\left\{\tau^{\mu_\ell} s^{\nu_\ell}\right\}_{\ell}}$ acting on spin-valley degrees of freedom, the projection of $\Delta$ onto the space spanned by $P$ can be written
\begin{equation}\begin{split}
    \mathbb{P}\left[\left\{\tau^{\mu_\ell} s^{\nu_\ell}\right\}_{\ell}\right] \Delta(\vb{k})
    = \frac{1}{4} \sum_\ell \tau^{\mu_\ell} s^{\nu_\ell} \otimes 
        \tr_{\tau s} \left[ \tau^{\mu_\ell} s^{\nu_\ell} \Delta(\vb{k}) \right],
\end{split}\end{equation}
where $\tr_{\tau s}$ denotes partial trace over spin-valley degrees of freedom. Then $\smash{\mathbb{P}\left[\left\{\tau^0 s^0, \tau^z s^0\right\}\right] \Delta(\vb{k})}$ gives the fully symmetric $\Delta$ with symmetry-breaking components removed. The $\tau^0 s^0$ component encodes overall filling of electron states, and $\tau^z s^0$ accommodates local valley polarization at each $\vb{k}$-point, which occurs naturally even in fully symmetric states, as the Fermi surfaces in the two valleys are mirror-images (reflected about $\vu{y}$) of each other and do not exactly coincide on the $\vb{k}$-grid. The global valley polarization traced over all $\vb{k}$-points, of course, vanishes for fully symmetric states, $\smash{\sum_{\vb{k}} \mathbb{P}\left[\left\{\tau^z s^0\right\}\right] \Delta(\vb{k}) = 0}$. We do not invoke this projection scheme for Hartree-Fock runs targeting other symmetry sectors.

\subsection{Momentum grid}
\label{app-sec:hartree-fock/momentum-grid}

The primitive lattice vectors $\vb{a}_j$ and reciprocal lattice vectors $\vb{b}_j$ for graphene can be chosen to be
\begin{equation}\begin{split}
    \vb{a}_1 = \left(\frac{a}{2}, \frac{\sqrt{3} a}{2}\right), \qquad 
    \vb{a}_2 = \left(\frac{a}{2}, -\frac{\sqrt{3} a}{2}\right), \qquad 
    \vb{b}_1 = \left(\frac{2 \pi}{a}, \frac{2 \pi}{\sqrt{3} a}\right), \qquad 
    \vb{b}_2 = \left(\frac{2 \pi}{a}, -\frac{2 \pi}{\sqrt{3} a}\right),
\end{split}\end{equation}
where the lattice constant $a = \SI{2.46}{\angstrom}$. An $L \times L$ grid of unit cells (in real space) then produces the microscopic momentum grid
\begin{equation}\begin{split}
    \vb{q} \in \left\{ \frac{m_1 \vb{b}_1}{L} + \frac{m_2 \vb{b}_2}{L} \right\},
\end{split}\end{equation}
where $m_1$, $m_2$ are integers that run from $-L/2$ to $L/2$, and $\vb{q} = \vb{0}$ is the $\Gamma$-point of the Brillouin zone. The sampled area is then $A = L^2 A_{\text{uc}}$ for unit cell area $A_{\text{uc}} \approx \SI{0.0524}{\nano\meter\squared}$. To capture the relevant low-energy physics, it suffices to retain only momenta near the two valley points where the Fermi surfaces lie. Imposing a circular momentum cutoff $\Lambda$, the retained valley-centered momentum grid is
\begin{equation}\begin{split}
    \vb{k} \in \left\{ \vb{k} = \frac{n_1 \vb{b}_1}{L} + \frac{n_2 \vb{b}_2}{L} : 
    \, n_1 \in \mathbb{Z}, 
    \, n_2 \in \mathbb{Z},
    \, \abs{\vb{k}} \leq \Lambda \right\}.
\end{split}\end{equation}

Since $\vb{K} = (\vb{b}_1 + \vb{b}_2) / 3$, we choose $L$ to be a multiple of $3$, such that the $\vb{k}$-grid maps onto the microscopic grid under translation---$\forall \vb{k} \, \exists \vb{q}: \vb{k} + \tau \vb{K} = \vb{q}$ for both valleys $\tau \in \{\pm 1\}$---and $\vb{k} = \vb{0}$ correspond to the valley points. Under this definition the local momentum grid around both valleys coincide, and each $\vb{k}$-point can encode the degrees of freedom of both valleys. That is, each $\vb{k}$-point carries a $(2\text{ valley}) \times (2\text{ spin}) \times (6\text{ sublattice}) = 24$-dimensional local Hilbert space.

The maximum Fermi momenta found in our interacting symmetry-restricted ground states do not exceed $k_{\text{F}} \approx 0.12 a^{-1}$, therefore in principle a momentum cutoff $\Lambda \approx 0.18 a^{-1}$ suffices. But as the Fermi surfaces can be much smaller in some cases, especially at low carrier densities, setting a universal large $\Lambda$ is wasteful. Instead we adopt a semi-adaptive scheme for our momentum grid. We precompute the non-interacting ground state $\Delta_0$ (using $\smash{\hat{H}_0}$) at each carrier density of interest and acquire the non-interacting Fermi momentum $k^0_{\text{F}}$, and set a carrier-density dependent $\Lambda = \max(\Lambda_0, r k^0_{\text{F}})$ for $\Lambda_0 \approx 0.12 a^{-1}$ and $r \approx 1.5$, such that the momentum cutoff scales with the size of the non-interacting Fermi surfaces. Then, given a target number $K \sim 10^3$ of $\vb{k}$-points, which is held fixed for a simulation sweep, appropriate $L$ can be chosen that adapts to carrier densities. Thus the computational cost of our self-consistent Hartree-Fock procedure is largely independent of carrier density and the size of resultant Fermi surfaces, and is dependent only on the momentum grid resolution $K$.

\subsection{Convergence and stability checks}
\label{app-sec:hartree-fock/convergence-stability-checks}

Inaccurate solutions from self-consistent Hartree-Fock can result from unsuitable momentum grid settings---namely, an insufficient number $K$ of $\vb{k}$-points, thus presenting a momentum grid too coarse to capture pertinent detail of the Fermi surfaces; or too small a momentum cutoff $\Lambda$, which introduces truncation error in the interacting Hamiltonians (that involve momentum transfer) and in extreme cases may cut off parts of Fermi surfaces. Separately, there is also a possibility that certain broken symmetries are missed in the set of Hamiltonian perturbations chosen, and as a result lower-energy solutions harboring those broken symmetries remain unobserved in the Hartree-Fock runs.

To verify that our Hartree-Fock calculations were performed on suitably large momentum grids, we ran subsets of our simulations across the different considered scenarios (\ie~with and without Hund's coupling, and with and without Ising SOC) on significantly larger momentum grids---in particular ${\sim} 1.5 \times$ larger $K$ and momentum cutoff $\Lambda$ than in our presented results (\cref{fig:results-wo-hunds-wo-soc,fig:results-w-hunds-wo-soc,fig:results-wo-hunds-w-soc,fig:results-w-hunds-w-soc}). The phase diagrams obtained from these verification runs matched with our presented results. To check that no lower-energy broken-symmetry ground state is missed in our Hartree-Fock calculations, we repeatedly impose random symmetry-breaking perturbations on each identified ground state and run again until convergence. In these checks, the randomly perturbed ansatzes do not exhibit energy advantages relative to the original ground state upon convergence.

\clearpage 
\pagebreak

\section{Further results and analyses}
\label{app-sec:results}

\subsection{Benchmarking against experiments}
\label{app-sec:results/benchmarking}

In \cref{sec:benchmarking} of the main text, we presented a benchmarking of our self-consistent Hartree-Fock phase diagrams against prior experiments~\cite{zhou2021half} at $\epsilon_{\text{r}} = 20$. Here in \cref{app-fig:results-benchmark-eps-15,app-fig:results-benchmark-eps-30}, we show additional comparisons with experiment phase boundaries at $\epsilon_{\text{r}} = 15$ and $\epsilon_{\text{r}} = 30$ and across a range of $J_{\text{H}}$ interaction strengths. 

\begin{figure}[!h]
    \centering
    \includegraphics[width = \linewidth]{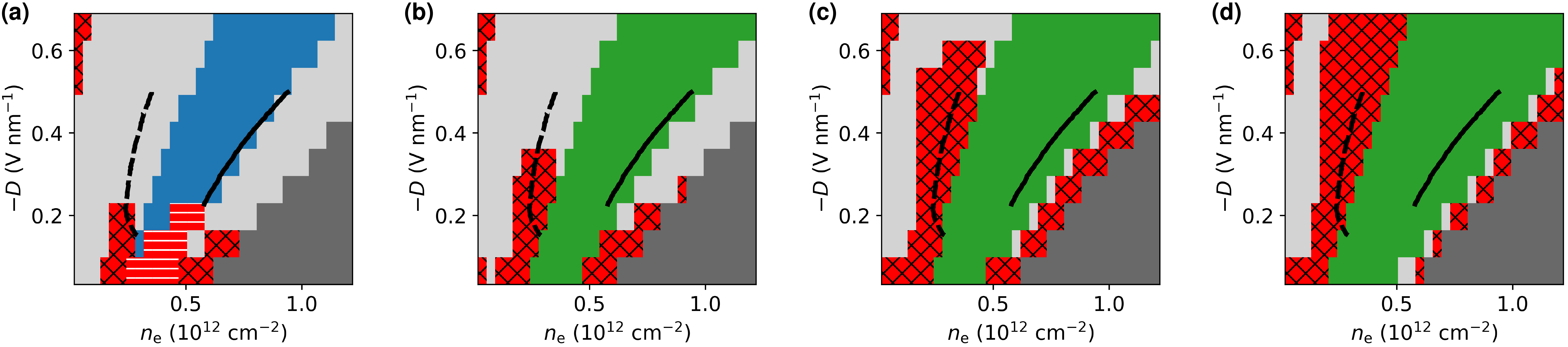}
    \phantomsubfloat{\label{app-fig:results-benchmark-eps-15-Jh-00}}
    \phantomsubfloat{\label{app-fig:results-benchmark-eps-15-Jh-04}}
    \phantomsubfloat{\label{app-fig:results-benchmark-eps-15-Jh-08}}
    \phantomsubfloat{\label{app-fig:results-benchmark-eps-15-Jh-12}}
    \vspace{-1.8\baselineskip}
    \caption{\textbf{Additional comparisons of Hartree-Fock phase diagrams to prior experiments.} Phase diagrams of electron-doped RTG in the electron density--displacement field ($n_{\text{e}}$--$D$) parameter space, at $\epsilon_{\text{r}} = 15$ and \textbf{(a)} $J_{\text{H}} = 0$, \textbf{(b)} $J_{\text{H}} = \SI{4}{\electronvolt\ucdot\unitcellarea}$, \textbf{(c)} $J_{\text{H}} = \SI{8}{\electronvolt\ucdot\unitcellarea}$, \textbf{(d)} $J_{\text{H}} = \SI{12}{\electronvolt\ucdot\unitcellarea}$. Dashed lines denote phase boundaries between a quarter-metal and a spin-polarized half-metal phase, and solid lines denote phase boundaries into the fully symmetric metal. Experiment data reproduced from Ref.~\cite{zhou2021half}. An out-of-plane $\epsilon_{\text{r}}^\perp = 4.4$ is used to convert between interlayer potential $\Delta_1$ and displacement field $D$ (see \cref{eq:Delta1-def}).}
    \label{app-fig:results-benchmark-eps-15}
\end{figure}

\begin{figure}[!h]
    \centering
    \includegraphics[width = \linewidth]{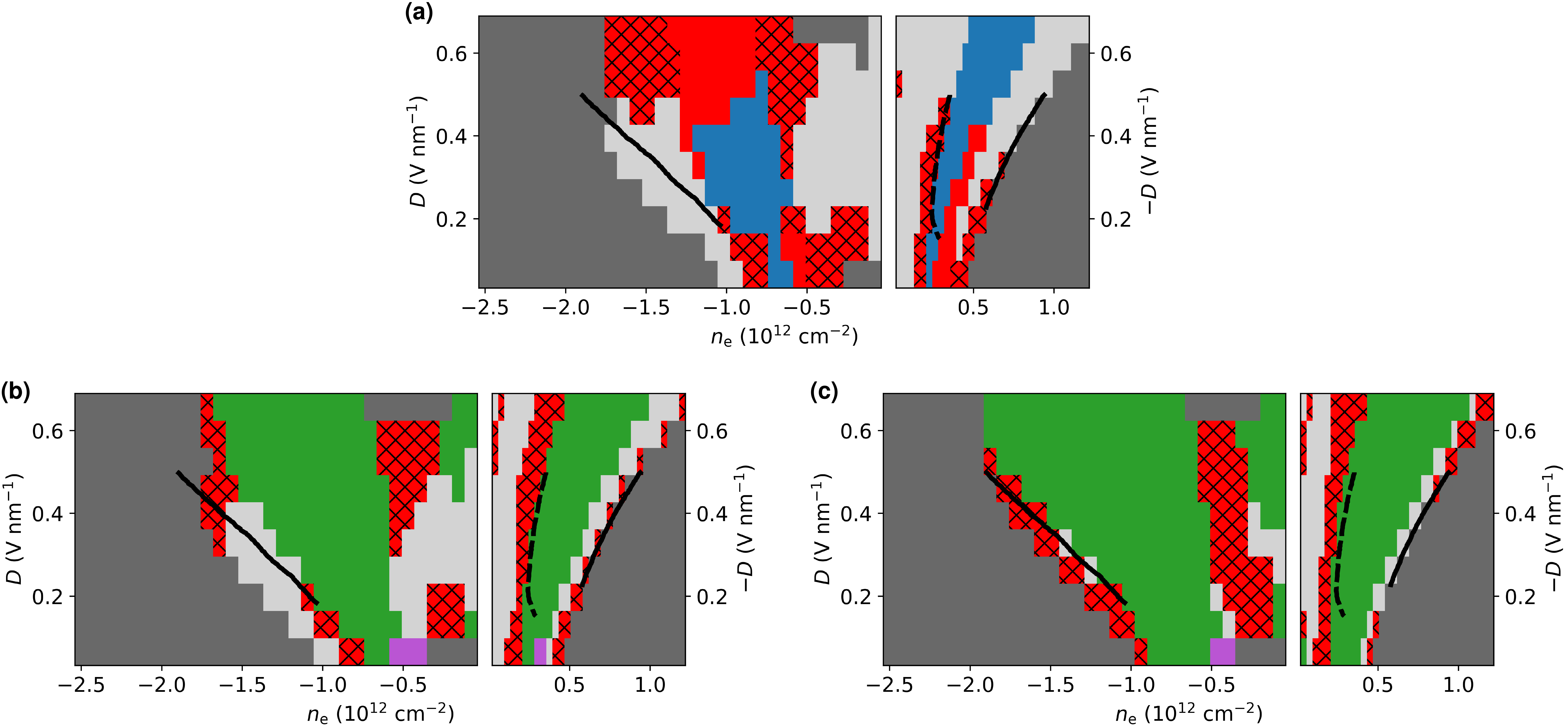}
    \phantomsubfloat{\label{app-fig:results-benchmark-eps-30-Jh-00}}
    \phantomsubfloat{\label{app-fig:results-benchmark-eps-30-Jh-04}}
    \phantomsubfloat{\label{app-fig:results-benchmark-eps-30-Jh-08}}
    \vspace{-1.8\baselineskip}
    \caption{\textbf{Additional comparisons of Hartree-Fock phase diagrams to prior experiments.} RTG phase diagrams in the electron density--displacement field ($n_{\text{e}}$--$D$) parameter space in hole- and electron-doped regimes, at $\epsilon_{\text{r}} = 30$ and \textbf{(a)} $J_{\text{H}} = 0$, \textbf{(b)} $J_{\text{H}} = \SI{4}{\electronvolt\ucdot\unitcellarea}$, \textbf{(c)} $J_{\text{H}} = \SI{8}{\electronvolt\ucdot\unitcellarea}$. In the hole-doped regime, solid lines denote experimental phase boundaries between a fully symmetric and a partially polarized $g = 2$ phase. In the electron-doped regime, dashed lines denote phase boundaries between a quarter-metal and a spin-polarized half-metal phase, and solid lines denote phase boundaries into the fully symmetric metal. Experiment data reproduced from Ref.~\cite{zhou2021half}. An out-of-plane $\epsilon_{\text{r}}^\perp = 4.4$ is used to convert between interlayer potential $\Delta_1$ and displacement field $D$ (see \cref{eq:Delta1-def}).}
    \label{app-fig:results-benchmark-eps-30}
\end{figure}

\end{widetext}

\end{document}